\documentclass{aastex631}

\usepackage{float}
\usepackage{amsmath}
\usepackage{tasks}
\usepackage{booktabs}
\usepackage{rotating}
\usepackage{makecell}
\usepackage{xspace}
\usepackage{enumitem}
\setlist[itemize]{noitemsep}

\shorttitle{Retrospective Search for Strongly Lensed Supernovae in the DESI Legacy Imaging Surveys}
\shortauthors{Sheu et al.}

\graphicspath{{./}{figures/}}

\newcommand{\IaBpeak}{\ensuremath{-19.25}\xspace}
\newcommand{\ipg}{\emph{g}\xspace}
\newcommand{\ipr}{\emph{r}\xspace}
\newcommand{\ipi}{\emph{i}\xspace}
\newcommand{\ipz}{\emph{z}\xspace}

\newcommand{\ipB}{\emph{B}\xspace}

\newcommand{\txr}{\textcolor{black}}
\newcommand{\txb}{\textcolor{black}}
\newcommand{\txg}{\textcolor{black}}

\newcommand{\trr}{\textcolor{black}}

\begin{document}

\title{Retrospective Search for Strongly Lensed Supernovae in the DESI Legacy Imaging Surveys}

\correspondingauthor{William Sheu, Xiaosheng Huang}
\email{wsheu@astro.ucla.edu, xhuang22@usfca.edu}

\author[0000-0003-1889-0227]{William Sheu}
\affiliation{Department of Physics \& Astronomy, University of California, Los Angeles \\ 430 Portola Plaza, Los Angeles, CA 90095, USA}
\affiliation{Physics Division, Lawrence Berkeley National Laboratory \\ 1 Cyclotron Road, Berkeley, CA 94720, USA}

\author[0000-0001-8156-0330]{Xiaosheng Huang}
\affiliation{Physics Division, Lawrence Berkeley National Laboratory \\ 1 Cyclotron Road, Berkeley, CA 94720, USA}
\affiliation{Department of Physics \& Astronomy, University of San Francisco \\ 2130 Fulton Street, San Francisco, CA 94117-1080, USA}

\author[0000-0001-7101-9831]{Aleksandar Cikota}
\affiliation{Gemini Observatory / NSF's NOIRLab \\ Casilla 603, La Serena, Chile}

\author[0000-0001-7266-930X]{Nao Suzuki}
\affiliation{Physics Division, Lawrence Berkeley National Laboratory \\ 1 Cyclotron Road, Berkeley, CA 94720, USA}
\affiliation{Kavli Institute for the Physics and Mathematics of the Universe, University of Tokyo \\ Kashiwa 277-8583, Japan}

\author[0000-0002-5042-5088]{David J. Schlegel}
\affiliation{Physics Division, Lawrence Berkeley National Laboratory \\ 1 Cyclotron Road, Berkeley, CA 94720, USA}

\author[0000-0002-0385-0014]{Christopher Storfer}
\affiliation{Physics Division, Lawrence Berkeley National Laboratory \\ 1 Cyclotron Road, Berkeley, CA 94720, USA}
\affiliation{Institute for Astronomy, University of Hawaii \\ 2680 Woodlawn Drive, Honolulu, HI 96822-1897, USA}

\begin{abstract}

The introduction of \txb{deep} wide-field surveys in recent years and the adoption of machine learning techniques \txg{have} led to the discoveries of $\mathcal{O}(10^4)$ strong gravitational lensing systems and candidates.  
However, the discovery of multiply lensed transients remains a rarity.  Lensed transients and especially lensed supernovae are invaluable tools to cosmology as they allow us to constrain cosmological parameters via lens modeling and the measurements of their time delays.  
\txr{\txg{In this paper, we} develop a pipeline to perform a \emph{targeted} lensed transient search.  We apply this pipeline to 5807 strong lenses and candidates, identified in the literature, in the DESI Legacy Imaging Surveys} \citep{DR9} \txr{Data Release 9 (DR9) footprint.  For each system,} we analyze every exposure in all observed bands (DECam \ipg, \ipr, and \ipz).  
Our pipeline finds, groups, and ranks detections that are in sufficient proximity temporally and spatially.  \txr{After the first round of inspection, \txb{for promising candidate systems,} we further examine the newly available DR10 data (with additional \ipi~and Y~bands).}  
\txb{Here} we present our \txr{targeted} lensed supernova search pipeline and seven new lensed supernova candidates, including a \txb{very likely lensed supernova --- probably a Type~Ia ---} \txr{in a system} with an Einstein radius of $\sim 1.5''$.

\end{abstract}



\section{Introduction} \label{sec:background}
The flat $\Lambda$CDM cosmological model is highly successful in describing our universe from the time of photon decoupling (at a redshift $z \approx$ 1100) to the present time.  According to this model, our universe has a flat geometry and is expanding at an accelerating rate (\citealp{riess98}, \citealp{perlmutter99}).  The present day expansion rate of the universe is known as the Hubble constant ($H_0$).  The inferred value for $H_0$ from the Planck CMB measurements is $67.4\pm 0.5$~km/s/Mpc \citep{planck2018}.  On the other hand, direct measurements of $H_0$ by using local distance ladders is higher by $\gtrsim 4\sigma$ (e.g., \citealp{reiss2021}).  Thus if this inconsistency is not due to systematic effects (e.g., \citealp{freedman2021}), then at a minimum, $\Lambda$CDM needs revision.

If a transient event were to occur in a strongly lensed background galaxy, it can be potentially observed multiple times, in each of the lensed images.  The time delay $\Delta t$ between the different images consists of a geometric component and a gravitational component (e.g., \citealp{naryanandbart96}).  Competitive $H_0$ constraint can be achieved if the time delays can be measured precisely and the lensing potential can be modelled accurately, providing an independent method of measuring the Hubble constant \citep[e.g.,][]{holismokes}.  If the observed transient is a lensed Type Ia supernova (L-SN~Ia), their standardizability can also significantly reduce the main systematic effect for \txb{lensing-based} $H_0$ measurements: the mass sheet degeneracy (e.g., \citealp{birrer2021}).

Currently, there have been seven confirmed lensed SNe: \citeauthor{quimby2014} (\citeyear{quimby2014}; PS1-10afx), \citeauthor{kelly2015} (\citeyear{kelly2015}; ``SN Refsdal"), \citeauthor{goobar} (\citeyear{goobar}; SN 2016geu), \citeauthor{rodney2021} (\citeyear{rodney2021}; ``SN Requiem"), \citeauthor{astronotekelly} (\citeyear{astronotekelly}; AT 2022riv), \citeauthor{astronotegoobar} (\citeyear{astronotegoobar}; ``SN Zwicky"), \txr{and} \citeauthor{kellysn2} (\citeyear{kellysn2}; C22).  \txr{Out of the seven, four are lensed by a galaxy cluster (``SN Refsdal", ``SN Requiem", AT 2022riv, C22).  Of the remaining \txb{three which were lensed by single galaxies,} two were found live (SN 2016geu, ``SN Zwicky").  Both happen to be Type Ia.}  However, both \trr{have sub-arcsec Einstein radii and thus} short time delays of $\lesssim 1$ day, making them not useful for $H_0$ measurements \citep{dhawan19, astronotegoobar, zwicky_pierel}.  
\trr{Some cluster lenses have very long time delays  (e.g., two decades for ``SN Requiem"), and they generally have larger modeling uncertainties, but with different systematics compared with galaxy-scale lenses. 
Lensed SNe found behind cluster with time delays at a reasonable timescale may still yield competitive $H_0$ measurements 
\citep[e.g.,][]{grillo2020}.
In recent years, thousands of new strong lenses have been discovered in imaging surveys \citep[e.g.,][]{jacobs2019, huang2020, canameras2020, canameras2021, huang2021, stein2022, storfer2022, shu2022}.  
Most of them are galaxy-scale lenses, with a small number of group/cluster lenses.
They are resolved in ground-based observations, and thus are expected to have time delays (on the order of days to weeks) that are useful for $H_0$ measurements.   
In this paper, we develop a targeted lensed transient pipeline to search among these systems in the DESI Legacy Imaging Surveys \citep{DR9} footprint.}

Applying our pipeline to the Legacy Surveys in a retrospective search, we have discovered seven new lensed supernova candidates.  
Along with the discoveries of many lensed quasars (which will be presented in a separate publication), we are confident our pipeline is capable of finding live lensed transients for present (e.g., Pan-STARRS; \citealp{panstarrs}) and future surveys (e.g., LSST and the Roman Space Telescope; \citealp{LSST} and \citealp{Roman} respectively).

\section{Observation} \label{sec:data}
\begin{figure}
\begin{center}
\includegraphics[width=160mm]{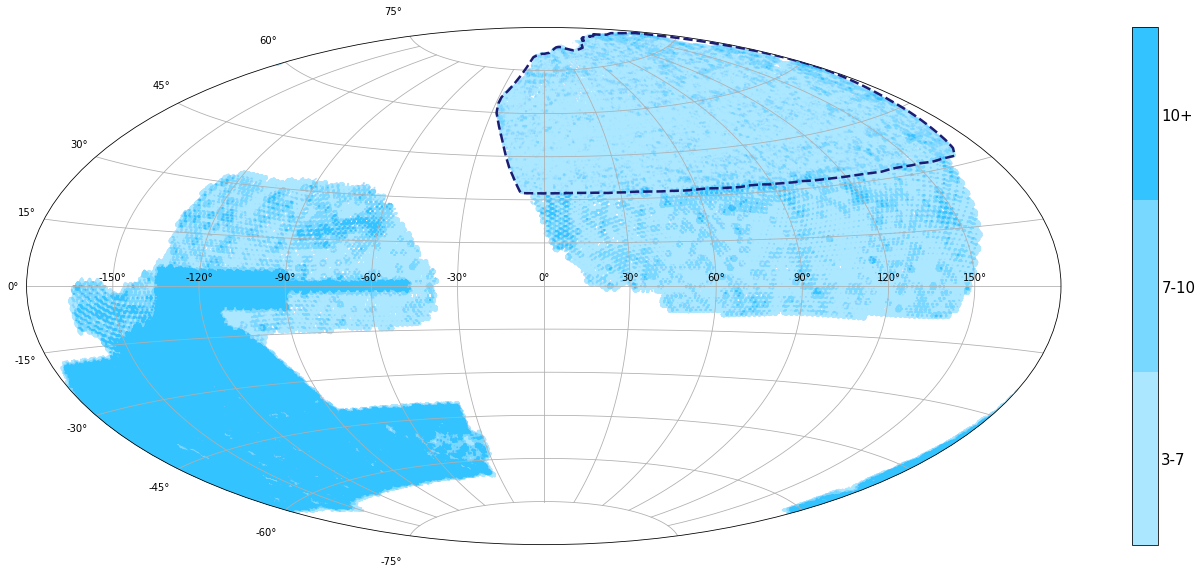}
\caption{The Legacy Imaging Surveys DR9 footprint, color coded by the \ipz~band \txr{observation} depth.  We perform a targeted search for lensed transients within the DECaLS region (dec $< 32^{\circ}$).  Exposures from the MzLS and BASS surveys (blue dashed outline) are excluded.}
\end{center}
\end{figure}

The \txr{DESI Legacy Imaging} Surveys is composed of three surveys: the Dark Energy Camera Legacy Survey (DECaLS), the Beijing Arizona Sky Survey (BASS), and the Mayall \ipz-band Legacy Survey (MzLS).  DECaLS is observed by the Dark Energy Camera \citep[DECam;][]{flaugher2015} on the 4-m Blanco telescope, which covers $\sim 9000 \text{ deg}^2$ of the sky in the range of $-18^{\circ} \lesssim \delta \lesssim +32^{\circ}$. BASS/MzLS are observed in the \ipg~and \ipr~bands by the 90Prime camera \citep{williams2004} on the Bok 2.3-m telescope and in the \ipz~band by the Mosaic3 camera \citep{dey2016} on the 4-m Mayall telescope. Together BASS/MzLS cover the same $\sim 5000 \text{ deg}^2$ of the northern subregion of the Legacy Surveys.  For this search, we exclude BASS and MzLS data as the number of exposures from each of the component surveys are fewer, with inferior seeing in $gr$~bands for reliable detection of transients.  Data Release 9 (DR9) contains additional DECam data reprocessed from the Dark Energy Survey \citep[DES;][]{abbott2016} for $\delta \lesssim -18^{\circ}$. This provides an additional $\sim 5000 \text{ deg}^2$, resulting in a total footprint of $\sim 19,000 \text{ deg}^2$.  \trr{Data release 10 (DR10) supplements DR9 with additional DECam exposures (including additional optical bands) from NOIRLab.  Both data releases are publicly available\footnote{\url{https://www.legacysurvey.org}}.} The DECam surveys will hereafter be referred to in its entirety as DECaLS, within which we distinguish DES and non-DES regions.  The pipeline deployment will mostly focus on the exposures in DECaLS, in \ipg, \ipr, and \ipz~filters, with a nominal DES exposure time of 90 seconds, and non-DES exposure time ranging from 60 to 200 seconds.  In addition, the Legacy Surveys contain deep field DECam observations (from surveys such as COSMOS, XMM-LSS, and SN-X3), with 800+ exposures for any given target in these fields.  Our pipeline has been applied to DECaLS and these deep \txb{observations}.

As we take the approach of a targeted search, we compile a database of 5807 strong \txb{lenses} and candidates found within 
DECaLS, \txb{with} the majority from \citealp{huang2020}, \citealp{huang2021}, and \citealp{storfer2022}, and the rest from \citealp{moustakas2012}, \citealp{carrasco2017}, \citealp{diehl2017}, \citealp{jacobs2017}, \citealp{pourrahmani2018}, \citealp{sonnenfeld2018}, \citealp{wong2018}, and \citealp{jacobs2019}.  Note that \citealp{storfer2022} only included C-grade or above candidates.  However on the project website\footnote{\url{https://sites.google.com/usfca.edu/neuralens}}, they also included D-grade candidates.  \txr{These receive numerical scores of 1 or 1.5\footnote{\txr{C. Storfer, private communication.}}.  Here we include those with the higher numerical score of 1.5, which in \txb{this} paper we will call D+.}

\section{Pipeline} \label{sec:pipeline}

The general framework of the pipeline consists of image reprojection and reference image generation (\S~\ref{subsec:image coadding}), image subtraction (\S~\ref{subsec:image subtraction}), and source detection and grouping (\S~\ref{subsec:source detection}) for each of 5807 \txb{lensing systems and candidates}. 

\subsection{Image Reprojection and Reference Image Generation} \label{subsec:image coadding}

The pipeline first collects all relevant exposures from DR9 for all targets.  The images are then reprojected onto the same World Coordinate System (WCS) orientation, with the system centered in each 801$\times$801 pixel ($216.27''\times 216.27''$) cutout.  For each filter, we use the median coadd as the reference image, in order to reduce the influence of a potential transient (and other time-dependent systematics such as cosmic rays).  These steps are done using the Montage software package \citep{montage}.  For exposures of the same band within 1.5 days of each other, we combine them as a mean coadd, as we do not expect significant change in flux \txr{for an astrophysical transient} in that short of a time frame, while increasing detection efficiency.

\subsection{Image Subtraction} \label{subsec:image subtraction}

To find transients, we perform image subtraction between each exposure and reference image for the same filter.  This pipeline uses two different image subtraction algorithms: that of \citeauthor{bramich2008} (\citeyear{bramich2008}; henceforth B08), and Saccadic Fast Fourier Transform (SFFT; \citealp{hu2021}).  

The B08 algorithm fits for a spatially varying kernel that attempts to convolve the reference image to appear comparable to the image of each exposure (``science" image).  B08 uses delta functions as its basis functions, and thus fits for every pixel in the kernel to minimize the $\chi^2$ of the difference image between the science image and the convolved reference image.  Because it fits for every pixel in the kernel, it makes no assumption on the functional form of the fitted kernel.  

SFFT is a fully Fourier implementation of image subtraction.  SFFT performs Fourier transforms on both reference and science images, and fits for a convolutional, also spatially varying kernel for the reference image (thus similar \txb{to B08}, but in the Fourier space).  

We have experimented with other well-known image subtraction algorithms (\citealp{alard2000}, \citealp{zogy}), but found that B08 and SFFT best suit this pipeline's application.  \citet[][henceforth A00]{alard2000}, implemented in the HOTPANTS package \citep{hotpants}, is widely used for large surveys and is very similar to B08.  The main difference is that A00 uses a set of Gaussian priors, and thus assumes a functional form of the convolutional kernel.  Rather than fitting for each pixel of the kernel (as with B08), A00 only has to fit for the Gaussian parameters, which leads to significant speed-up for large scale surveys.  But since we are conducting a targeted search, our pipeline can afford to use the slower, more flexible B08 algorithm.  \citealp{zogy} (also known as ZOGY) takes a different approach to the image subtraction problem, by utilizing the concept of cross-filtering (\txb{i.e.,} two separate convolutional kernels) for the difference image generation and solving for both kernels in Fourier \txg{space.}  We opt to use the SFFT algorithm, as \txg{their results indicate an improvement to addressing photometric mismatch within image subtraction over ZOGY}.  Despite appearing as a \txr{publication} only recently, SFFT has been extensively applied to time-domain observations \footnote{L. Wang, private communication.}.  For almost all cases, we find that the SFFT algorithm \txr{produces} cleaner and more accurate image subtraction compared to B08.  Thus, though both algorithms are used for transient detection, we use the SFFT difference images for all subsequent photometry and detection presentation in this paper. 

\subsection{Source Detection and Grouping} \label{subsec:source detection}

After generating difference images with both image subtraction algorithms, we use a Python implementation (SEP; \citealp{sep}) of the source extraction algorithm from \citet{bertin1996} to detect any potential sources in all difference images, with thresholds ranging from 1.0 to 2.5$\sigma$ in \txr{0.25} increments as determined by SEP (with detections $>$ 2.5$\sigma$ treated as the same detection level as 2.5$\sigma$).  Henceforth, we denote a detection in a single difference algorithm as a ``sub-detection" (a given transient \txr{event in} \txb{a single} exposure can generate two sub-detections, by being detected in difference images produced \txr{by both the B08 and SFFT algorithms}).  All sub-detections (from both subtraction algorithms, across all bands and across all exposures) are then grouped together spatially and temporally.  These groupings contain all sub-detections that are within three pixels ($0.8''$) of one another, and are within 50 days of other sub-detections in the group.  If a given group has less than three sub-detections, the pipeline disregards them.  If a group has three or more difference image sub-detections, it is labelled as a possible transient detection.  That is, an event must be observed \txr{by DECaLS} at least twice, in at least two separate exposures, to be labelled as a possible transient detection\footnote{\txr{\txb{For example}, at the threshold (inclusive), a group with three sub-detections can correspond to two or three detections (i.e., \txb{in} two or three exposures).  In the case of two detections, one of them is detected by both subtraction algorithms.  In the case of three detections, each are detected by a single subtraction algorithm.}}.  This is to reduce the number of false detections (from noise, cosmic rays, CCD artifacts, etc.) and inconclusive events, and improve the overall quality of detections that will be visually examined.  \txr{Had we conducted} this search live, \txr{$\mathcal{O}(100)$} \txb{2+ sub-detection} groups would have been \txr{identified, most of which would warrant follow-up observations.} 

The entire process typically takes about two hours to run per system, with approximately one hour for data collection and reprojection and one hour for image subtraction.  To run this pipeline on 5807 systems (approximately 120,000 individual exposure cutouts), parallelization is necessary. The full deployment is performed on the National Energy Research Scientific Computing Center (NERSC) Cori supercomputer.  Using 20 nodes, 32 CPUs per node, and 1 thread per CPU, this requires that each thread run 9 to 10 systems, taking a total of 18 to 20 hours.  SFFT is capable of being run on GPUs with significant speed gain.  We will take advantage of this capability in our pipeline in the near future.

\txb{Twenty-five} of the 5807 candidate systems lies within a deep field survey footprint, and thus each has 800+ individual exposures.  \txr{For these systems, the amount of memory and time \txb{required} makes naively running the pipeline infeasible.}  Instead, we opt to split up the exposures from these systems into smaller groups of temporally-similar exposures, and running the pipeline on each possible pair of groups\txr{, for all permutations}.  The groups were created such that every exposure appears in at least three groups, as to ensure that the pipeline does not miss a possible lensed transient in any exposure.  \txr{We do not find any lensed transients within the 25 deep field systems.}

\subsection{Selection Criteria} \label{sec:analysis}
Human inspection of the pipeline results is necessary to validate the pipeline detections.  \txb{For first round} visual inspection, for each system, two \txb{initial} grades are assigned regarding its most convincing group of detections.  \txr{Firstly, we assign a location grade to assess how close the detection location lies relative to any putative lensed features.  We use this grade to filter out transient candidates that are clearly not lensed.  Secondly, a transient grade is given by how likely the detection is a transient.  }

Most systems were not given any grades; i.e.,~there were no convincing detections for these systems.  As our pipeline's focus is to detect transients that are lensed, very obvious transients far from any lensing features are given high transient grades, but low location grades.  By using these metrics, we identified the most promising lensed transient candidates.  

\txr{We apply our pipeline to these select candidates a second time,} with the change of removing all exposures that \txr{contain} the suspected transient detection from the median coadd, in order to generate a reference free of possible transient light.  \trr{Then-preliminary} Legacy Surveys DR10 data are \txr{included} in this second run.  The new data also includes observations from the DECam \ipi~and Y~bands.  Using Point Spread Functions (PSFs) modelled from isolated stars within the entire CCD exposure brick\footnote{\url{https://www.legacysurvey.org/dr10/description/}}, PSF photometry is then applied to all SFFT difference images.  \txr{For exposures that we are not able to fit with a PSF at the detection location (i.e., non-detections that take place well before or after the peak of the suspected transient event),} aperture photometry is applied to the location of the possible transient in order to establish a baseline flux for light curve fitting. 
For promising candidates, we apply our own final set of criteria to \txr{select} the candidates with the \txr{highest} potential of being a L-SN:
\begin{enumerate}
  \item \emph{Strong Lensing Plausibility} - \txr{Because the lens candidates included in the search are from multiple publications and different search efforts, we determine how likely a candidate is a strong lensing system.  The criteria we use are similar to those in} \citet{huang2021}.  
  \item \emph{Asteroid Filtering} - If there are only a few detections minutes apart in a given night, and no detections after that night, this is an indication that transient is \txb{possibly} an asteroid.  To confirm this, we can approximate the speed of the asteroid between detections (using PSF fitting to precisely locate the punitive asteroid), and compare it to the speed of a typical main-belt asteroid.
  \item \txr{\emph{Location Consideration} - In combination with the location grade above, we also take into consideration surrounding objects (in some cases, modeling their light profiles) and assess the overall probability of the transient candidate being lensed based on the detection location.}
  \item \emph{Light Curve Fit Quality} - To narrow the possible identities of the detection, we fit a SALT3 \citep{salt3} SN~Ia light curve model, and 161 different core-collapse (CC) SN models to the observed photometry.  As the survey data is sparse and the search is retrospective, we use a photometric redshift prior \txr{or a spectroscopic redshift} in the fitting process.  \txr{A fit with low $\chi^2/$DOF is an indication of a possible identity.  When a light curve model fits the photometry well, we also assess whether the best-fit light curve model parameters are reasonable (e.g., the SALT2 \txb{$x_1$ parameter is generally between $-3$ and $3$}).}
  \item \emph{Amplification/Hubble Diagram Residual} - From the \txr{best-fit} light curve models, we can further deduce whether the transient is amplified, and if the amplification is reasonable given the system configuration.  For the case of a SN~Ia, we \txr{can find its Hubble residual and determine the amplification (if any).}
\end{enumerate}

\trr{While unlikely, there is the possibility that one or some of our lensed transient candidates are rare instances of microlensed high-redshift stars, e.g. \citealp{kelly2018} and \citealp{welch2022}.  However, this would be extremely coincidental, as most of our targets and candidates are single-galaxy scale lenses.  In contrast, all discovered microlensed high-redshift stars are lensed by galaxy clusters, which boasts much higher magnification, allowing a significantly larger strong lensing cross-section of near-infinite magnification.  With only the ground-based Legacy Imaging Surveys data present for our candidates, we are unable to accurately model these lensing systems and thus cannot tell if a posited transient lies near the critical curve.  Lastly, the photometric data from DR9 and DR10 is too sparse to support this claim.  Therefore, while we do not disqualify the possibility of a microlensed high-redshift star, we will not further entertain the postulation.}


\section{Expectations} \label{sec:expectations}
\begin{figure}
\begin{center}
\includegraphics[width=180mm]{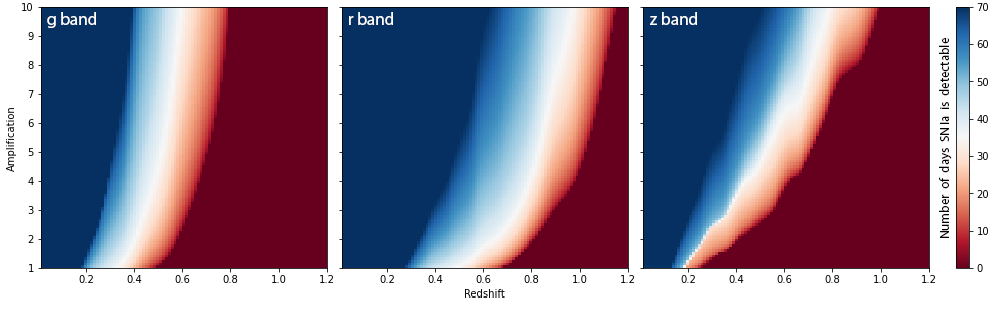}
\caption{Number of days when a L-SN~Ia is detectable as a function of redshift and lensing amplification in DECam \ipg, \ipr, and \ipz~bands.  Every pixel in each plot represents a simulated SALT3 light curve model, color-coded by the number of days it exceeds the corresponding filter's $5\sigma$ PSF detection limit.}\label{fig:just}
\end{center}
\end{figure}

To assess the feasibility of finding \txb{L-SNe~Ia} in the DESI Legacy Imaging Surveys, we can simulate SNe~Ia light curves at various redshifts and lensing amplifications of a lensed SN, and calculate the amount of time a given simulated SNe will be detectable in the DECam \ipg, \ipr, and \ipz~bands.  We will be using the following values: 23.47, 23.43, and 21.63 for \ipg, \ipr, and \ipz~bands respectively, based on the 5$\sigma$ PSF detection thresholds from \citet{DR9}, adjusted to the nominal time of 90 seconds per exposure in DR9.  We use SNCosmo \citep{sncosmo} to simulate SNe~Ia light curves, based on the SALT3 model \citep{guyt2007, salt3}.  We assume $x_1=0$ and $c=0$ for the SALT3 parameters.  Figure~\ref{fig:just} illustrates the length of time that a L-SN~Ia is detectable in the DECam \ipg, \ipr, and \ipz~filters across a range of reasonable redshifts and amplifications.  \txb{This initial investigation indicated that L-SN~Ia are discoverable in the Legacy Surveys and motivated this search.}

As a careful forecast of lensed SN rates is beyond the scope of this paper, we use the formulation of \citeauthor{shu2018} (\citeyear{shu2018}, \citeyear{shu2021}; henceforth S18) to provide a first-order estimation.  Following S18, we simulate the star formation rate for a source galaxy at a given redshift in order to sample SN rates.  

\begin{figure}
\begin{center}
\includegraphics[width=170mm]{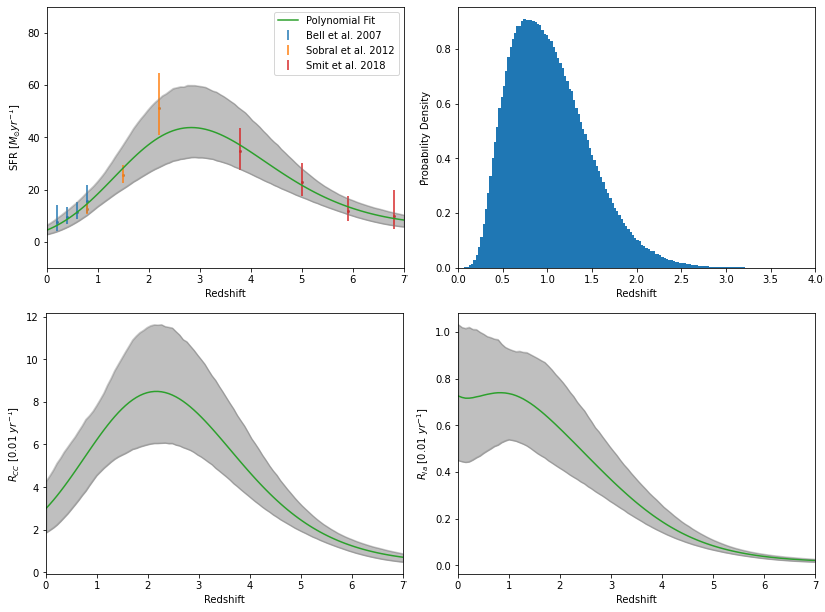}
\caption{Distributions used to simulate necessary parameters for calculating the expected results of our pipeline's deployment.  Top Left: star formation rate versus redshift, with the polynomial fit (degree of 3) to the data (\citealp{bell2007}, \citealp{sobral2012}, \citealp{smit2012}).  The uncertainty bounds (shown in grey) is generated using the the covariance matrix of the resulting fit.  Top Right: the assumed source redshift distribution (see text).  Bottom Left: CC SNe rates verses redshift.  The same polynomial fit and uncertainty bounds are used to calculate these rates, according to equation~\ref{eq:1} and \ref{eq:2}.  Bottom Right: SNe Ia rates versus redshift.  The same polynomial fit and uncertainty bounds are used to calculate these rates, according to equations~\ref{eq:2}, \ref{eq:3}, and \ref{eq:4}.  }\label{fig:dist}
\end{center}
\end{figure}

We can now estimate the number of lensed Type Ia and CC SNe we expect to find in our retrospective search.  We start by simulating the source redshift of a given lensing system.  As most of our lens candidates are from \citet{huang2020}, \citet{huang2021}, and \citet{storfer2022}, we generalize the lens galaxy spectroscopic or photometric redshifts from those candidates as the lens galaxy redshift distribution in our simulations.  We then multiply this with a truncated normal distribution $\mathcal{N}(2, 0.5)$, \txb{with a lower bound at} 1, to represent the source galaxy redshift distribution.  Figure~\ref{fig:dist} shows the source galaxy redshift distribution from which we sample for our simulations.  

Using SFR estimations of varying redshifts (from $z=0.2$ to $z=6.8$) from \citet{bell2007}, \citet{smit2012}, and \citet{sobral2012}, we fit a polynomial function (degree of three, using Numpy's polyfit algorithm) to $\log_{10}($SFR$)$, with the uncertainties given by the polynomial fit covariance matrix.  We sample SFRs at a given source redshift from this polynomial model.  

From S18, we can convert SFRs to CC SNe rates:
\begin{equation}\label{eq:1}
    R_{CC} = 0.0068M^{-1}_\odot \frac{\text{SFR}}{1+z_S} [\text{yr}^{-1}]
\end{equation}
The CC~SNe rates are the broken down into sub-rates for the different types, based on the percentages in Table~\ref{table:ccrates}.

\txb{We} can estimate the SFH from the functional form \citep{madau2014}, normalized by the recent SFR (S18):
\begin{equation}\label{eq:2}
    \text{SFR}(t(z)) = \text{SFR}\times\left(\frac{1+z(t)}{1+z_S}\right)^{2.7}\left(\frac{1+\left( \frac{1+z_S}{2.9} \right) ^{5.6}}{1+\left( \frac{1+z(t)}{2.9} \right) ^{5.6}}\right) 
\end{equation}
From S18, by assuming \txb{a} delay time ($t_D$) distribution, $f_D$:
\begin{equation}\label{eq:3}
    f_D(t_D) \propto t_D^{-1.07} ,
\end{equation}
we can estimate SN~Ia rates:
\begin{equation}\label{eq:4}
    R_{Ia} = 0.00084M^{-1}_\odot\frac{ \int^{t(z_S)}_{0.1} \text{SFR}(t(z_S)-t_D)f_D(t_D) \,dt_D }{(1+z_S) \int_{0.1}^{t(z=0)}f_D(t_D)\,dt_D} [\text{yr}^{-1}],
\end{equation}

Each system is assumed to have two or four lensed images (with probabilities of 0.7 and 0.3 respectively; \citealp{oguri2010}), and each image has a magnification sampled from a lognormal distribution \txr{(mean = 1.5 and standard deviation = 0.35)},  with an expected magnification of 4.765.  As context, \citet{shu2021} used a constant magnification of 5 for their targeted rates estimations, whereas \citet{craig2021} performs a targeted estimate on a set of 40 strong lenses (from \citealp{shu2017}) with magnifications ranging [2, 105], with a median of 6.5.  We sample time delays between each lensed image from $\mathcal{N}(36, 4)$ days \citep{craig2021}.

We do not simulate the times of exposures, but rather use the true exposure times of our 5807 targets, observed in the DECam \ipg, \ipr, and \ipz~bands.  Conservatively, we assume 90 seconds for each exposure (since while occasionally an exposure can be as low as 60 seconds, the vast majority of the exposures are 90 seconds or longer).
\begin{deluxetable*}{cccccc}
\tablecaption{\txb{Supernova simulation percentages and brightness parameters}\label{table:ccrates}}
\tablewidth{0pt}
\tablehead{
\colhead{SNe Type} & \colhead{$\overline{M_B}$} & \colhead{$\sigma$} & \multicolumn{1}{p{2.5cm}}{\centering \% Rate of\\CC Occurrence} & \colhead{SNCosmo Template} & \colhead{Template}}
\decimalcolnumbers
\startdata
IIp & $-16.75 \pm 0.37$ & 0.98 & 55.83 & nugent-sn2p & \citet{nugent}\\
Ic & $-17.66 \pm 0.40$ & 1.18 & 17.00 & nugent-sn1bc & \citet{levan2005}\\
IIb & $-16.99 \pm 0.45$ & 0.92 & 12.43 & v19-2006t-corr & \citet{vincenzi2019}\\
Ib & $-17.45 \pm 0.33$ & 1.12 & 9.00 & nugent-sn1bc & \citet{levan2005}\\
IIL & $-17.98 \pm 0.34$ & 0.86 & 3.34 & nugent-sn2l & \citet{nugent}\\
IIn & $-18.53 \pm 0.32$ & 1.36 & 2.40 & nugent-sn2n & \citet{nugent}\\
\\
Ia & $\IaBpeak \pm 0.20$ & 0.50 & - & salt3 & \citet{salt3}\\
\enddata
\tablecomments{This table shows the supernova-related parameters used in the pipeline simulated results.  As with \citet{craig2021}, the rate for \txb{87A}-like \txb{SNe} (1\%) is uniformly distributed across IIp, IIb, and IIL \txb{SNe}.  All $\overline{M_B}$ and $\sigma$ values are reported in \citet{richardson2014}.  CC~SNe \txb{percentages} are reported in \citet{eldridge2013}.}
\end{deluxetable*}

Assuming all 5807 systems in our catalog are real lensing systems, we simulate the expected results for each system, using their time of exposures in the DECam \ipg, \ipr, and \ipz~bands for our search.  We sample their source redshifts, number of lensed images, lensing amplifications, lensing time delays, and star formation rates.  Using these sampled values, we calculate the Ia and CC \txb{SN} rates.  Based on these rates, we simulate \txb{SNe} across the duration of the DR9 \txb{observation} range (from the date of the first exposure $-$100 days to the date of the last exposure +100 days; the time frame being significantly larger than the typical width of Ia and CC light curve widths).  In Table~\ref{table:ccrates} are the SNCosmo light curve models used during simulations, as well as the absolute \ipB~band magnitude distributions.  For simulated SNe Ia, we sample the following parameters as: $c\sim \mathcal{N}(0, 0.1)$ and $x_1\sim \mathcal{N}(0, 1)$ \citep{guy2010, scolnic2022}.  For CC \txb{SNe}, we subdivide and simulate them as Ib, Ic, IIn, IIp, IIb, or IIL, with their \txr{$M_B$} sampled from Table~\ref{table:ccrates}.  For all simulated \txb{SNe}, we assume a small amount of host galaxy dust (\txb{$E(B-V)\lesssim 0.02$}).  Finally, we check if \txb{a} lensed SN image for a given system in a given band \txb{is above the} detection \txb{limit}. 

\begin{figure}
\begin{center}
\includegraphics[width=180mm]{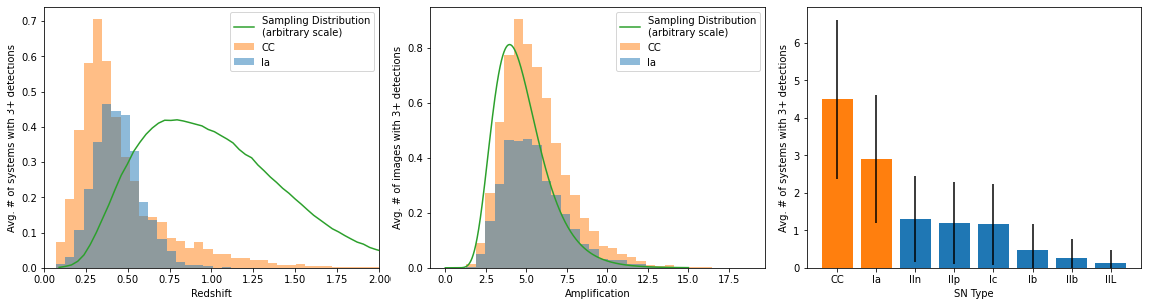}
\caption{\txb{Results} of the 1000 simulated runs for finding lensed \txb{SNe} in the 5807 \txb{targeted systems}.  Left: A histogram illustrating the average number of systems (out of 5807) with three or more detections verses redshift, as well as the scaled sampled source redshift distribution outlined in green.  The \txr{redshift distribution of detectable systems clearly is lower than the distribution we sample from, due to detection limits.}  The \txb{integrals} of the distributions are shown in Table~\ref{tab:expectation_results}.  Center: A histogram illustrating the number of systems with three or more detections verses magnification, as well as \txr{the magnification distribution we sample} from outlined in green.  There is a slight bias towards higher magnifications, \txr{as would be} expected.  Right:  A bar diagram illustrating the representation of each type of simulated \txb{SNe} in the systems with three or more detections.  }\label{fig:expect-results}
\end{center}
\end{figure}

The full simulation of our pipeline (on the 5807 target systems) was parallelized and run 1000 times on NERSC to estimate the expected SNe rates.  The final results are shown in Figure~\ref{fig:expect-results} and Table~\ref{tab:expectation_results}.  \txr{In order to reliably find transients, we require a threshold of three sub-detections or more.  As mentioned earlier (\S~\ref{subsec:source detection}), depending on the system, this means at least two or three detections, corresponding to the highlighted rows in Table~\ref{tab:expectation_results}.}

\begin{deluxetable*}{ccc}
\tablecaption{Expected \txb{numbers of lensed SN detections} \label{tab:expectation_results}}
\tablewidth{0pt}
\tablehead{
\colhead{Number of detections} & \colhead{\txb{L-SNe~Ia}} & \colhead{\txb{L-CC~SNe}}
}
\decimalcolnumbers
\startdata
1 or more & $16.49 \pm 4.14$ & $22.74 \pm 4.85$ \\
\textbf{2 or more} & $\mathbf{6.58 \pm 2.57}$ & $\mathbf{9.35 \pm 3.08}$ \\
\textbf{3 or more} & $\mathbf{2.90 \pm 1.68}$ & $\mathbf{4.53 \pm 2.12}$ \\
4 or more & $1.31 \pm 1.16$ & $2.43 \pm 1.55$ \\
5 or more & $0.63 \pm 0.80$ & $1.42 \pm 1.16$ \\
6 or more & $0.30 \pm 0.53$ & $0.82 \pm 0.87$ \\
7 or more & $0.14 \pm 0.36$ & $0.48 \pm 0.68$ \\
\enddata
\tablecomments{This table shows the final results of the 1000 simulated for finding lensed supernovae in the 5807 \txb{lenses and} candidates.  \txb{To compare} this forecast \txb{with} our targeted search, we highlight \txr{the rows for two and three} detections or more (see text).} 
\end{deluxetable*}

\newpage
\section{Testing Detection and Photometry Pipelines on Known SNe~Ia} \label{sec:testing}
To test the performance of our pipeline, we apply it to photometric data from known SNe Ia discovered in DES \citep{Andrea2018, DESDR2}.  From the DES \txb{SNe~Ia}, we select well observed and modelled \txb{SNe} that had at least two DR10 exposures within $-$15 to +30 days of the time of peak brightness in \txb{\ipB~band (or $t_0$)}.  This results in a set of 32 SNe~Ia.  The SNe were previously modelled using SALT2 \citep[extended by][]{hounsell2017} parameterization\footnote{\url{https://github.com/sam-dixon/sncosmo\_lc\_fits}}, with a host galaxy dust extinction model \citep{fitzpatrick1999}.  Modelled SALT2 parameters include redshift $z$, \txb{$t_0$}, the normalization factor $x_0$ (normalized so that the peak \ipB~band apparent magnitude is $10.5$ when $x_0 = 1$\txr{, per SNCosmo}), the ``stretch" factor $x_1$, and the color parameter $c$.  We plot our photometry \txb{together with} previously observed photometry and the SALT2 models.  All coadded RGB images are made with the Legacy Surveys' RGB image generation scheme\footnote{\url{https://github.com/legacysurvey/imagine/blob/main/map/views.py}}.  Below we present the results for \txb{four} SNe Ia systems at different redshifts ($z=[0.18, 0.40, 0.53, 0.69]$).  Results for the full 32 DES SNe~Ia test systems are shown in Appendix~\ref{appendix_kSNe}.  All light curves presented are in the observer frame.


\begin{figure}[H]
\begin{center}
\includegraphics[width=180mm]{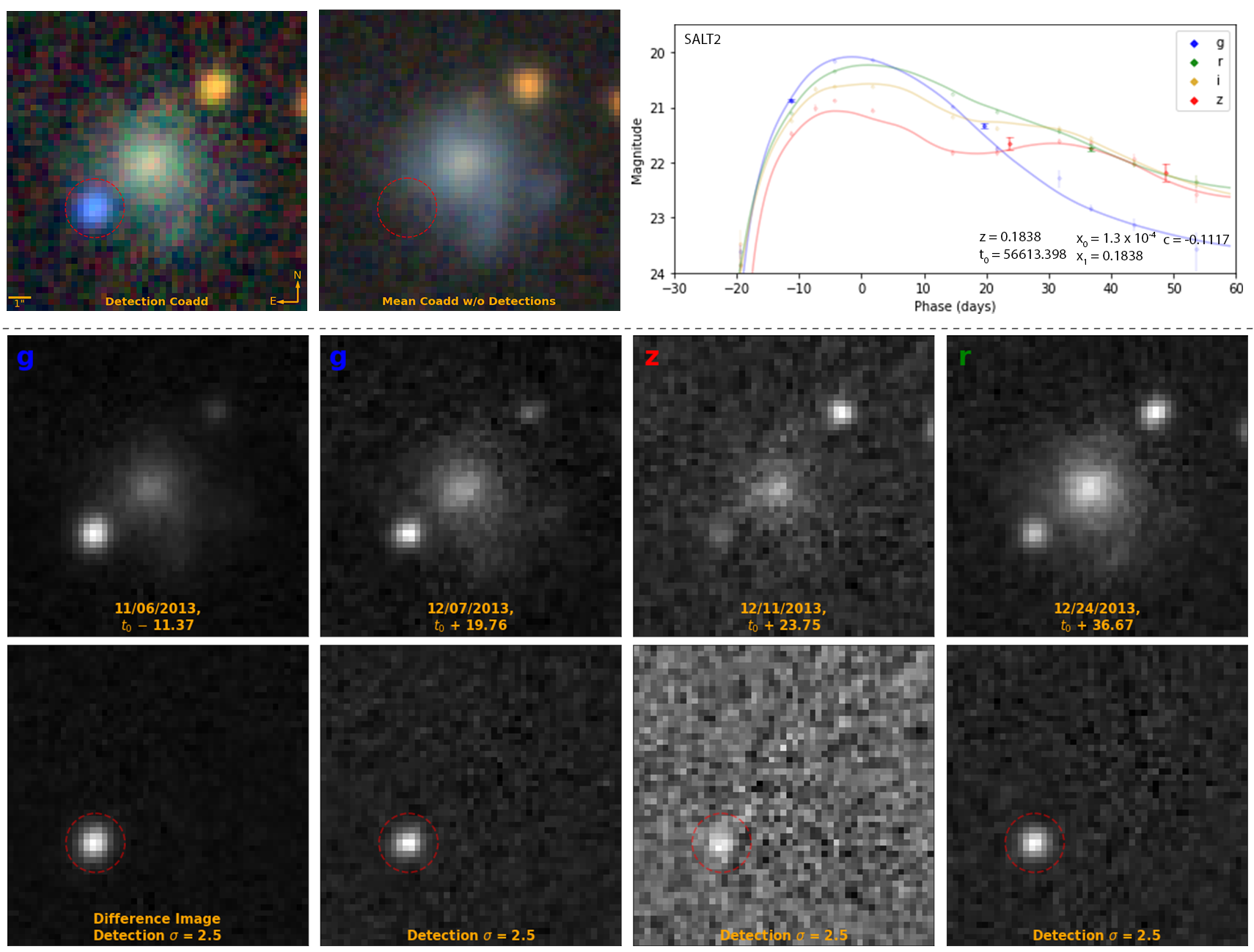}
\caption{Detection and photometry results of our pipeline, for a known SN~Ia at (RA, dec) = (7.6006, $-$42.2977), $z=0.1838$. 
 Top Left: \txb{The} mean coadded RGB (generated from \ipg, \ipr, \ipi, and \ipz~bands; see text) image, generated using exposures that include the SN event.  Top Middle: \txb{The} mean coadded RGB image, generated using exposures that exclude the supernova event.  The dotted red circle indicates the location of the \txb{SN} Ia event.  Top Right: Photometry for the detected SN from the SFFT difference images (solid points), plotted with DES photometry (fainter points) for this SN and the best-fit DES light curve.  We note the good agreement between our results and DES.  Below the dotted line: examples of single band images in chronological order, alongside its corresponding SFFT difference image (bottom) and SN detection, labelled with band, date of exposure, \txb{the phase}, and the $\sigma$-level of detection.}\label{fig:known-snia-figure}
\end{center}
\end{figure}

\begin{figure}[H]
\begin{center}
\includegraphics[width=180mm]{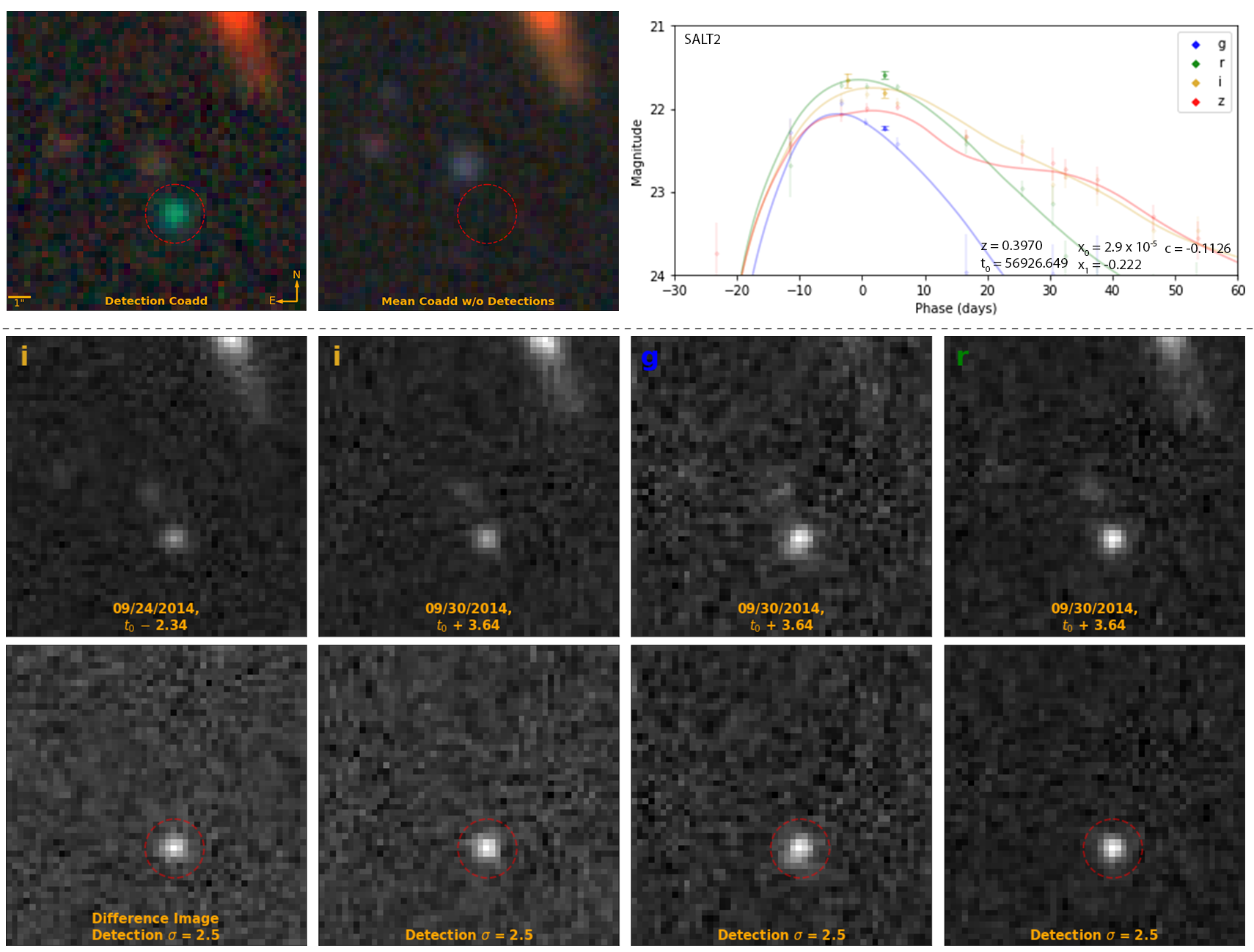}
\caption{Detection and photometry results of our pipeline, for a known SN~Ia at (RA, dec) = (41.0802, $-$0.4498), $z=0.397$.  For the arrangement of panels, see the caption of Figure~\ref{fig:known-snia-figure}.}
\end{center}
\end{figure}

\begin{figure}[H]
\begin{center}
\includegraphics[width=180mm]{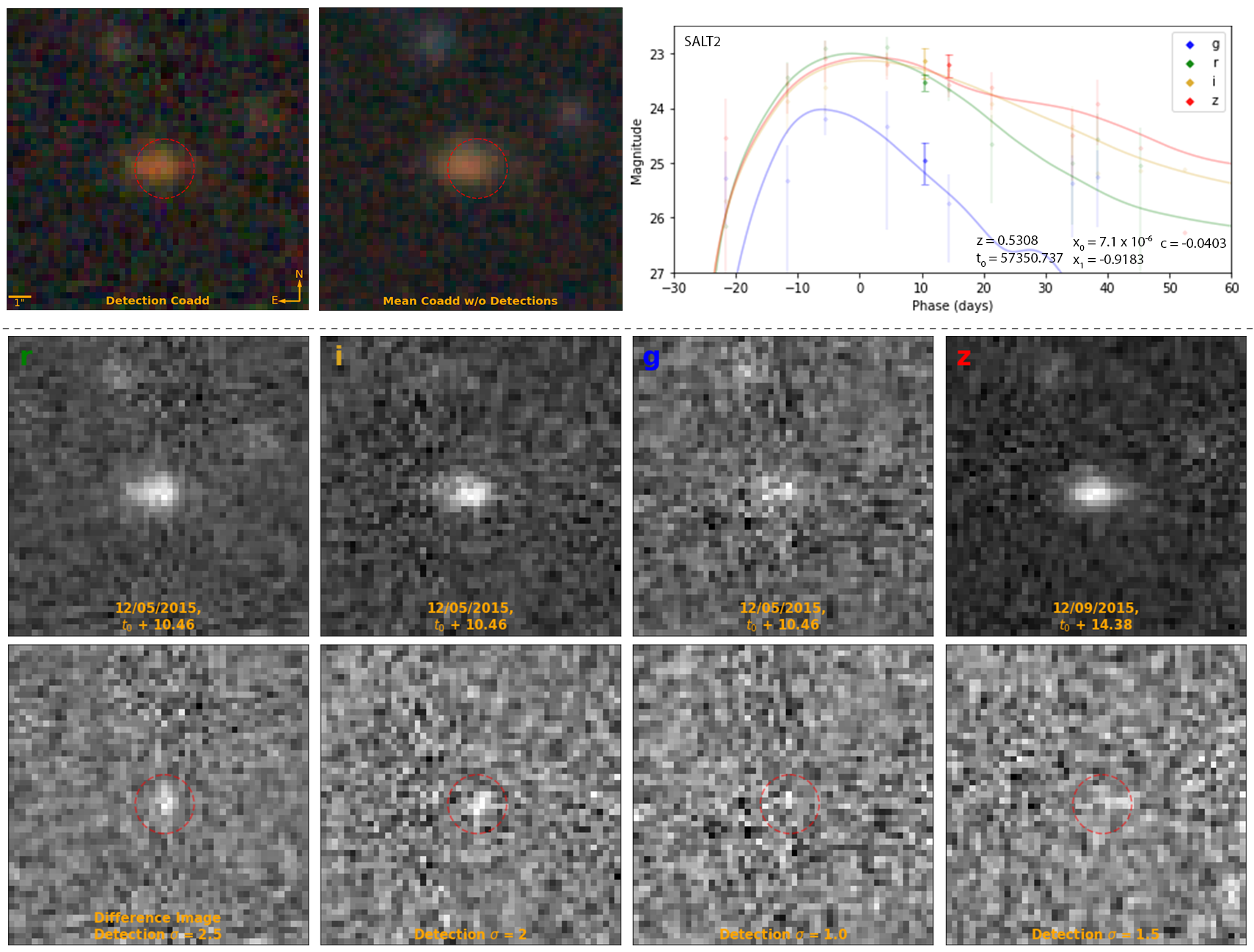}
\caption{Detection and photometry results of our pipeline, for a known SN~Ia at (RA, dec) = (41.2919, $-$1.5649), $z=0.5308$.  For the arrangement of panels, see the caption of Figure~\ref{fig:known-snia-figure}.}\label{fig:sn33}
\end{center}
\end{figure}

\begin{figure}[H]
\begin{center}
\includegraphics[width=180mm]{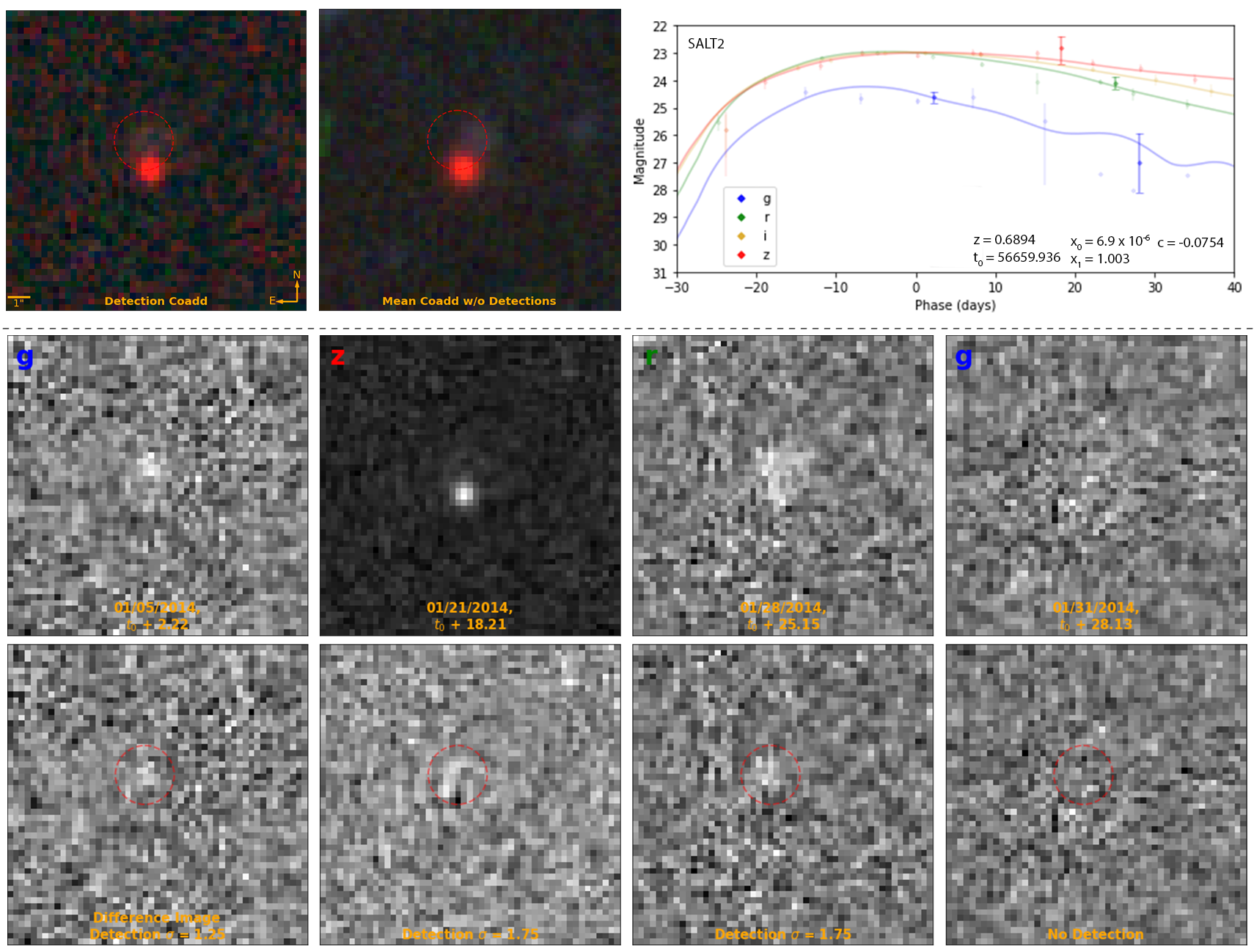}
\caption{Detection and photometry results of our pipeline, for a known SN~Ia at (RA, dec) = (53.3699, $-$28.4430), $z=0.6894$.  For the arrangement of panels, see the caption of Figure~\ref{fig:known-snia-figure}.  \txb{Note that there is no detection for the last exposure.  We therefore performed aperture photometry at the detected location of the transient.}}\label{fig:known-snia-figure-last}
\end{center}
\end{figure}



From these results, we found that our detection pipeline can detect known SN events within the DR9/DR10 data, with a 100\% detection rate for the 32 test SNe~Ia from DES.  Furthermore, the new photometry points from our pipeline are consistent with the DES photometry for these SNe~Ia.  

\section{Results and Discussion} \label{sec:results}
We have identified seven lensed SN candidates, one unlensed SN, and two asteroids detections with our pipeline.  This section will focus \txb{on one} Grade A and two Grade B lensed SN candidates, with the lower grade lensed SN candidates and the unlensed \txb{SN} in Appendix~\ref{appendix_SNC}, and the asteroid detections in Appendix~\ref{appendix_asteroid}.  We would also note that all eight SN detections (summarized in Table~\ref{tab:sncandidates}) would have warranted additional follow-up observations if found live.

\begin{longrotatetable}
\tabletypesize{\fontsize{6}{7}\selectfont}
\setlength\extrarowheight{0pt}
\setlength{\tabcolsep}{-3pt}
\movetabledown=0.8cm
\begin{deluxetable*}{cccccccccccccc}
\tablecaption{Lensed and Unlensed Supernova Candidates\label{tab:sncandidates}}
\tablewidth{0pt}
\tablehead{
    \multicolumn{1}{p{3cm}}{\centering System Name}&
    \multicolumn{1}{p{1.25cm}}{\centering Overall\\Grade}&
    \multicolumn{1}{p{2cm}}{\centering Total Number\\of Exposures\\(\ipg, \ipr, \ipi, \ipz, Y)} & 
    \multicolumn{1}{p{2.2cm}}{\centering Number of PSF\\Photometry Exposures\\(\ipg, \ipr, \ipi, \ipz, Y)\\}&
    \multicolumn{1}{p{1.5cm}}{\centering Distance\\Grade}&
    \multicolumn{1}{p{1.5cm}}{\centering Postulation}&
    \multicolumn{1}{p{1.25cm}}{\centering Shown}&
    \multicolumn{1}{p{2.05cm}}{\centering Redshift\\used in\\LC Prior}&
    \multicolumn{1}{p{2.15cm}}{\centering Best-fit\\$z_{SN}$}&
    \multicolumn{1}{p{1.25cm}}{\centering $\chi^2 / $DOF}&
    \multicolumn{1}{p{1.5cm}}{\centering Hubble\\Residual}&
    \multicolumn{1}{p{2.2cm}}{\centering Required\\Amplification}&
    \multicolumn{1}{p{1.5cm}}{\centering Model\\Fitting\\Grade}&
    \multicolumn{1}{p{1.25cm}}{\centering Reason}
    }

\decimalcolnumbers
\startdata
 & & & & & & & & & & & & & \\                      
\hyperref[subsubsec:DESI-344.6252-048.8977]{DESI-344.6252-48.8977} & A &                          (10, 8, 7, 7, 7) &                               (1, 2, 1, 0, 1) &              A &     uL-SN~Ia &     Y &         0.374 $\pm$ 0.053 & 0.299 $\pm$ 0.021 &      5.47 &  1.10 $\pm$ 0.24 &                                               - &     D & [1] [2]  \\
&                                    &       &                                               &                &      L-SN~Ia &     Y &         1.188 $\pm$ 0.255 & 0.833 $\pm$ 0.042 &       2.81 &  $-$2.29 $\pm$ 0.30 &                                              $8.23^{+2.61}_{-1.98}$ &     A &          \\  
&                                           &       &                                        &                &     uL-CC~SN &       &         0.374 $\pm$ 0.053 & 0.393 $\pm$ 0.026 &       4.93 &  - &                                     - &     D &     [1]  \\ 
&                    &                       &                                               &                &      L-CC~SN &     Y &         1.188 $\pm$ 0.255 & 0.731 $\pm$ 0.049 &       2.75 &  - &                                      $40.28^{+52.80}_{-22.89}$ &     B &          \\
\hyperref[subsubsec:DESI-058.6486-030.5959]{DESI-058.6486-30.5959} & B &                         (10, 11, 9, 9, 6) &                               (2, 2, 0, 0, 0) &         \txr{B} &     uL-SN~Ia &     Y &                0.1702 &        0.1702 &       1.73 &  0.66 $\pm$ 0.20 &                                           - &     B$-$ &   \\ 
&   &                                        &                                               &                &      L-SN~Ia &       &                0.6735 &        0.6735 &       4.61 &  $-$3.00 $\pm$ 0.04 &                                              $15.82^{+0.54}_{-0.52}$ &     C &     [1]  \\ 
&     &                                      &                                               &                &     uL-CC~SN &       &                0.1702 &        0.1702 &        3.40 &  - &                            - &     C &     [1]  \\
&                            &               &                                               &                &      L-CC~SN &     Y &                0.6735 &        0.6735 &       2.43 &  - &                                    $27.91^{+82.66}_{-20.92}$ &     B+ &         \\
\hyperref[subsubsec:DESI-308.7726-048.2381]{DESI-308.7726-48.2381} & B &                       (12, 12, 10, 15, 8) &                               (0, 1, 1, 1, 0) &              A &     uL-SN~Ia &       &         0.473 $\pm$ 0.032 & 0.448 $\pm$ 0.034 &       4.59 & $-$1.54 $\pm$ 0.61 &                                              - &     D &    [1] [2]  \\
&     &                                      &                                               &                &      L-SN~Ia &     Y &                     - & 0.869 $\pm$ 0.021 &       2.05 & $-$2.91 $\pm$ 0.32 &                                             $14.62^{+5.02}_{-3.74}$ &     B+ &          \\
&      &                                     &                                               &                &     uL-CC~SN &     Y &         0.473 $\pm$ 0.032 & 0.465 $\pm$ 0.026 &       4.88 & - &                                    - &     C &     [1]  \\
&                       &                    &                                               &                &      L-CC~SN &     Y &                     - & 0.828 $\pm$ 0.057 &       2.98 & - &                                     $15.61^{+45.94}_{-11.66}$ &     B$-$ &    
\\  \\
\hline \\
\hyperref[subsubsec:DESI-034.3625-035.3563]{DESI-034.3625-35.3563} & C &                        (11, 11, 10, 8, 9) &                              (2, 2, 1, 0, 0)  &         A &     uL-SN~Ia &     Y &         0.240 $\pm$ 0.008 &  0.238 $\pm$ 0.011 &       1.31 & 0.20 $\pm$ 0.25 &                                         - &           A &          \\  &      &                                     &                                               &                &      L-SN~Ia &       &                     - & 0.294 $\pm$ 0.026 &       1.64 & $-$0.09 $\pm$ 0.24 &                                            $1.09^{+0.27}_{-0.22}$ &     D & [3]  \\
&     &                                      &                                               &                &     uL-CC~SN &       &         0.240 $\pm$ 0.008 & 0.242 $\pm$ 0.011 &       1.71 & - &                                     - &     B &          \\  &                     &                      &                                               &                &      L-CC~SN &     Y &                     - & 0.530 $\pm$ 0.218 &       1.97 & - &                                    $35.16^{+50.83}_{-20.79}$ &     C &     [1]  \\ \hyperref[subsubsec:DESI-035.1374+00.4676]{DESI-035.1374+00.4676} & C &                        (21, 16, 7, 14, 9) &                              (1, 0, 1, 1, 0)  &              \txr{C} &     uL-SN~Ia &     Y &         0.269 $\pm$ 0.007 & 0.269 $\pm$ 0.008 &       1.20 &  0.27 $\pm$ 0.33 &                                        - &     A &          \\   &   &                                        &                                               &                &      L-SN~Ia &       &         0.776 $\pm$ 0.118 & 0.727 $\pm$ 0.056 &        1.25 &  $-$1.90 $\pm$ 0.41 &                                                $5.74^{+2.63}_{-1.81}$ &     B &       \\
&     &                                      &                                               &                &     uL-CC~SN &       &         0.269 $\pm$ 0.007 & 0.269 $\pm$ 0.005 &        1.84 & - &                                  - &     B &          \\
&                               &            &                                               &                &      L-CC~SN &     Y &         0.776 $\pm$ 0.118 & 0.795 $\pm$ 0.050 &       0.44 & - &                                    $18.97^{+55.29}_{-14.13}$ &     A &          \\
\hyperref[subsubsec:DESI-052.0083-037.2049]{DESI-052.0083-37.2049} & C &                        (43, 11, 64, 9, 4) &                               (3, 2, 1, 2, 0) &         A &     uL-SN~Ia &     Y &         0.292 $\pm$ 0.029 & 0.333 $\pm$ 0.023 &       1.48 & 0.47 $\pm$ 0.16 &                                               - &     B &          \\
&                      &                     &                                               &                &      L-SN~Ia &       &                     - & 0.357 $\pm$ 0.013 &       1.46 & 0.37 $\pm$ 0.10 &                                           $0.71^{+0.07}_{-0.07}$ &     D &     [3]  \\
&        &                                   &                                               &                &     uL-CC~SN &       &         0.292 $\pm$ 0.029 & 0.316 $\pm$ 0.018 &       2.07 & - &                                      - &     D &     [1]  \\
&                        &                   &                                               &                &      L-CC~SN &     Y &                     - & 0.450 $\pm$ 0.018 &       1.54 & - &                                    $12.60^{+18.20}_{-7.45}$ &     B &          \\
\hyperref[subsubsec:DESI-084.8493-059.3586]{DESI-084.8493-59.3586} & D &                          (9, 10, 6, 9, 3) &                               (3, 3, 0, 2, 0) &              \txr{D} &     uL-SN~Ia &     Y &         0.361 $\pm$ 0.015 & 0.365 $\pm$ 0.008 &       1.61 & $-$0.15 $\pm$ 0.09 &                                              - &     B &          \\
&     &                                      &                                               &                &      L-SN~Ia &      &         0.593 $\pm$ 0.146 & 0.391 $\pm$ 0.011 &       1.41 & $-$0.20 $\pm$ 0.12 &                                            $1.20^{+0.12}_{-0.11}$ &     D &   [3]       \\
&          &                                 &                                               &                &     uL-CC~SN &       &         0.361 $\pm$ 0.015 & 0.406 $\pm$ 0.006 &       3.64 & - &                                    - &     D & [1]  \\
&                             &              &                                               &                &      L-CC~SN &   Y   &         0.593 $\pm$ 0.146 & 0.819 $\pm$ 0.064 &       2.02 & - &                                     $86.05^{+153.15}_{-55.10}$ &     C &     [1]  \\
\hyperref[subsubsec:DESI-015.8465-050.5450]{DESI-015.8465-50.5450} & N/A &                           (9, 7, 4, 9, 5) &                               (1, 1, 0, 0, 0) &              N/A &     uL-SN~Ia &     Y &         0.301 $\pm$ 0.026 & 0.373 $\pm$ 0.131 &       0.58 & $-$0.91 $\pm$ 1.52 &                                         - &          B &          \\
&   &                                        &                                               &                &      L-SN~Ia &       &                     - &             - &          - & - &                                         - &        - &           \\
&           &                                &                                               &                &     uL-CC~SN &     Y &         0.301 $\pm$ 0.026 & 0.289 $\pm$ 0.036 &       1.30 &                                - &        - &     B &          \\
&    &                                       &                                               &                &      L-CC~SN &       &                     - &             - &          - & - &                                          - &     - &       \\ \\
\enddata
\tablecomments{\txr{Summary table for our eight lensed and unlensed supernova candidates.  \txr{Column~2: A wholistic grade on how likely a detection is a L-SN, based on the criteria in \S~\ref{sec:analysis}.}  Column~5: The distance grade, based on the location of the transient (for first round visual inspection, \S~\ref{sec:analysis}).  Column~6: The four postulations for each candidate, uL-SN~Ia, L-SN~Ia, uL-CC~SN, L-CC~SN.  Column~7: \txb{Only light curve models for postulations marked with ``Y” in this column} are shown (in \S~\ref{subsubsec:ABCandidates} and Appendix~\ref{appendix_SNC}, above and below the horizontal line, respectively).  Column~8: For redshift prior in the fitting process, we use photometric redshifts from \citet[][shown with three decimal places and uncertainties]{zhou2020}, or fix them to be the spectroscopic redshifts from A. Cikota et al. (in prep, shown with four decimal places and no uncertainties).  Columns~14 and 15: Each postulation is given a grade based on the light curve fitting and/or the Hubble residuals, as well as reasons for a C or D grade.  The reasons correspond to the following: [1] poor fit in comparision to the other postulations, [2] large Hubble residual in the case of uL-SN~Ia, and [3] best-fit $z_{SN}$ too close to the lens photo-$z$, and hence inconsistent with a lensing postulation (Note that, to be complete, we nevertheless include the best-fit $z_{SN}$ and magnification information).}}
\end{deluxetable*}
\end{longrotatetable}

\newpage
 \subsection{Grade A \& B Lensed Supernova Candidates}\label{subsubsec:ABCandidates}
 
The first seven systems (of eight) in Table~\ref{tab:sncandidates} are identified by the pipeline and \txb{determined by visual inspection} to be lensed \txb{SN} candidates.  For the \txb{eighth} system, we believe that \txb{it} is almost certainly not a strongly lensed system, and have given it an overall grade of ``N/A".  In this section, we will present the best three lensed \txb{SN} candidates, with the remaining four (and one unlensed \txb{SN}) candidates presented in Appendix~\ref{appendix_SNC}.

For each system, we attempt to narrow the identity of the transients by fitting different light curve models to the photometry.  We test for \txb{four} different postulations for each system: unlensed SN~Ia (uL-SN~Ia), lensed SN~Ia (L-SN~Ia), unlensed CC~SN (uL-CC~SN), and unlensed CC~SN (uL-CC~SN).  In this paper, we present figures only for the most probable scenarios (systems with ``Y" in the ``Shown" column 7 of Table~\ref{tab:sncandidates}).

We account for Milky Way extinction according to \citet{schlafly2011}.  When fitting a SN~Ia light curve model (for uL-SN~Ia and L-SN~Ia), we use the SALT3 model (\citealp{salt3}), and fit for the parameters: redshift $z$, time of \ipB~band peak $t_0$, the normalization factor \txb{$x_0$,} the ``stretch" factor $x_1$, and the color parameter $c$.  We do not fit for host-galaxy dust, as the $c$ parameter would be largely degenerate with small amounts of reddening.  We use the following priors in the fitting process: $c \sim \mathcal{N}(0, 0.1)$ and $x_1 \sim \mathcal{N}(0, 1)$.  We then use the best-fit SALT parameters for ``stretch" and color corrections, and \txb{find} the Hubble residual for the given model.  This is plotted together with SNe~Ia from \citet{suzuki2012}.  \txb{Though this data is modelled with SALT2, we expect negligible differences in Hubble residuals of $7 \pm 11$ mmag \citep{salt3}}.  To model CC~SNe light curves, we fit for 161 separate CC~SNe templates (as provided by SNCosmo).  All \txb{CC} templates are parameterized by only $t_0$, $z$, and amplitude (a scaling term with arbitrary units), allowing for a small amount of host galaxy dust.  \txb{In} all the \txg{CC~SN} light curves shown in this paper, the best-fit $E(B-V)$ \txb{values are} very small ($\lesssim 0.01$), and therefore we do not report the reddening parameters.  \txr{Except for one system, redshift priors used for both CC~SNe and SNe~Ia postulations (see column 8 of Table~\ref{tab:sncandidates}) are photometric redshifts of objects identified by the forward modeling source extraction algorithm, the Tractor \citep{lang2016}.  \txb{All photometric redshifts in this paper are from \citet{zhou2020}}.  For DESI-058.6486-30.5959, we fix the redshifts to be the spectroscopic redshifts (A. Cikota et al. in prep).}

In the figures below, if a lensed scenario is \txb{postulated}, we estimate the amplification, $\mu$.  If a CC~SN scenario is \txb{postulated}, the expected peak \ipB~band magnitude for a given SN type (``X") is represented with ``$\emph{E}(\text{M}_\text{B}~|~\text{X})$," where the values from Table~\ref{table:ccrates} are used.

\subsubsection{DESI-344.6252-48.8977}
 \label{subsubsec:DESI-344.6252-048.8977}

\begin{figure}[H]
\begin{center}
\includegraphics[width=175mm]{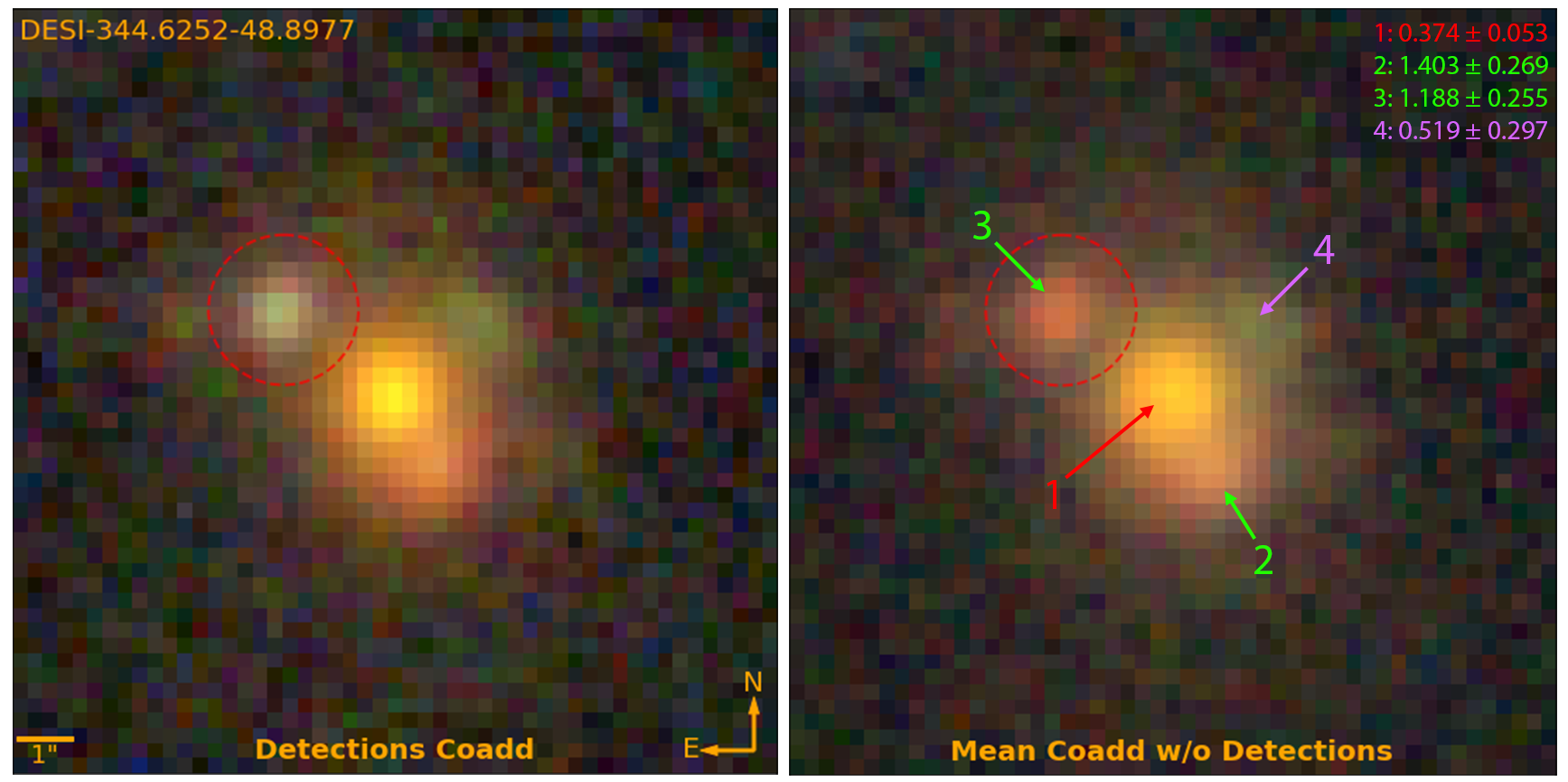}
\caption{Coadded RGB images (using \ipg, \ipr, \ipi, and \ipz~bands) of L-SN~Ia candidate DESI-344.6252-48.8977 with and without the transient detection \txb{(red dotted circle)} exposures.  The arrows are color-coded as the following postulated scenario: red for the lens galaxy, green for the lensed source galaxy, and purple for a possible second lensed source or an interloper.  Photometric redshifts are displayed on the top right.  \trr{The posited lens galaxy has a photo-$z$ of $0.374\pm 0.053$, and the posited lensed images have photo-$z$'s of $1.188\pm 0.255$ and $1.403 \pm 0.269$.}}\label{fig:3760rgb}
\end{center}
\end{figure}

DESI-344.6252-48.8977 is a strong lensing candidate discovered in \citet{storfer2022} \txb{and} assigned a C-grade.  \txr{In our analysis, we find this }relatively-low grade was given because in the Legacy Surveys coadded image, the arc and counterarc (objects 2 and 3 respectively, in Figure~\ref{fig:3760rgb}) appear to have \txr{somewhat} different colors due to the transient.  However, with an improved analysis of the color of the arc and counterarc by the criteria laid out in \txb{\citet{huang2021},} taking into account of the presence of the transient and photo-$z$ (see below), we now regrade this system as an A-grade lensing candidate.
 
\begin{figure}[H]
\begin{center}
\includegraphics[width=175mm]{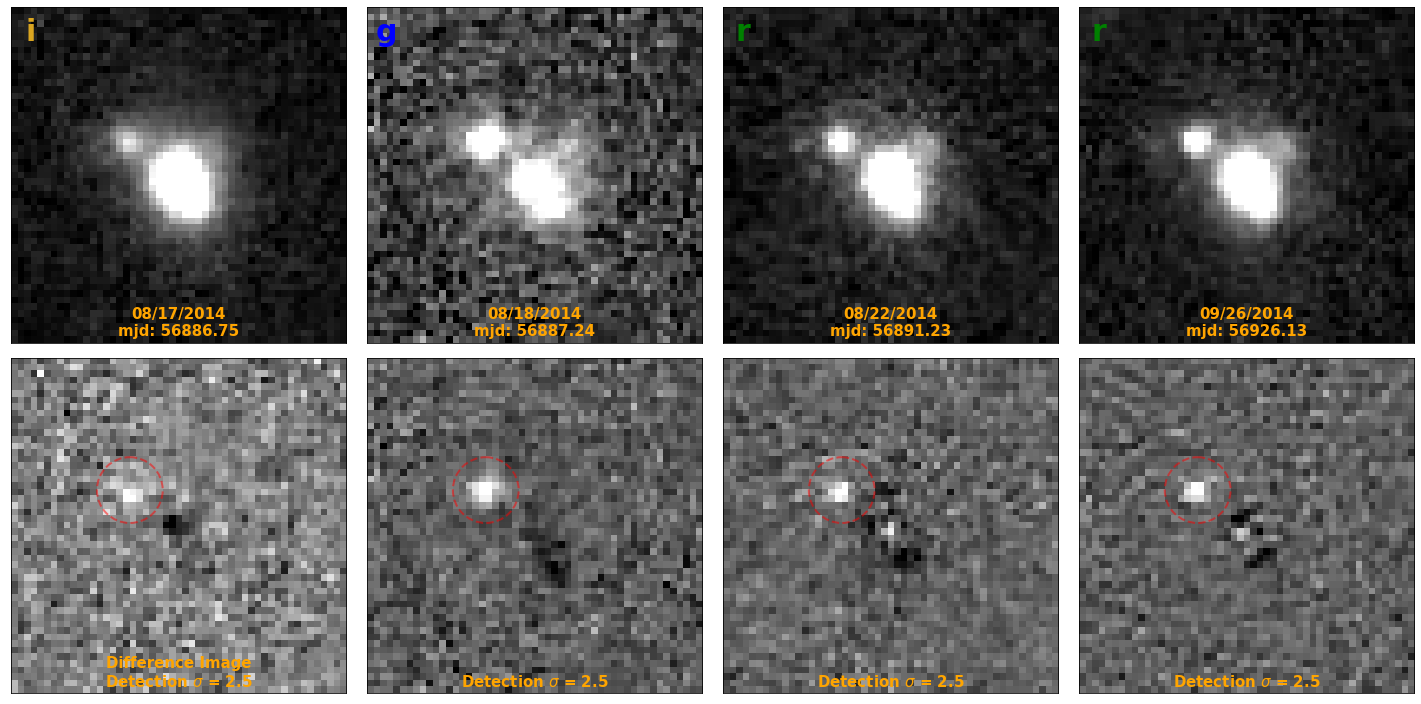}
\caption{Detection exposures for the transient in DESI-344.6252-48.8977 in chronological order.  Top row: single pass images.  Bottom row: corresponding differencing images from SFFT, where the red circle indicates the location of the detected transient.}\label{fig:3760dets}
\end{center}
\end{figure}

The lensed arc in DESI-344.6252-48.8977 is located Southwest of the lens, with its counterimage appearing Northeast of the lens.  The color and photo-$z$'s of the two lensed images agree with each other (within uncertainties), with both photo-$z$'s \trr{($z=1.403 \pm 0.269$ and $1.188 \pm 0.255$)} being significantly higher than the lens photo-$z$ \trr{($z=0.374 \pm 0.053$)}.  \txr{We note that for object 3, as \txb{the majority} of the exposures do not contain the transient, its light is unlikely to significantly affect the photo-$z$ of object 3.}  As the lens and source galaxies appear to be red elliptical galaxies, we \txb{consider their photo-$z$'s} to be reliable.  The image-counterimage arrangement indicates a strong lensing configuration.  The detected transient lies directly on the counterimage.  As with all other light curves presented, the light curves below are constrained by both detection and non-detection exposures.

\begin{figure}[H]
\begin{center}
\includegraphics[width=180mm]{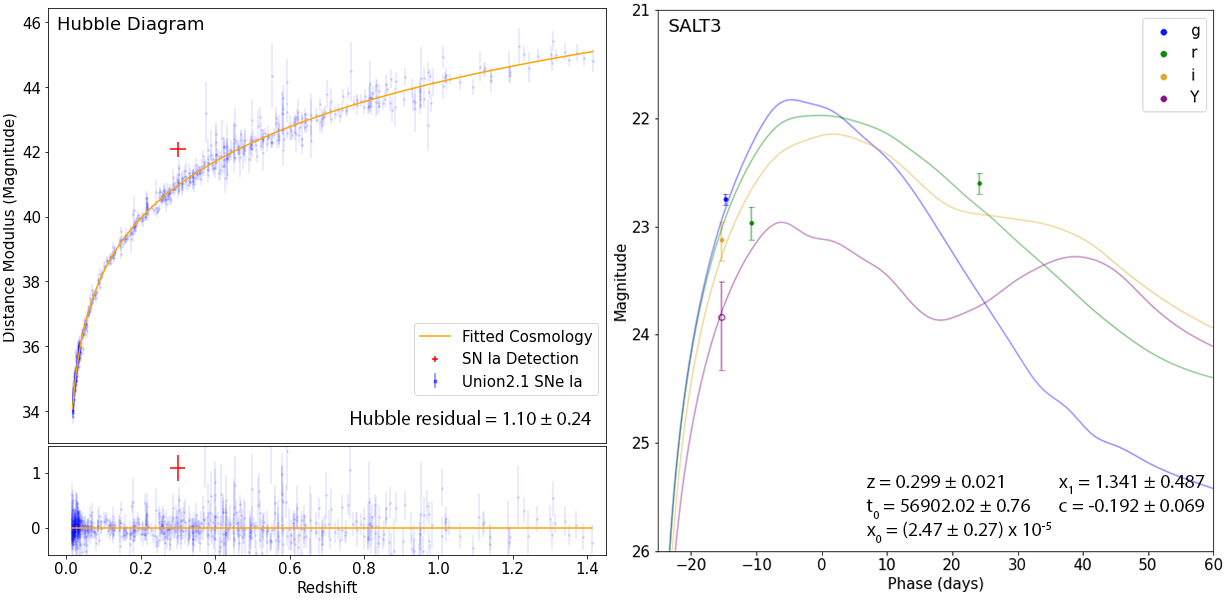}
\caption{Best-fit SALT3 model for DESI-344.6252-48.8977 with a lens photo-$z$ redshift prior of $0.374 \pm 0.053$.  For this and Figures~\ref{fig:3760p2} and \ref{fig:3760p3}, solid photometry points correspond to the detection passes shown in Figure~\ref{fig:3760dets}.}\label{fig:3760p1}
\end{center}
\end{figure}

\paragraph{Postulation 1: uL-SN~Ia} Figure~\ref{fig:3760p1} shows the best-fit light curve model for the uL-SN~Ia scenario.  We see that it is not a good fit ($\chi^2/\text{DOF} = 5.47$), especially in the \ipr~band.  Additionally, the resulting SALT3 model has a statistically significant Hubble residual \trr{of $1.10 \pm 0.24$}.  Thus, we believe this detection is unlikely to be an uL-SN~Ia.  Not shown is the uL-CC~SN scenario, which also (as with the uL-SN~Ia scenario) has a large $\chi^2/\text{DOF}$ (see Table~\ref{tab:sncandidates}).

\begin{figure}[H]
\begin{center}
\includegraphics[width=180mm]{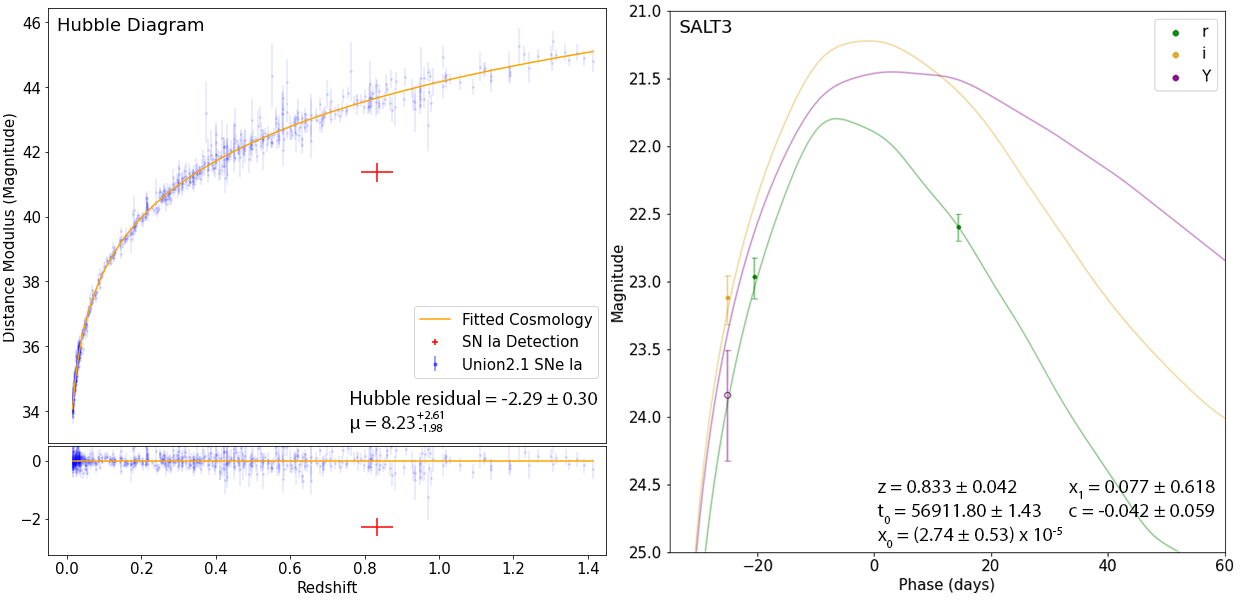}
\caption{Best-fit SALT3 model for DESI-344.6252-48.8977 with a source photo-$z$ redshift prior of $1.188 \pm 0.255$.}\label{fig:3760p2}
\end{center}
\end{figure}

\newpage\paragraph{Postulation 2: L-SN~Ia} Figure~\ref{fig:3760p2} shows the best-fit light curve model for the L-SN~Ia scenario.  Due to the high redshift, SALT3 cannot model the \ipg~band observation and it is ignored in the fitting process.  For the three redder bands, the best-fit SALT3 curve model agrees well with the photometric data.  Based on the Hubble residual \trr{of $-2.29 \pm 0.30$}, the implied amplification is $8.23^{+2.61}_{-1.98}$.  This is consistent with the expectation of a multiply imaged SN by a galaxy scale lens (e.g., \citealp{shu2018}).  \txb{Finally, the best-fit SN redshift, \trr{$z_{SN}=0.833 \pm 0.042$}, is consistent with the photo-$z$ of the source galaxy.  We note that this redshift value is in line with our preliminary investigation of the feasibility of our search (see Figure~\ref{fig:just}).}

\begin{figure}[H]
\begin{center}
\includegraphics[width=110mm]{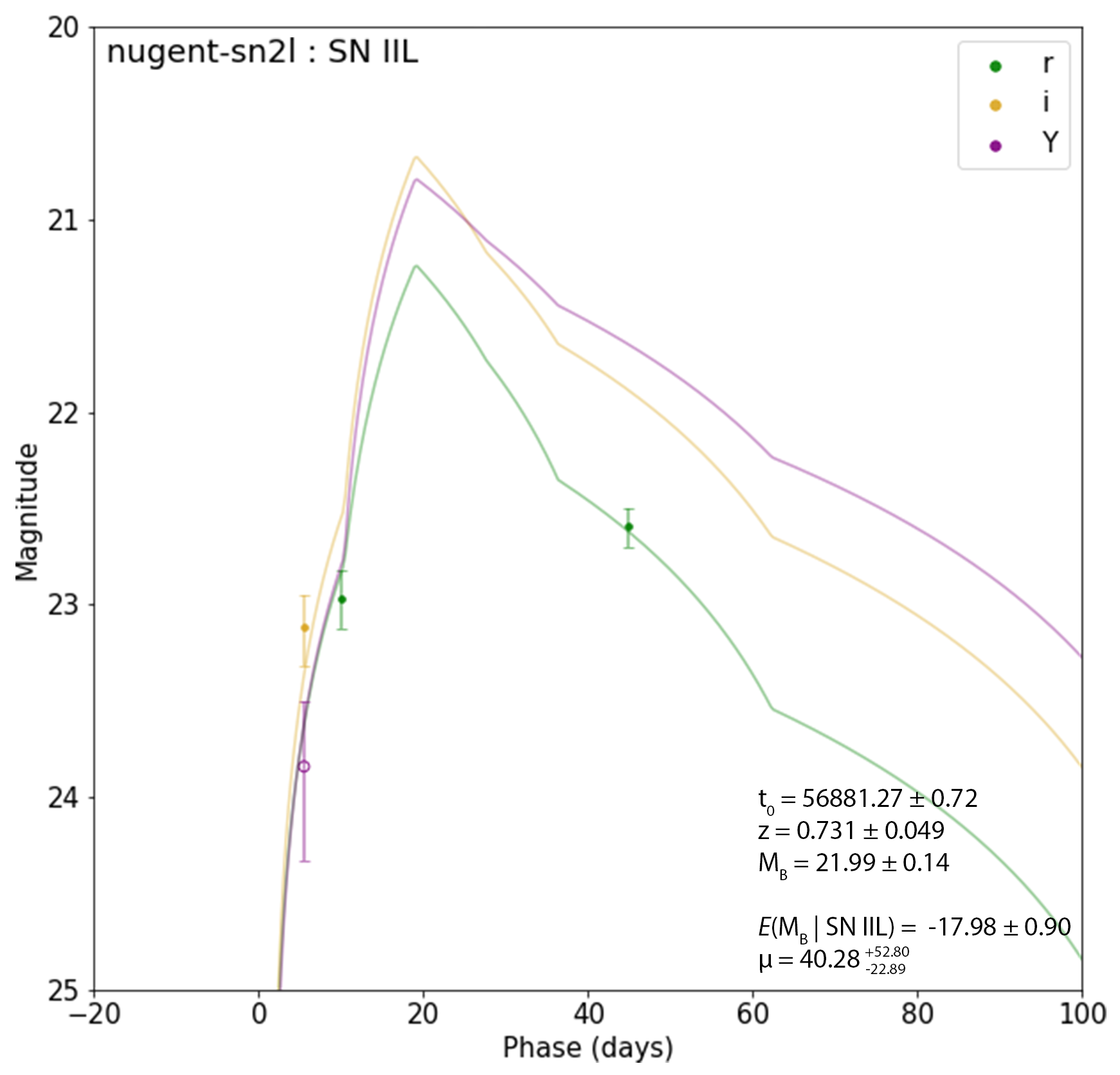}
\caption{Best-fit core collapse template model for DESI-344.6252-48.8977 with a lensed source photo-$z$ redshift prior $1.188 \pm 0.255$.}\label{fig:3760p3}
\end{center}
\end{figure}

\newpage\paragraph{Postulation 3: L-CC~SN} Figure~\ref{fig:3760p3} shows the best-fit light curve model for the L-CC~SN scenario.  As with the previous postulation, this model cannot fit the \ipg~band datapoint due to the high redshift, and thus is ignored in the fitting process.  The best-fit model (the ``nugent-sn2l" SN~IIL template) is in good agreement with the photometric data from the redder bands.  However, the implied magnification is quite large.  Despite this, we note that 1.) such a magnification is not impossible \citep[e.g., ][]{quimby2014}, and 2.) since CC~SNe have a large range of $M_B$, the uncertainties of the amplification is large.  Therefore the L-CC~SN scenario is still plausible for this system.

\paragraph{Conclusion} \txr{Any unlensed SN postulation seems unlikely, due to the poor agreement between the light curve model and the data.  Additionally, for the uL-SN~Ia scenario, the Hubble residual would be too high.  With all the evidence considered, this is very likely a lensed SN.  A L-SN~CC is possible.  But, we believe it is most likely a L-SN~Ia.  This conclusion is based on} the following:
\begin{enumerate}
  \item The red color and morphology seem to indicate that the putative lensed source is an elliptical galaxy.  The foreground galaxy is clearly an elliptical galaxy.  Thus, the photo-$z$'s for the punitive lens and source are both likely reliable, with the later being significantly higher than the former.  Based on this, combined with the classic image-counterimage configuration, we regard this system as a grade-A lens candidate.
  \item The putative SN is situated directly on the counterimage.
  \item Given the source galaxy is likely an elliptical galaxy, the SN is more likely a Ia than CC.  
  \item \txr{The SN~Ia light curve model is a good fit to the photometry and the Hubble residual is most consistent with it being lensed.  For this scenario, the amplification is also consistent with galaxy-scale strong lensing.}
\end{enumerate}
\txr{If it is indeed a lensed SN, this would \txb{be} the first galaxy-scale strongly lensed SN resolved by ground-based observations.  Furthermore, either for the case of L-CC~SN or L-SN~Ia, it is at a significantly higher redshift \txb{($z_{SN} \gtrsim 0.8$)} than the other two \txb{resolved} galaxy-scale strongly lensed SNe (\citealp{goobar}, \citealp{astronotegoobar}).  Given the Einstein radius is $\sim 1.5''$, the expected time delay would be on the order of weeks.  If caught live, it could have resulted in a $H_0$ measurement competitive with those from lensed quasars \citep[e.g.,][]{wong2019}.  \txr{Additionally, if it were a L-SN~Ia, the systematic effect of the mass sheet degeneracy could be significantly reduced due to the standardizability of its brightness} \citep[e.g.,][]{birrer2021}.  Therefore, if discovered live, this system would make a strong case for high resolution imaging and spectroscopic follow-up observations.\footnote{At the present, it is still possible to obtain the source galaxy spectra to measure its star formation rate.  This in turn would \txb{more precisely quantify} the SN~Ia likelihood.}}

\subsubsection{DESI-058.6486-30.5959}
 \label{subsubsec:DESI-058.6486-030.5959}

DESI-058.6486-30.5959 was discovered in \citet{huang2021}, as a grade-A strong lensing candidate.

  \begin{figure}[H]
\begin{center}
\includegraphics[width=175mm]{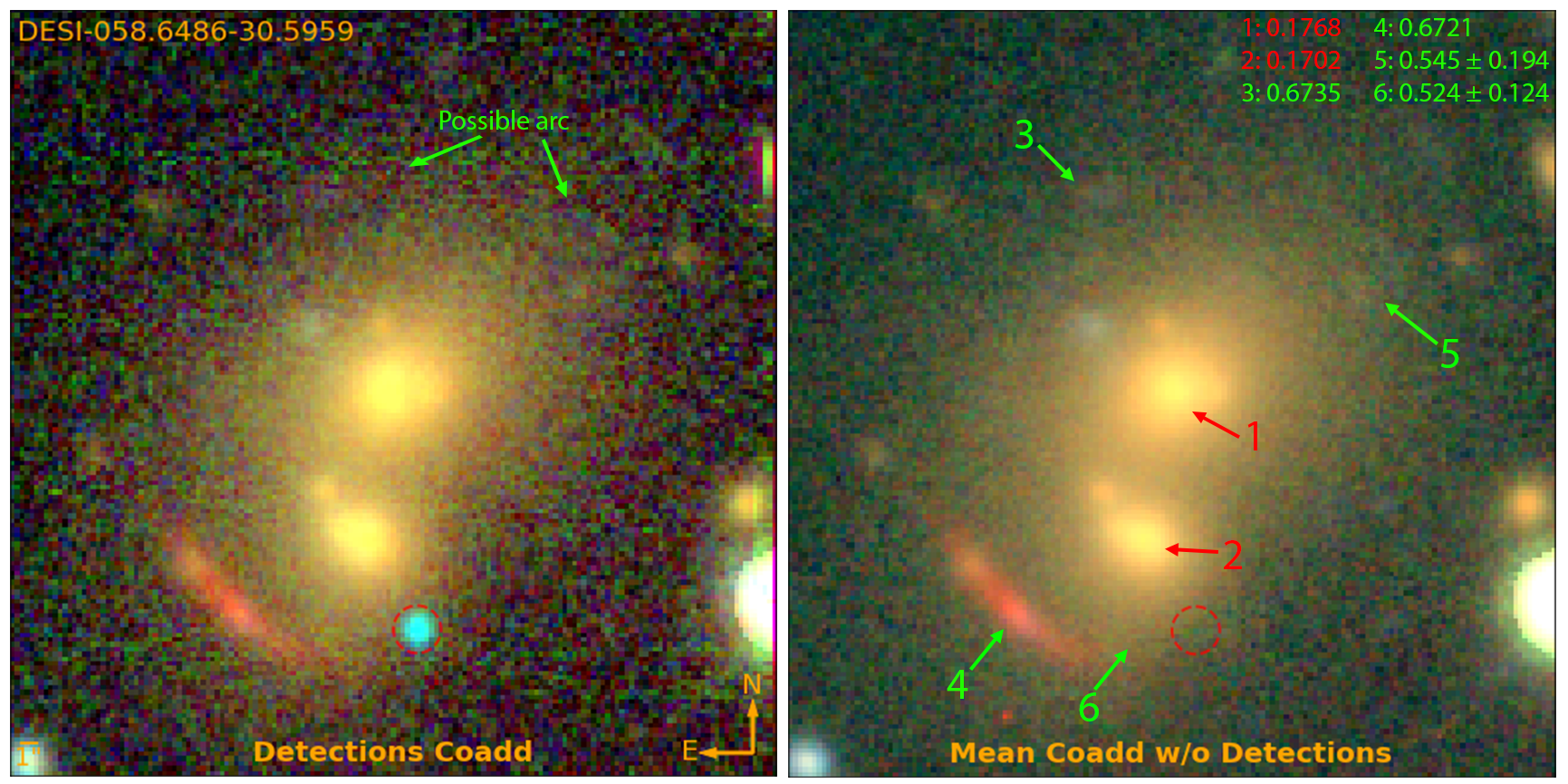}
\caption{Coadded RGB images (using \ipg, \ipr, \ipi, and \ipz~bands) of DESI-058.6486-30.5959 with and without the transient detection \txb{(red dotted circle)} exposures.  Labelled objects are color-coded as the following postulated scenario: red as the lenses and green as the source galaxy.  Photometric (shown with uncertainties) and spectroscopic (shown without uncertainties; A. Cikota et al. in prep) redshifts for the labelled objects are displayed on the top right.  \trr{The posited lens galaxy has a spectroscopic redshift of $0.1702$, and the posited lensed images have spectroscopic redshift of $0.6735$ and $0.6721$.}}\label{fig:411rgb}
\end{center}
\end{figure}

  \begin{figure}[H]
\begin{center}
\includegraphics[width=170mm]{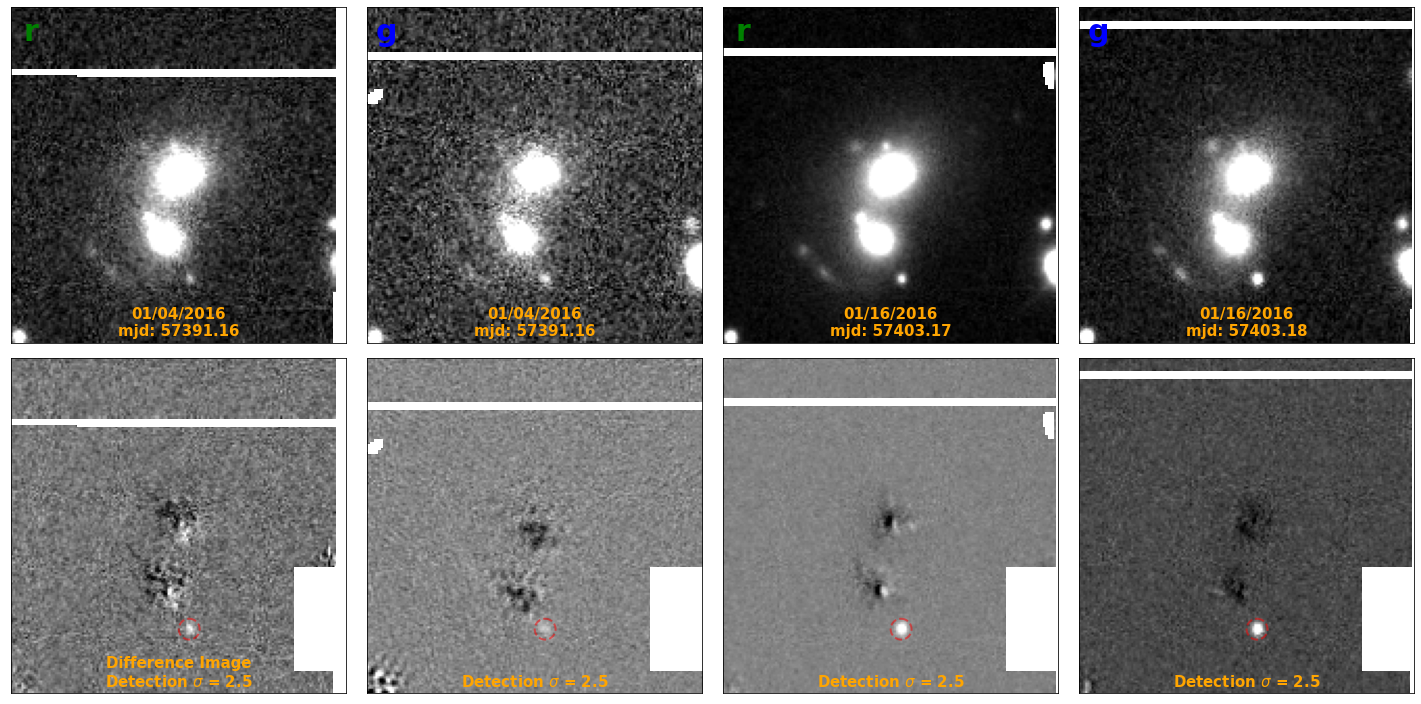}
\caption{Detection exposures for the transient in DESI-058.6486-30.5959 in chronological order. See caption of Figure~\ref{fig:3760dets} for the full description.}\label{fig:411dets}
\end{center}
\end{figure}

  \begin{figure}[H]
\begin{center}
\includegraphics[width=110mm]{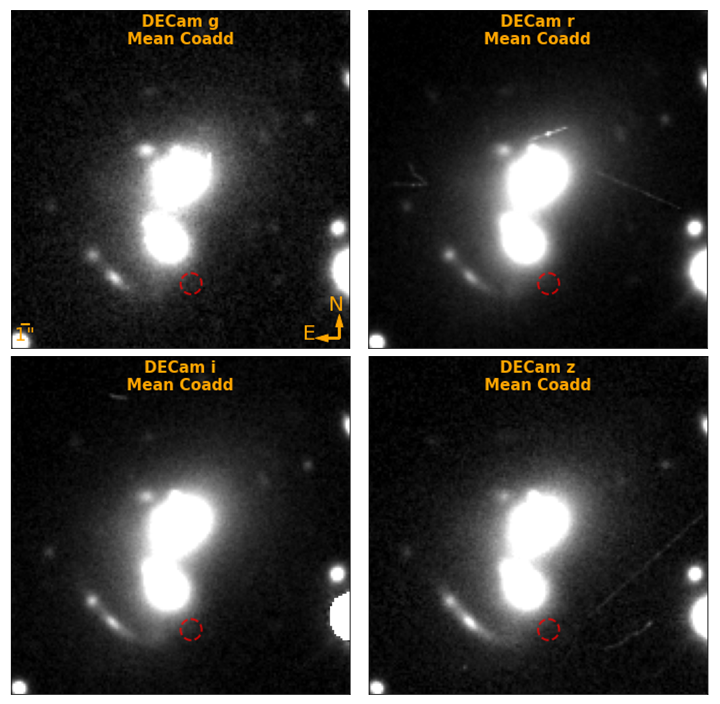}
\caption{Mean coadded images of DESI-058.6486-30.5959 across different observed bands.  These \txr{deeper} images (especially the \ipr~and \ipi~bands) seem to support the possibility of a faint lensed image between the detection (red dotted circle) and the arc.}\label{fig:411coadds}
\end{center}
\end{figure}

DESI-058.6486-30.5959 is a galaxy-group \txr{lensing} system, with a prominent red arc Southeast of the lens.  Including additional observations from DR10, there appears to be a faint, highly magnified blue arc Northwest of the lens as well, somewhat further from the estimated center of mass for the foreground galaxy group than the red arc.  For this system, we have obtained VLT MUSE spectroscopy with preliminary 
\txb{redshifts (A. Cikota et al. in prep) for objects~1, 2, 3, and 4 in Figure~\ref{fig:411rgb}}.  The clearly detected transient lies $\sim 3''$ from the tip of the red arc \txr{(object~4, spectroscopically confirmed to be in the background)} in the deep \ipi~band image (Figure~\ref{fig:411coadds}, lower left panel).  The deep \ipz~band image (Figure~\ref{fig:411coadds}, lower right panel) seems to show that the arc curves towards the direction of the transient.  There also seems to be a very faint galaxy \txr{(object~6)} between the red arc and the transient location.  
\txb{The source extraction code for the Legacy Surveys, the Tractor, also identifies this object.}  
\txr{\txb{It has} similar photo-$z$ \txb{\citep{zhou2020}} as the aforementioned blue arc (object~3, also spectroscopically confirmed to be in the background), and therefore could possibly be its counter-image.}

\begin{figure}[H]
\begin{center}
\includegraphics[width=180mm]{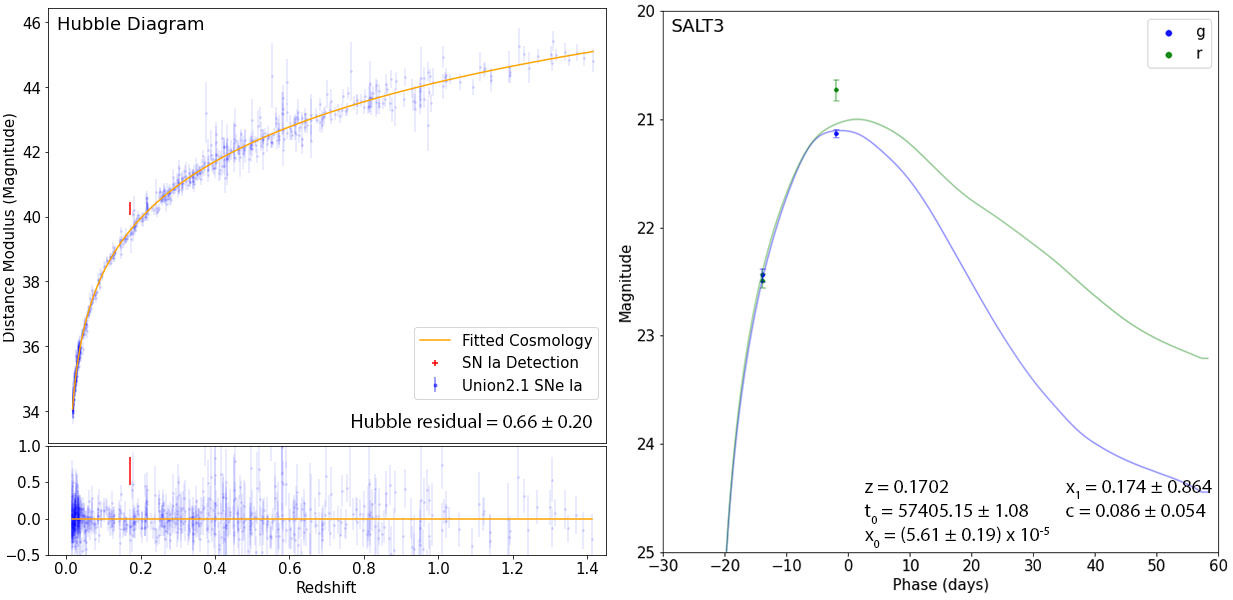}
\caption{Best-fit SALT3 model for DESI-058.6486-30.5959 with a lens redshift of $0.1702$.  For this and Figure~\ref{fig:411p3}, solid photometry points correspond to the detection passes shown in Figure~\ref{fig:411dets}.}\label{fig:411p1}
\end{center}
\end{figure}

\paragraph{Postulation 1: uL-SN~Ia} Figure~\ref{fig:411p1} shows the best-fit light curve model for the uL-SN~Ia scenario.  We see that the second \ipr~band photometry point is not well accounted for in this SALT3 model, but not at an unreasonable level.  Additionally, the Hubble residual indicates that this scenario is unusually faint.  

\begin{figure}[H]
\begin{center}
\includegraphics[width=110mm]{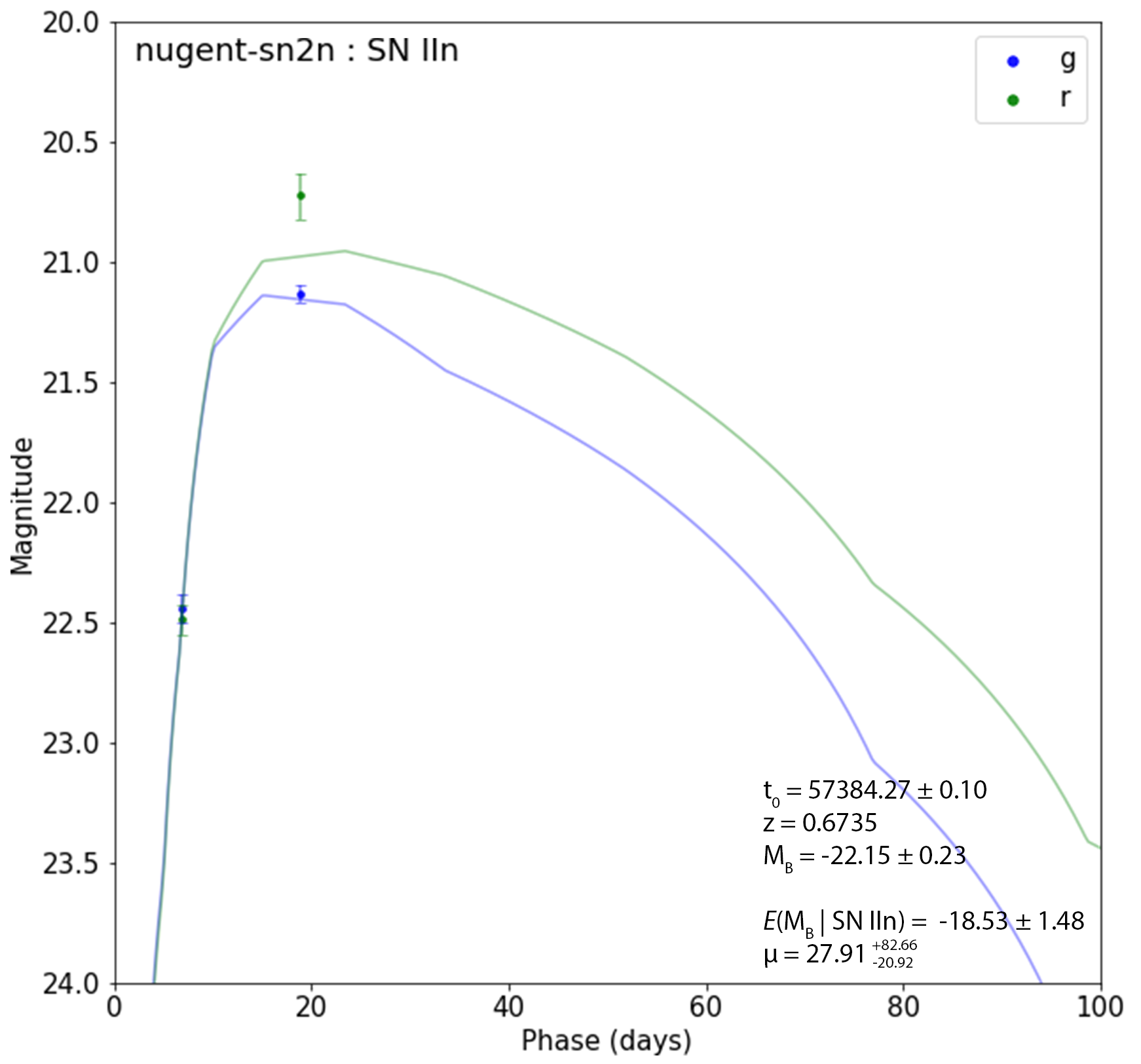}
\caption{Best-fit core collapse template model for DESI-058.6486-30.5959 with a source photo-$z$ redshift of $0.6735$.}\label{fig:411p3}
\end{center}
\end{figure}

\newpage\paragraph{Postulation 2: L-CC~SN} Figure~\ref{fig:411p3} shows the best-fit light curve model (with the best-fitting parameters of the best-fitting CC~SN templates) for the L-CC~SN scenario.  The best CC~SN template to the data is the ``nugent-sn2n" SN~IIn model.  This model seems to provide the best overall fit.  The implied amplification would be $27.91^{+82.66}_{-20.92}$.  Depending on the location of the SN relative to the \txb{lensing} critical curve, a large amplification for a group-scale lens is not impossible (e.g., SN Refsdal; \citealp{kelly2016} and \citealp{rodney2016}).  We also note the large uncertainty, as typical for CC~SNe.

\paragraph{Conclusion} 
The rise time for the transient in DESI-058.6486-30.5959 is consistent with it being a \txb{SN}.  If so, there are three possibilities for the host galaxy -- objects 2, 4, or 6.  Object 2 is an elliptical galaxy \trr{(A. Cikota et al. in prep)}.  Thus, if it hosts a SN, it is \txr{more likely a Ia than CC}.  Figure~\ref{fig:411p1} shows that the SALT3 fit for a SN~Ia at its spectroscopic redshift cannot account well for the \ipr~band photometry near maximum light.  Furthermore, the Hubble residual of \trr{$0.66 \pm 0.20$}, is unusually large at $> 3\sigma$.  Object 4 appears to be at the greatest angular separation from the transient.  However, given the high degree of distortion due to lensing, without lens modeling, it is difficult to meaningfully assess how far away the SN is from object 4 ---~\txb{e.g., in terms of half-light radius or directional light radius \citep{sako2014} if the delensed source is highly elliptical}.  Object 6 has the least angular separation from the transient.  By color, location, and photo-$z$, it appears to be the possible counterarc of the large arc (spectroscopically confirmed) to the Northwest of the lens.  The L-CC~SN postulation is also consistent with object 6 appearing to be a blue, and therefore likely star-forming, galaxy.  Given the sparsity of the photometric data, it is difficult to be certain.  All factor considered, we assigned a grade of B to this transient as a lensed \txb{SN (more likely a CC than Ia)}.  Additional spectroscopic observation of object 6 can test whether it is a lensed counterimage.  We also note that if this transient was detected live, real time photometric and spectroscopic \txb{follow-up} observation could be triggered to determine the nature of this transient.

 \subsubsection{DESI-308.7726-48.2381}
 \label{subsubsec:DESI-308.7726-048.2381}

\begin{figure}[H]
\begin{center}
\includegraphics[width=175mm]{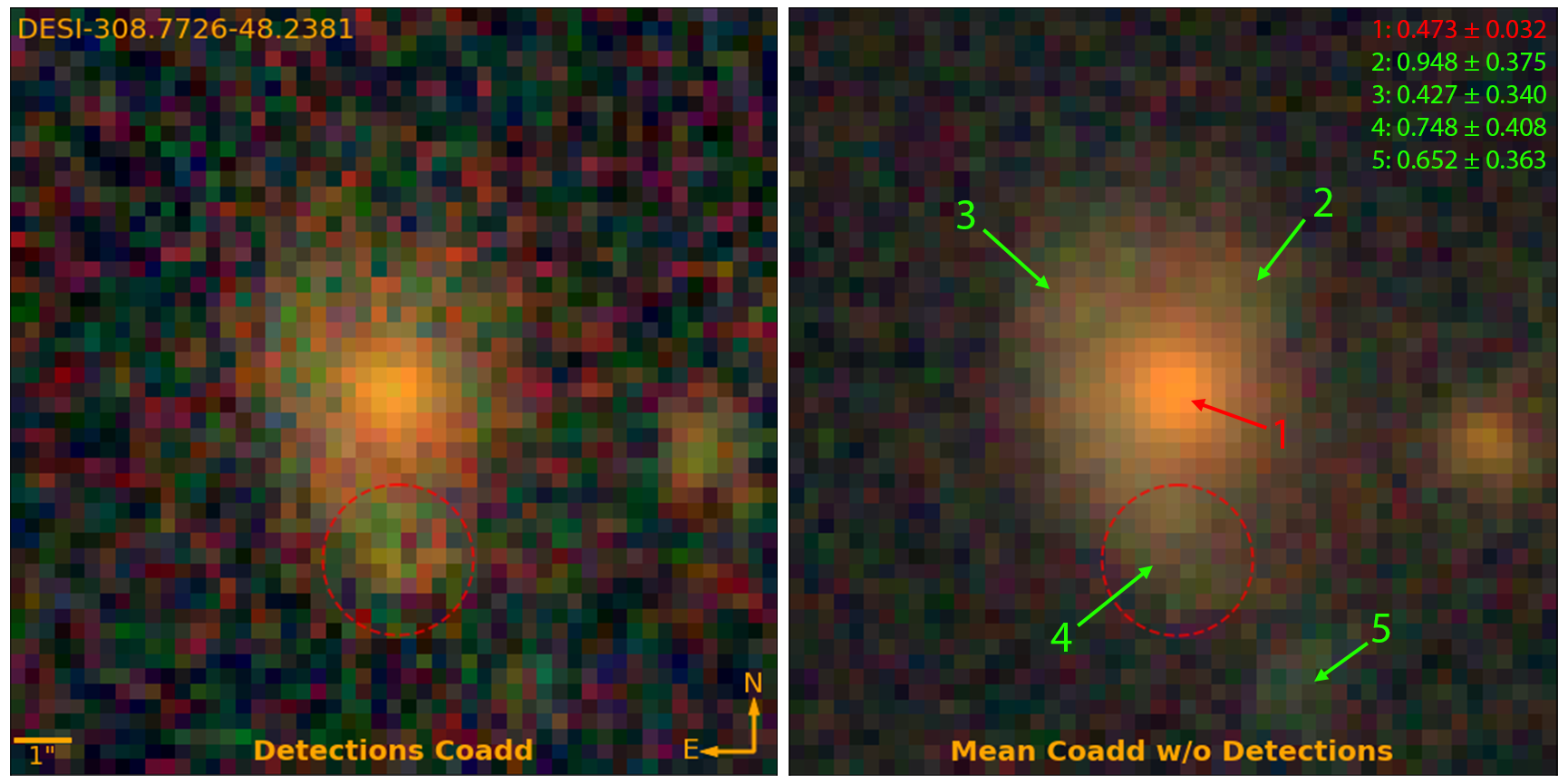}
\caption{Coadded RGB images (using \ipg, \ipr, \ipi, and \ipz~bands) of DESI-308.7726-48.2381 with and without the transient detection \txb{(red dotted circle)} exposures.  Labelled objects are color-coded as the following postulated scenario: red as the lens galaxy and green as the source galaxy.  Photometric redshifts are displayed on the top right.  \trr{The posited lens galaxy has a photo-$z$ of $0.473\pm 0.032$, and the posited lensed images have photo-$z$'s of $0.948\pm 0.375$, $0.427 \pm 0.340$, $0.748 \pm 0.408$, and $0.652 \pm 0.363$.}}\label{fig:5531rgb}
\end{center}
\end{figure}

DESI-308.7726-48.2381 was discovered in \citet{storfer2022}\txr{, and is given a} D+ grade strong lensing candidate (see \S~\ref{sec:data}).  If it turns out to be a lensing system, the location of the detection would lie directly on the arc (Figure~\ref{fig:5531rgb}).  

\begin{figure}[H]
\begin{center}
\includegraphics[width=90mm]{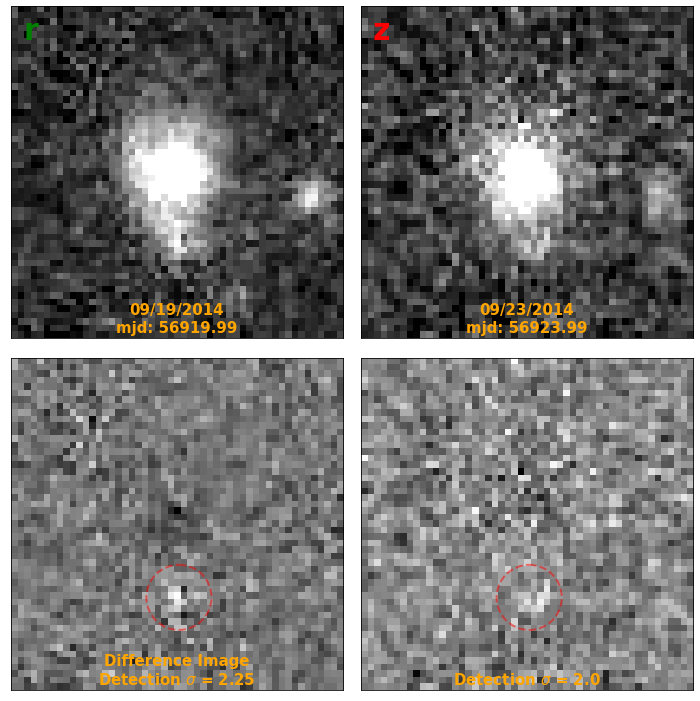}
\caption{Detection exposures for the transient in DESI-308.7726-48.2381 in chronological order. See caption of Figure~\ref{fig:3760dets} for the full description.}\label{fig:5531dets}
\end{center}
\end{figure}

\begin{figure}[H]
\begin{center}
\includegraphics[width=180mm]{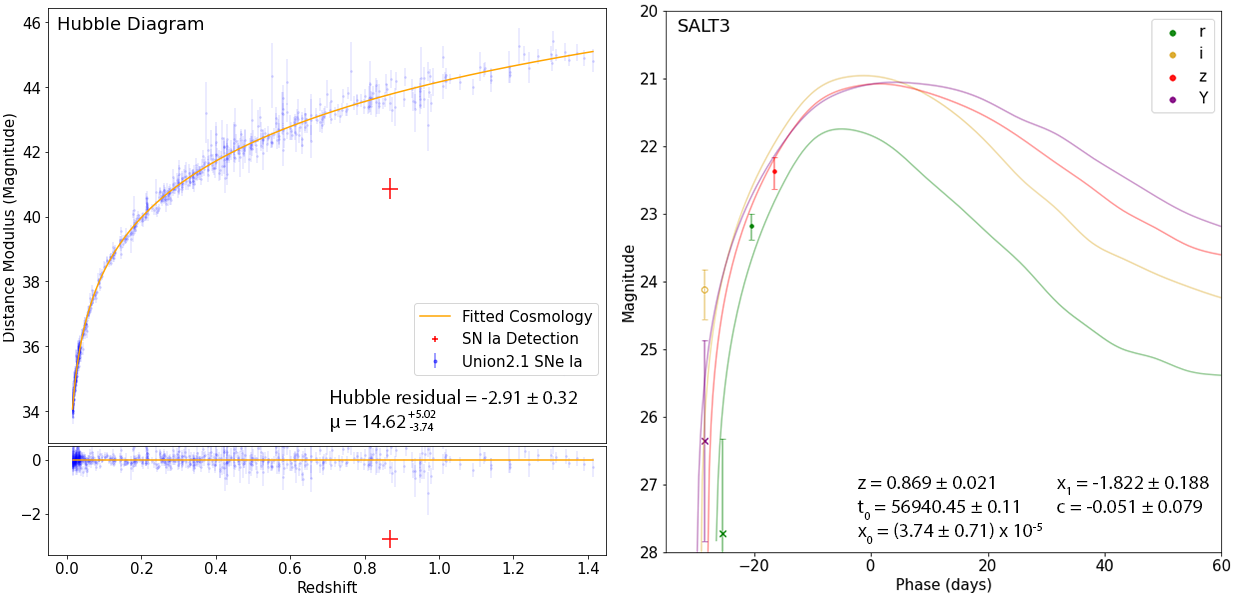}
\caption{Best-fit SALT3 model for DESI-308.7726-48.2381 with no redshift prior. 
 \txr{For this and Figures~\ref{fig:5531p2} and \ref{fig:5531p3}, solid photometry points correspond to the detection passes shown in Figure~\ref{fig:5531dets}, hollow points correspond to other exposures with PSF photometry, and crosses correspond to measurements using aperture photometry (\S~\ref{sec:analysis}).}}\label{fig:5531p1}
\end{center}
\end{figure}

\begin{figure}[H]
\begin{center}
\includegraphics[width=110mm]{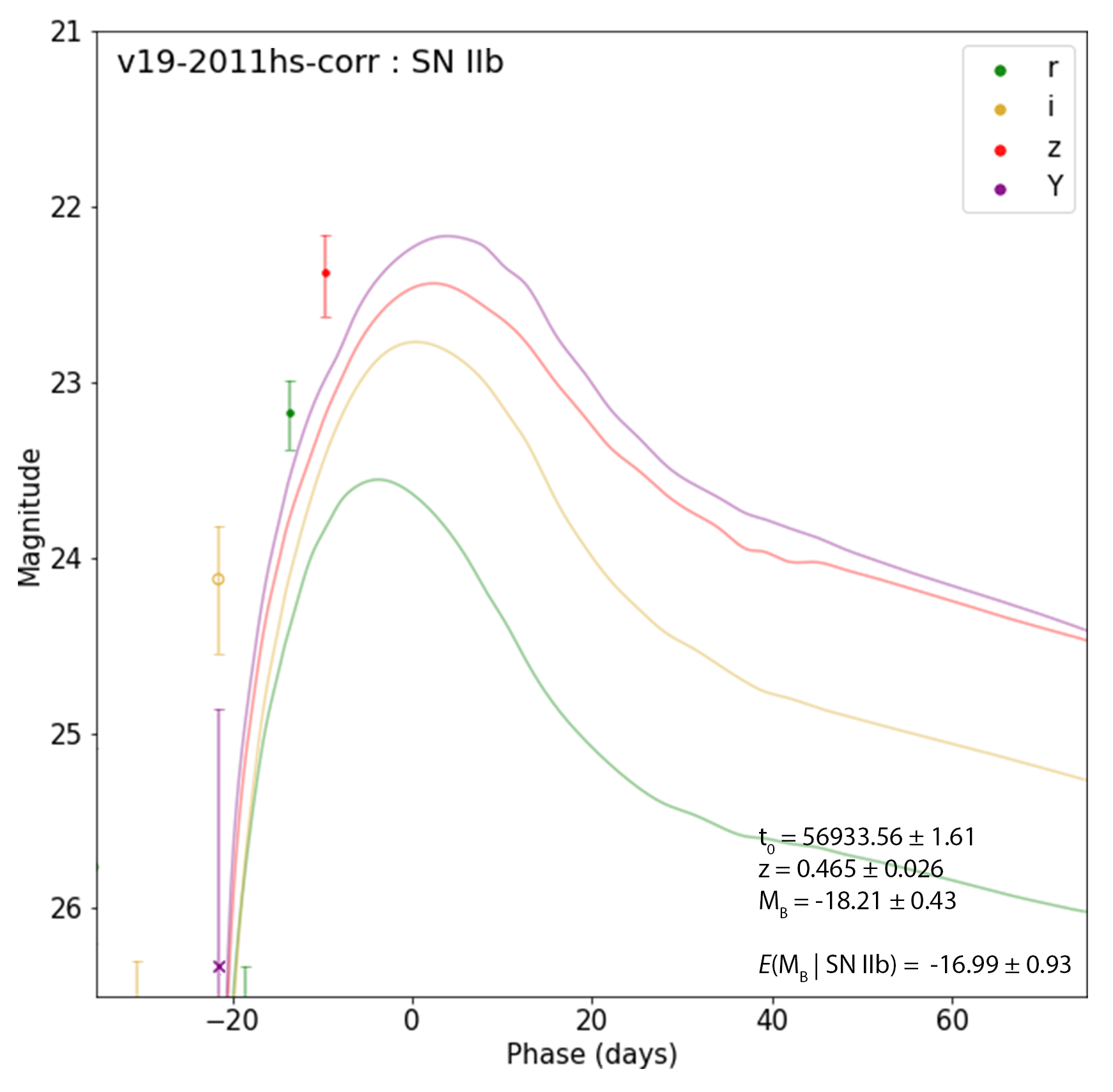}
\caption{Best-fit core collapse template model for DESI-308.7726-48.2381 with a lens photo-$z$ redshift prior of $0.473 \pm 0.032$.}\label{fig:5531p2}
\end{center}
\end{figure}

\begin{figure}[H]
\begin{center}
\includegraphics[width=110mm]{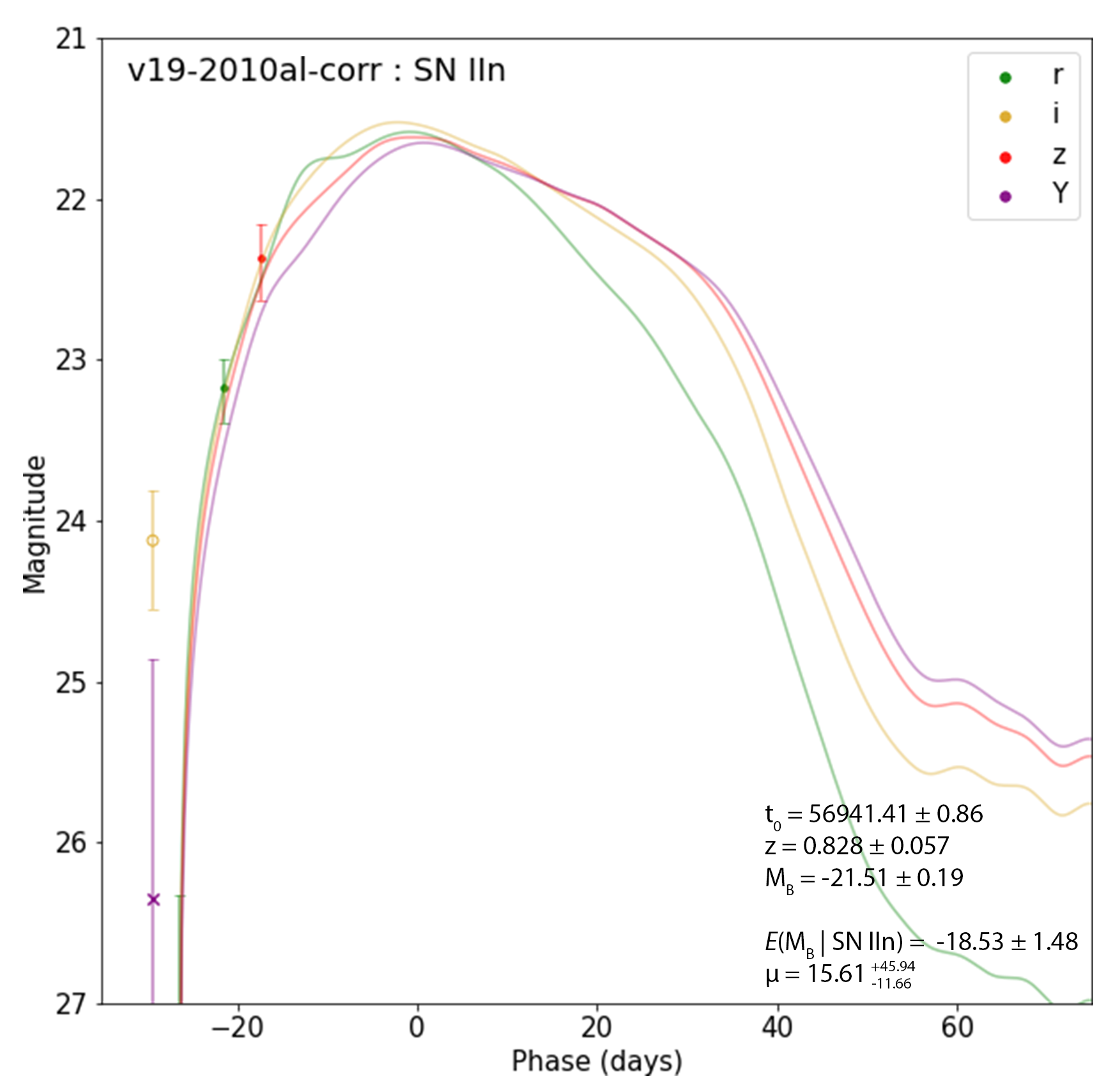}
\caption{Best-fit core collapse template model for DESI-308.7726-48.2381 with no redshift prior.}\label{fig:5531p3}
\end{center}
\end{figure}

For DESI-308.7726-48.2381, there is a possible arc stretching from object 3 to 4 (in Figure~\ref{fig:5531rgb}), with object 2 as a counterimage; all three objects are identified by \txb{the} Tractor.  The transient candidate is only detected twice: in the \ipr-band, and four days later in the \ipz-band.  This system serves as an example of how our pipeline is able to detect transients with only two detections, as the event was captured in at least three sub-detections (see \S~\ref{subsec:source detection}).  These detections are visually comparable to the difference and detection images of known high-redshift SNe~Ia in the \S~\ref{sec:testing} (e.g., Figures~\ref{fig:sn33} and \ref{fig:known-snia-figure-last}).  While there are only two detections, we are reasonably confident that this is an astrophysical transient, \txb{as forced photometry in other bands at the detection location supports this postulation (Figures~\ref{fig:5531p1} to~\ref{fig:5531p3}).  \trr{Lastly, the photo-$z$ measurements of the posited lens and source galaxies ($0.473\pm 0.032$ and $0.748\pm 0.408$, respectively) are consistent with DESI-308.7726-48.2381 being an instance of strong lensing.}}   

\paragraph{Postulation 1: L-SN~Ia} Figure~\ref{fig:5531p1} shows the best-fit light curve model for the L-SN~Ia scenario with no prior on the redshift (given how broad the photometric redshifts of the punitive lensed images are).  The SALT3 parameters are all reasonable with a best-fit redshift of 0.869.  The resulting SALT3 model seems to fit the photometric data reasonably well, with an amplification of $14.62^{+5.02}_{-3.74}$.  

\newpage\paragraph{Postulation 2: uL-CC~SN} Figure~\ref{fig:5531p2} shows the best-fit light curve model for the uL-CC~SN scenario, which appears to be far worse compared with Postulation 1.  To reiterate, this is the best-fitting CC~SN template model of 161 templates supplied by SNcosmo.

Not shown is the uL-SN~Ia fit, \txr{which results} in a model too bright for what is expected of a SN~Ia (see Table~\ref{tab:sncandidates}).

\newpage\paragraph{Postulation 3: L-CC~SN} Figure~\ref{fig:5531p3} shows the best-fit light curve model for the L-CC~SN scenario.  The best-fit redshift is at 0.828, and the required amplification is 15.61, albeit with a large uncertainty, as is typical for CC~SNe.  The data appear to rise more rapidly than the model, but this scenario may still be possible.  

\paragraph{Conclusion}  The transient found in DESI-308.7726-48.2381 \txb{is likely a lensed SN (more likely a Ia than CC).  As with the previous two candidates,} to be more confident of the lensing nature of this system, higher resolution and/or spectroscopic observations are needed.

\section{Conclusion} \label{sec:conclusion}
We have developed a pipeline for a \emph{targeted} search for lensed transients.  \txb{For 5807 strong lensing systems and candidates observed by the DESI Legacy Imaging Surveys, this pipeline first generates a median coadd for each observed band as a reference image.  It then employs two image subtraction algorithms to identify transient detections that are in close proximity both spatially and temporally}.  By applying this pipeline to the DESI Legacy Imaging Surveys DR9/10, we have found seven lensed SN candidates, one unlensed \txr{SN}, and two asteroids.  We have also confirmed the variability of a large number of lensed quasars, which we will present in a subsequent paper (Sheu et al. in prep).  Of the seven lensed SN candidates, \txr{the one in} DESI-344.6252-48.8977 is \txb{very} likely \txr{a} galaxy-scale strongly lensed SN\txb{, probably a Type~Ia}.  \txr{Follow-up high resolution imaging and spectroscopy, as well as lens modeling, can help reach} a more definitive conclusion on whether some of these \txr{transient candidates} are lensed.

\txr{\txb{Of our grade A and B candidates, the transients in DESI-344.6252-48.8977 and DESI-308.7726-48.2381 are likely L-SNe~Ia, whereas the transient in DESI-058.6486-30.5959 is likely a L-CC~SN.  Preliminary results indicate that} half of the 5807 systems, for which we \txb{have conducted the} search, are actually strong lenses (S. Tabares-Tarquinio et al. in prep).  \txb{Since} the uncertainties of our forecast results in Table~\ref{tab:expectation_results} are Poisson in nature, \trr{we adjust our estimates by $1/2$ and their uncertainties by $\sqrt{1/2}$.  And so, the number of L-SNe~Ia and L-CC~SNe with two or more detections becomes} $3.29 \pm 1.82$ and $4.68 \pm 2.18$, respectively, and the corresponding numbers for three or more detections \trr{becomes} $1.45 \pm 1.19$ and $2.27 \pm 1.50$.  The results from our grade A and B candidates are broadly consistent with these forecasts.}

We believe that these \txb{results demonstrate} the very promising viability of our pipeline and its applicability to future surveys such as the Vera C. Rubin Observatory Legacy Survey of Space and Time (LSST) and the Nancy Grace Roman Space Telescope (RST) to find \txb{live lensed SNe and other types of transients, as well as lensed quasars}.  Assuming \txb{the trend of three high grade} lensed SN candidates for every \txb{$5807/2 \approx 3000$ systems found in our search}, \trr{we have reached an approximate rate of $1$ lensed SN per $1000$ lensing systems in our targeted search.  The Legacy Imaging Surveys DR9 spanned $\sim 5$ years, with an \ipr~band limiting magnitude of $23.43$ and an average cadence of $\sim 90$ days (note that it was not intended to look for transients).  Given that LSST and RST will have significantly greater depth and higher cadence, we can expect this rate to be a lower bound for lensed SN discoveries in future targeted searches within those surveys.  A targeted search strategy requires prior knowledge of the locations of lenses and lens candidates.  
However, for resolvable lensing systems, 
we anticipate that this will impose little limitation, as lens search pipelines are becoming increasingly efficient and fast (on the order of days).  Thus, iterative lens searches can be rapidly carried out as the observational coverage expands and depth increases (e.g., \citealp{huang2020, huang2021, storfer2022})
for LSST and RST.   
Targeted searches for lensed transients can then quickly follow.}
Lensed transient discoveries in these future surveys will likely realize the potential to dramatically improve lens modeling and possibly resolve the $H_0$ tension.  

\section*{Acknowledgments} \label{sec:acknowledgments}



This work was supported in part by the Director, Office of Science, Office of High Energy Physics of the US
Department of Energy under contract No. DE-AC025CH11231. This research used resources of the National Energy
Research Scientific Computing Center (NERSC), a U.S. Department of Energy Office of Science User Facility operated
under the same contract as above and the Computational HEP program in The Department of Energy’s Science Office
of High Energy Physics provided resources through the “Cosmology Data Repository” project (Grant \#KA2401022).  X. Huang acknowledges the University of San Francisco Faculty Development Fund.

This paper is based on observations at Cerro Tololo Inter-American Observatory, National Optical Astronomy Observatory (NOAO Prop. ID: 2014B-0404; co-PIs: D. J. Schlegel and A. Dey), which is operated by the Association of Universities for Research in Astronomy (AURA) under a cooperative agreement with the National Science Foundation.

This project used data obtained with the Dark Energy Camera, which was constructed by the Dark Energy Survey collaboration. Funding for the DES Projects has been provided by the U.S. Department of Energy, the U.S. National Science Foundation, the Ministry of Science and Education of Spain, the Science and Technology Facilities Council of the United Kingdom, the Higher Education Funding Council for England, the National Center for Supercomputing Applications at the University of Illinois at Urbana-Champaign, the Kavli Institute of Cosmological Physics at the University of Chicago, the Center for Cosmology and Astro-Particle Physics at the Ohio State University, the Mitchell Institute for Fundamental Physics and Astronomy at Texas A\&M University, Financiadora de Estudos e Projetos, Fundação Carlos Chagas Filho de Amparo à Pesquisa do Estado do Rio de Janeiro, Conselho Nacional de Desenvolvimento Científico e Tecnológico and the Ministério da Ciência, Tecnologia e Inovacão, the Deutsche Forschungsgemeinschaft, and the Collaborating Institutions in the Dark Energy Survey. The Collaborating Institutions are Argonne National Laboratory, the University of California at Santa Cruz, the University of Cambridge, Centro de Investigaciones Enérgeticas, Medioambientales y Tecnológicas-Madrid, the University of Chicago, University College London, the DES-Brazil Consortium, the University of Edinburgh, the Eidgenössische Technische Hochschule (ETH) Zürich, Fermi National Accelerator Laboratory, the University of Illinois at Urbana-Champaign, the Institut de Ciències de l’Espai (IEEC/CSIC), the Institut de Física d’Altes Energies, Lawrence Berkeley National Laboratory, the Ludwig-Maximilians Universität München and the associated Excellence Cluster Universe, the University of Michigan, the National Optical Astronomy Observatory, the University of Nottingham, the Ohio State University, the OzDES Membership Consortium the University of Pennsylvania, the University of Portsmouth, SLAC National Accelerator Laboratory, Stanford University, the University of Sussex, and Texas A\&M University.

The work of Aleksandar Cikota is supported by NOIRLab, which is managed by the Association of Universities for Research in Astronomy (AURA) under a cooperative agreement with the National Science Foundation.

We thank Alex Kim at the Lawrence Berkeley National Laboratory for insightful discussions on difference image photometry, as well as Saul Perlmutter and Greg Aldering for general commentary on our paper results.



\software{Astropy \citep{astropy2013, astropy2018},  
          Montage \citep{montage},  
          SEP \citep{bertin1996, sep},
          SNCosmo \citep{sncosmo}, 
          NumPy \citep{numpy}, 
          Matplotlib \citep{matplotlib}
          }

\newpage
\appendix
\section{Photometry on Previously Discovered SNe~Ia}\label{appendix_kSNe}
 
\begin{figure}[H]
\begin{center}
\includegraphics[width=180mm]{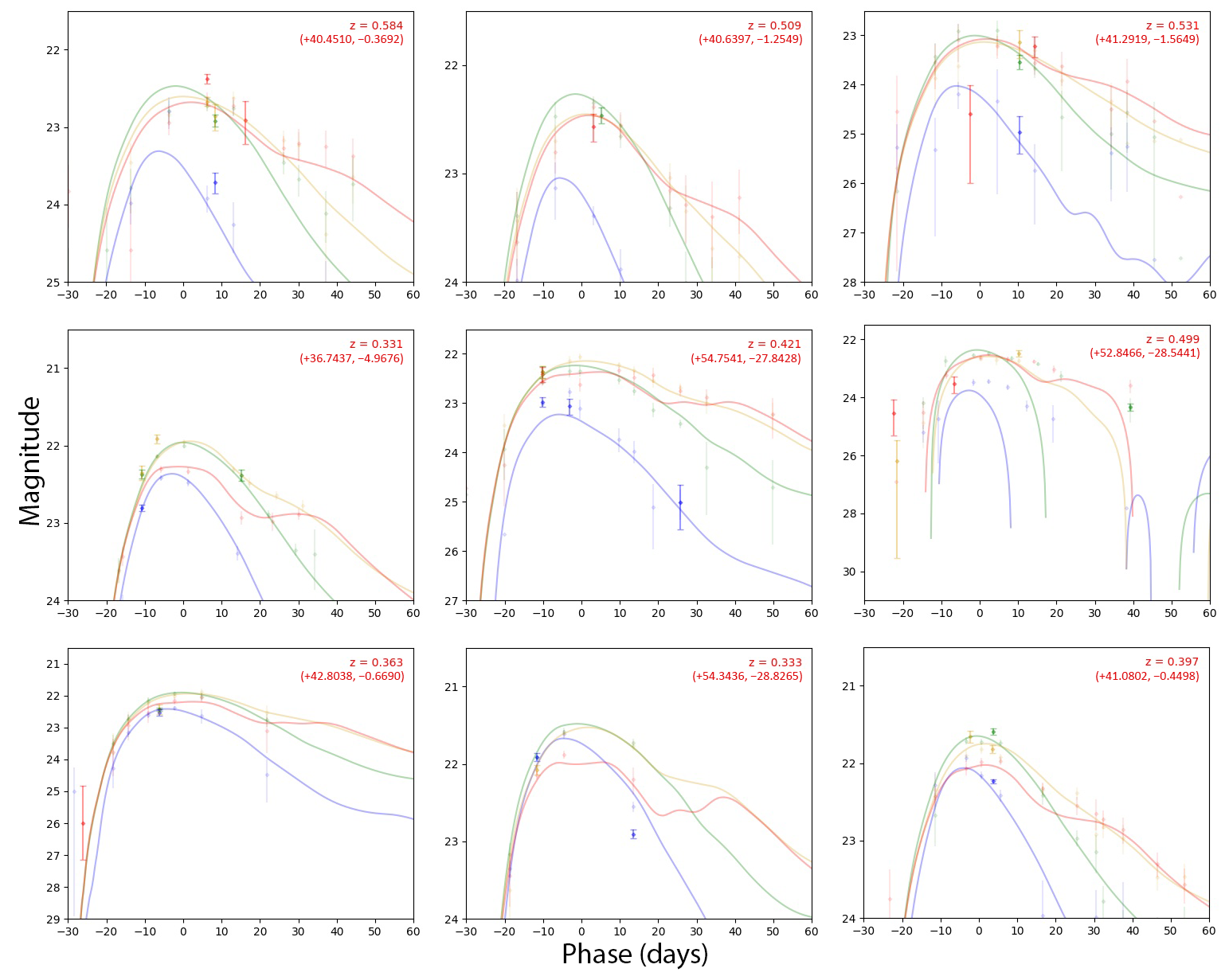}
\caption{Our photometry (solid data-points) for new detections of 9 (of 32; for the rest see Figures~\ref{fig_app_2} and~\ref{fig_app_3}) known SNe~Ia from DES, shown along with DES photometry (faint points) and their best-fit SALT2 light curve\footnote{\url{https://github.com/legacysurvey/imagine/blob/main/map/views.py}}.  For all panels, we follow the color scheme of \txr{blue=\ipg~band, green=\ipr~band, yellow=\ipi~band, and red=\ipz~band.}  The redshift and the (RA, dec) is given on the top right of every plot.  \txr{Note that our measurements match well with DES photometry, and provide additional photometry points for these SNe~Ia.}}\label{fig_app_f}
\end{center}
\end{figure}

\begin{figure}[H]
\begin{center}
\includegraphics[width=180mm]{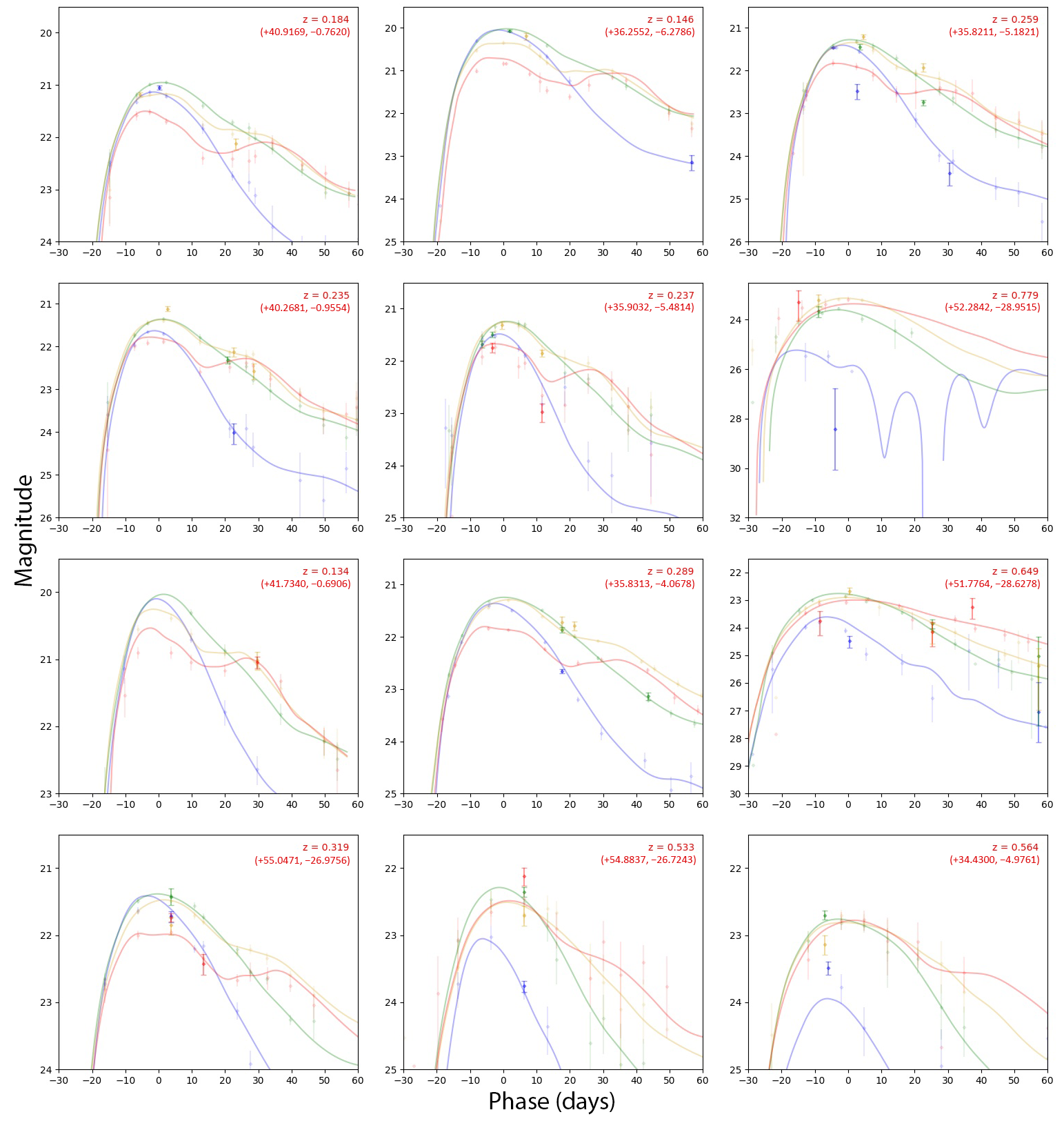}
\caption{Our photometry (solid data-points) for new detections of 12 (of 32) known SNe~Ia from DES.  For additional details, see the caption of Figure~\ref{fig_app_f}.}\label{fig_app_2}
\end{center}
\end{figure}

\begin{figure}[H]
\begin{center}
\includegraphics[width=180mm]{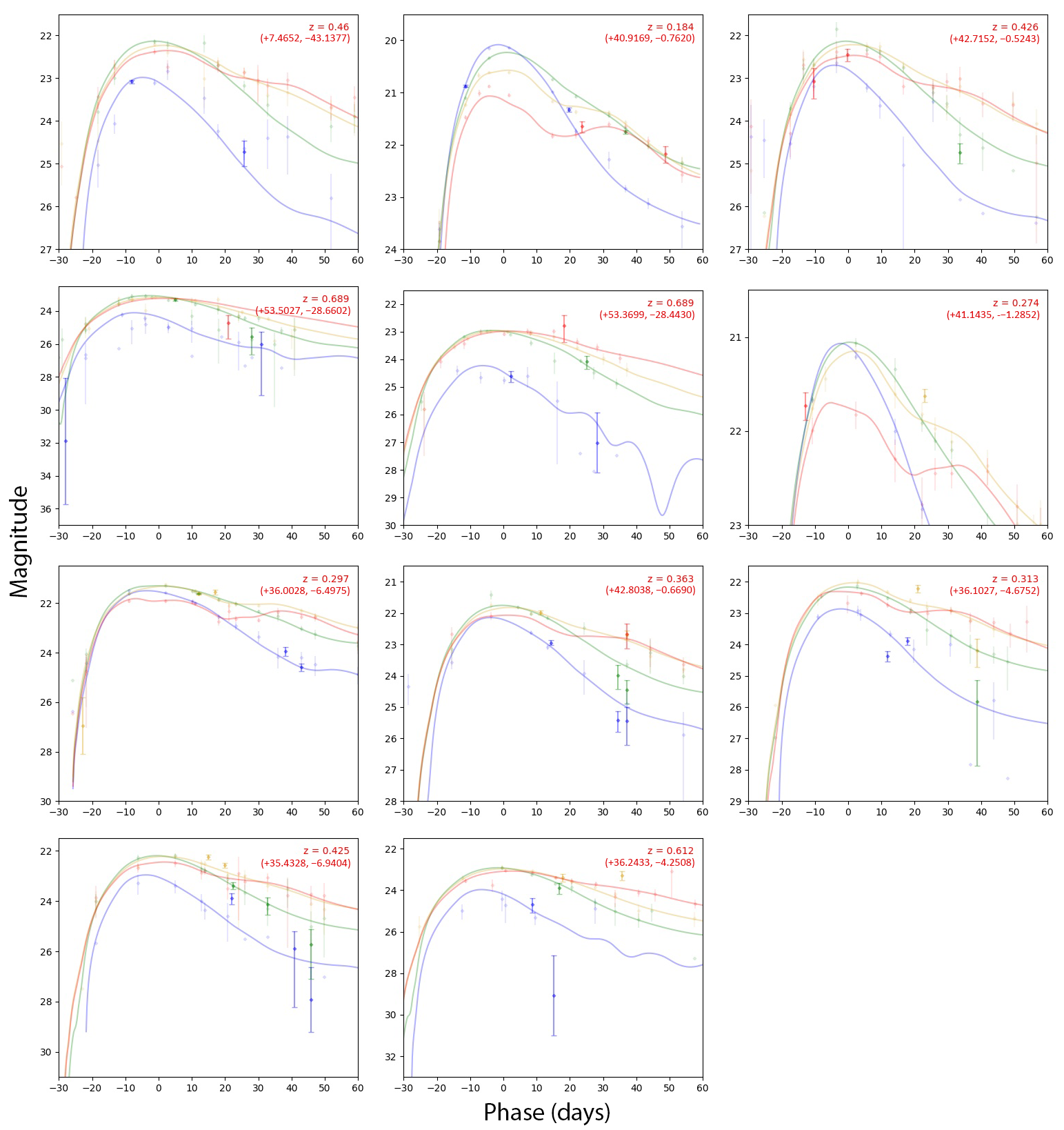}
\caption{Our photometry (solid data-points) for new detections of 11 (of 32) known SNe~Ia from DES.  For additional details, see the caption of Figure~\ref{fig_app_f}.}\label{fig_app_3}
\end{center}
\end{figure}

\newpage
\section{Grade C \& D Lensed Supernova and Unlensed Supernova Candidates}\label{appendix_SNC}

\subsection{DESI-034.3625-35.3563}
 \label{subsubsec:DESI-034.3625-035.3563}

DESI-034.3625-35.3563 is a C-grade strong lensing candidate, discovered in \citet{storfer2022}.

  \begin{figure}[H]
\begin{center}
\includegraphics[width=175mm]{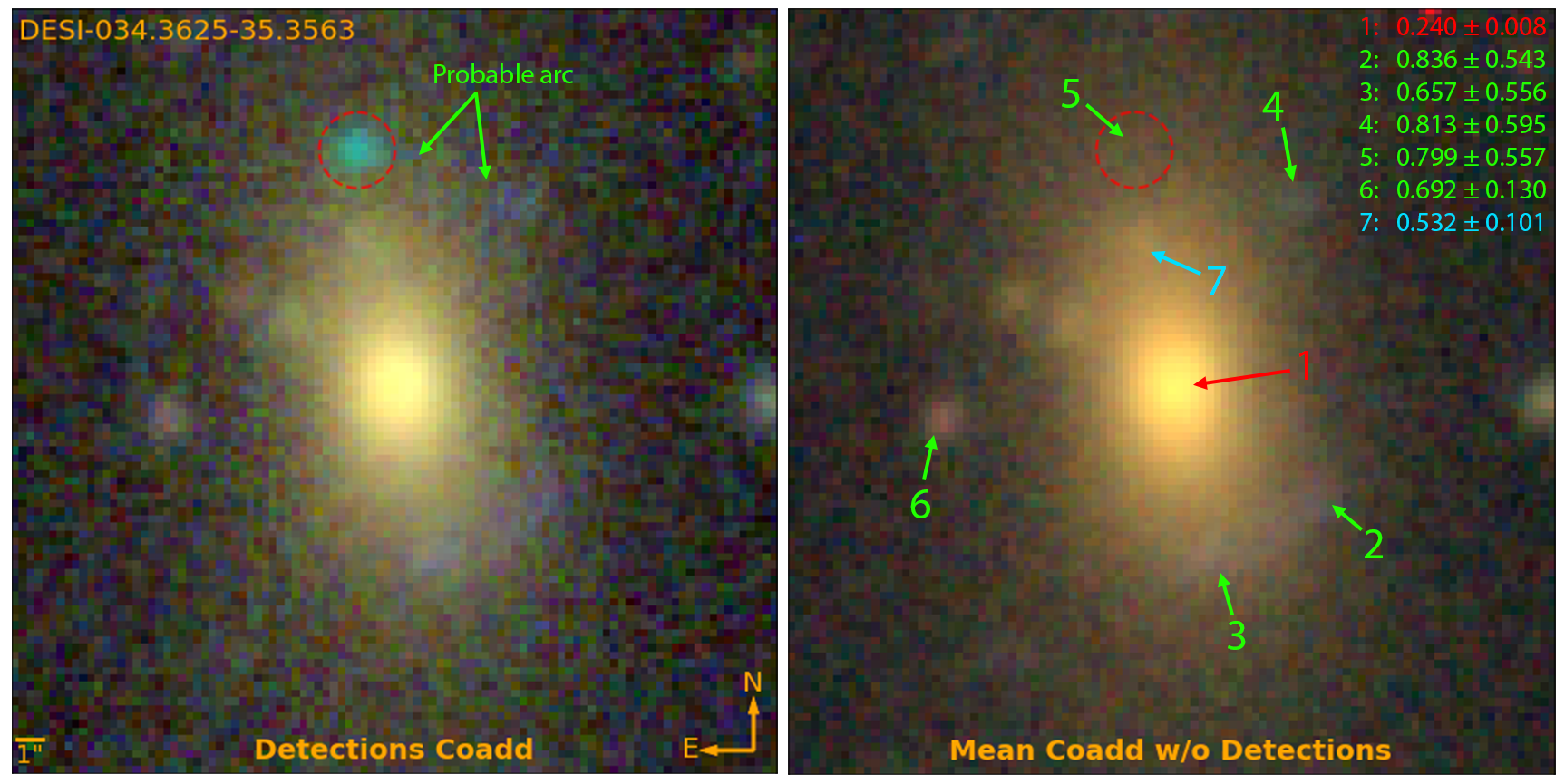}
\caption{Coadded RGB images (using \ipg, \ipr, \ipi, and \ipz~bands) of DESI-034.3625-35.3563 with and without the transient detection \txb{(red dotted circle)} exposures.  Labelled objects are color-coded as the following postulated scenario: red as the lens galaxy, green as the source galaxy, and cyan as an interloper or a second source.  Photometric redshifts are displayed on the top right.  \trr{The posited lens galaxy has a photo-$z$ of $0.240\pm 0.008$, and the posited lensed images have photo-$z$'s of $0.836\pm 0.543$, $0.657 \pm 0.556$, $0.813 \pm 0.595$, $0.799 \pm 0.557$, and $0.692 \pm 0.130$.}}\label{fig:3032rgb}
\end{center}
\end{figure}

\begin{figure}[H]
\begin{center}
\includegraphics[width=175mm]{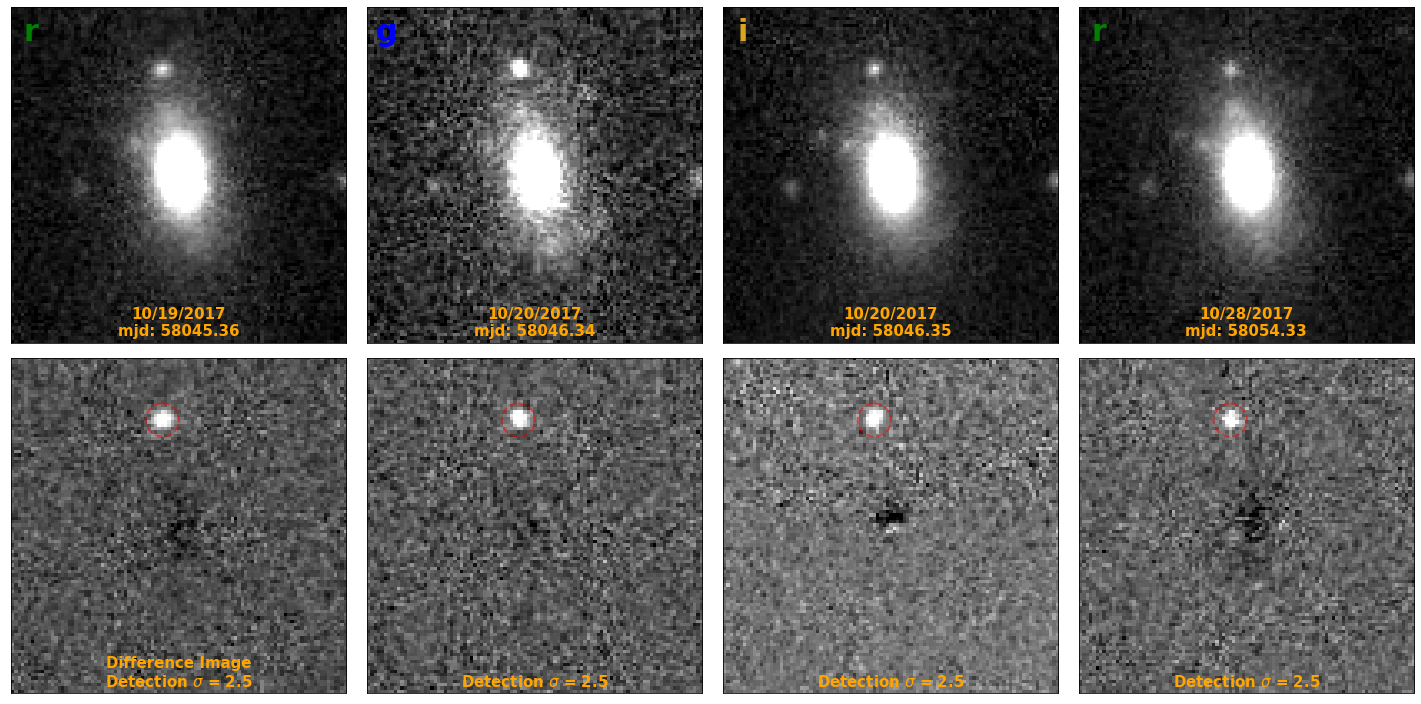}
\caption{Detection exposures for the transient in DESI-034.3625-35.3563 in chronological order. See caption of Figure~\ref{fig:3760dets} for the full description.}\label{fig:3032dets}
\end{center}
\end{figure}

\begin{figure}[H]
\begin{center}
\includegraphics[width=75mm]{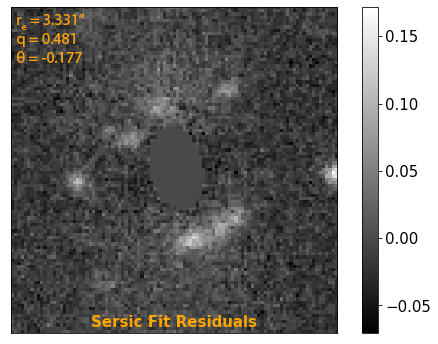}
\caption{Residual image after lens light subtraction for DESI-034.3625-35.3563, with the Sérsic light parameters (half-light radius, axis ratio, semi-major axis orientation from the y axis) in the top left (modelled in DECam \ipg~filter)\txr{, with the core of the lens galaxy masked out.}}\label{fig:3032sersic}
\end{center}
\end{figure}

DESI-034.3625-35.3563 is \txb{a} strong lensing candidate system with a single massive galaxy as the main lens.  There appear to be two faint arcs, located North (identified by Tractor as objects 4 and 5 in Figure~\ref{fig:3032rgb}) and South (objects 2 and 3) of the foreground galaxy, \txb{at} approximately four arcseconds away.  \trr{Given the similarities in morphology, color, and photo-$z$}, they quite possibly correspond to the same background source.  The transient lies directly at the east end of the first arc (object 5).  This transient is also only about 2.5 effective radii (Figure~\ref{fig:3032sersic}) away from the lens, and so it is possible that the foreground galaxy is the host. 

\begin{figure}[H]
\begin{center}
\includegraphics[width=180mm]{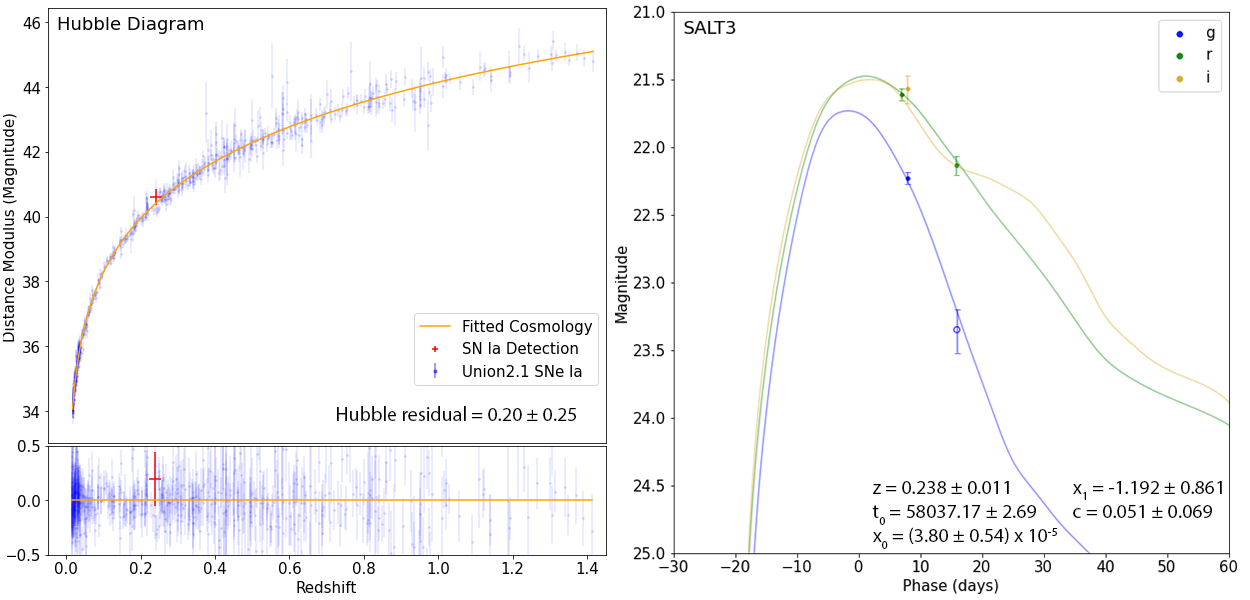}
\caption{Best-fit SALT3 model for DESI-034.3625-35.3563 with a lens photo-$z$ redshift prior of $0.240 \pm 0.008$.  For this and Figure~\ref{fig:3032p2}, solid photometry points correspond to the detection passes shown in Figure~\ref{fig:3032dets}.}\label{fig:3032p1}
\end{center}
\end{figure}

\newpage \paragraph{Postulation 1: uL-SN~Ia} Figure~\ref{fig:3032p1} shows the best-fit SALT3 light curve model for the uL-SN~Ia scenario.  This model agrees well with the data, with reasonable light curve parameters. 
The inferred absolute magnitude is consistent with the expectation for a SN~Ia at the redshift of the foreground galaxy.  Therefore, this seems to be a likely identity the transient.

\begin{figure}[H]
\begin{center}
\includegraphics[width=110mm]{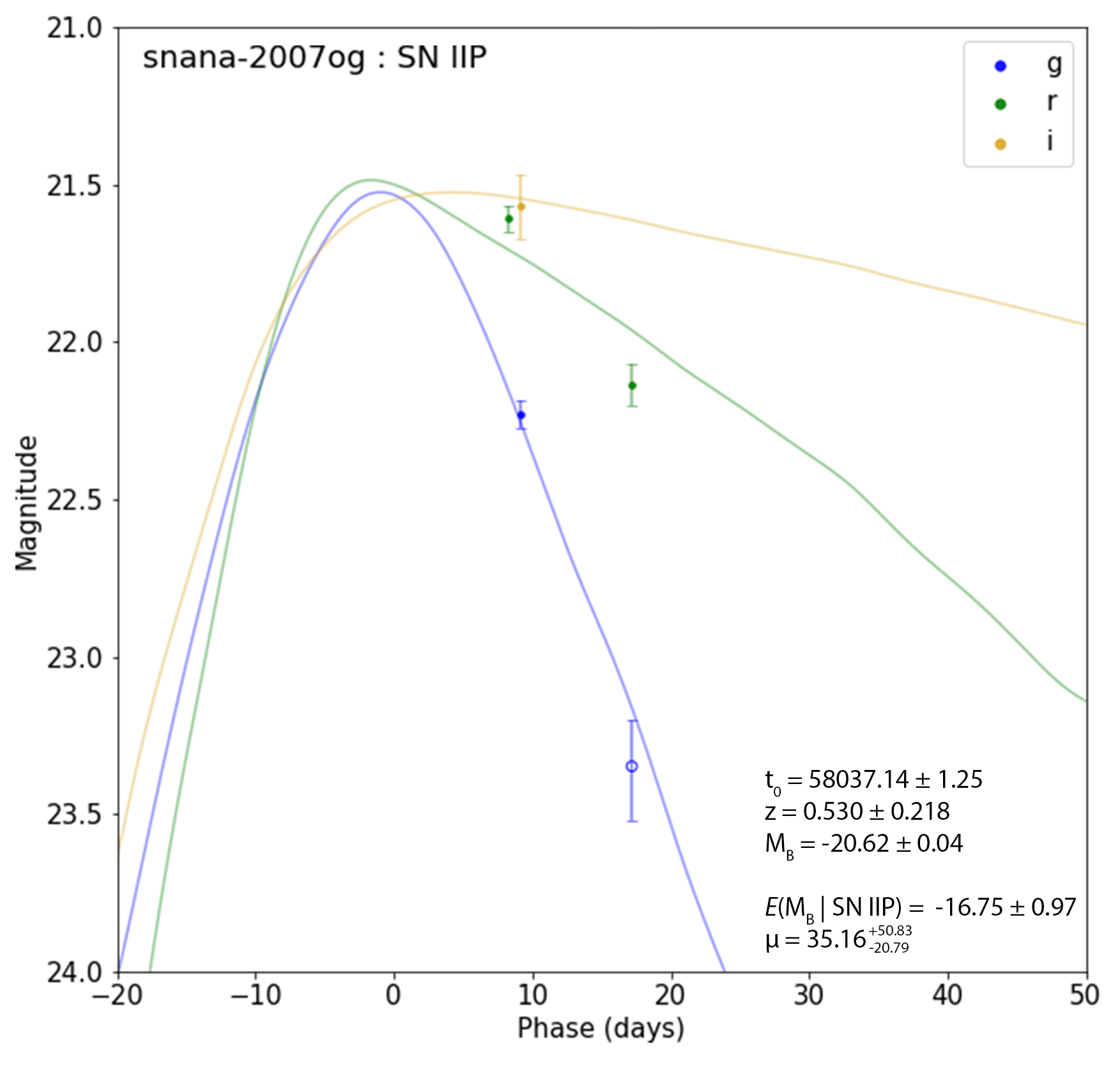}
\caption{Best-fit core collapse template model for DESI-034.3625-35.3563 with no redshift prior.}\label{fig:3032p2}
\end{center}
\end{figure}

\newpage \paragraph{Postulation 2: L-CC~SN} Figure~\ref{fig:3032p2} shows the best-fit light curve model for the L-CC~SN scenario.  This SN~IIP template fit has a slightly worse \txb{$\chi^2/$DOF} compared to Postulation 1 (see Table~\ref{tab:sncandidates}),  but this scenario is nevertheless possible.  \txb{The} model does require a fairly high amplification of $35.16^{+50.83}_{-20.79}$, albeit with large uncertainties.  

\paragraph{Conclusion} The photometry seems to suggest that this detection is an uL-SN~Ia.  However, if additional high resolution and/or spectroscopic observation can confirm the faint lensed arc North of the lens, the data would strongly support the postulation of a L-CC~SN.  This points to the importance of timely follow-up if this were a live detection, as both the lensed and unlensed scenarios are possible, given the location.

 \newpage
\subsection{DESI-035.1374+00.4676}
 \label{subsubsec:DESI-035.1374+00.4676}

\begin{figure}[H]
\begin{center}
\includegraphics[width=180mm]{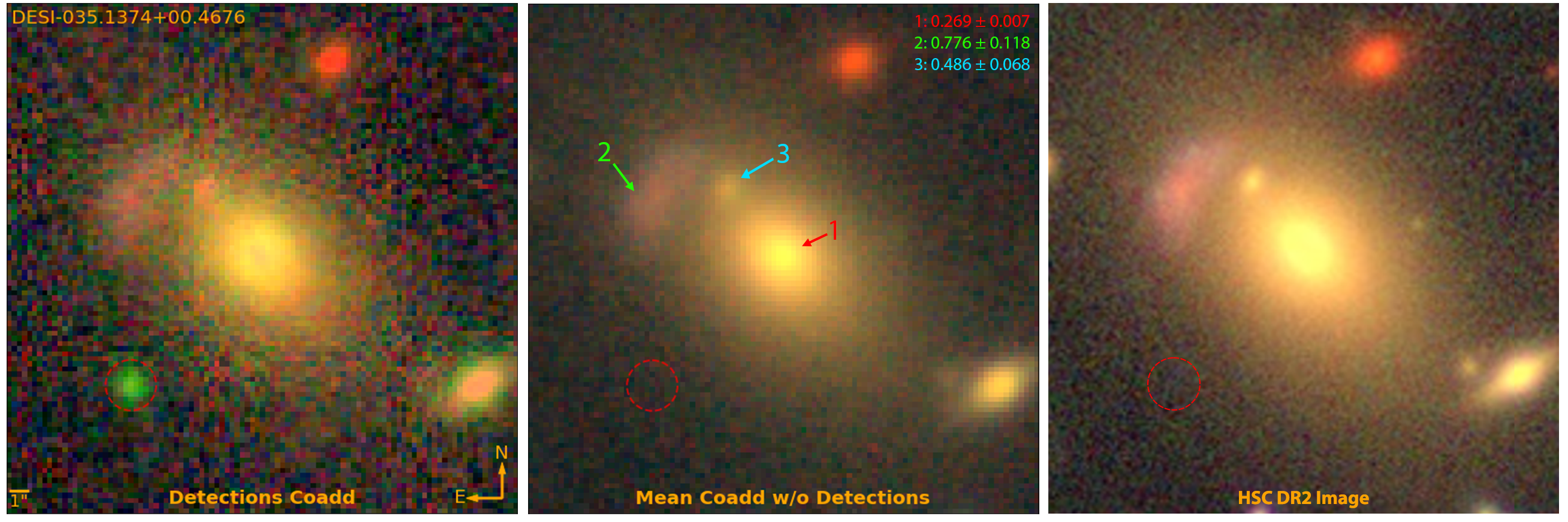}
\caption{Coadded RGB images (using \ipg, \ipr, \ipi, and \ipz~bands) of DESI-035.1374+00.4676 with and without the transient detection \txb{(red dotted circle)} exposures, as well as the HSC DR2 image.  Labelled objects are color-coded as the following postulated scenario: red as the lens galaxy, green as the source galaxy, and cyan as an interloper or \txr{a member galaxy of the foreground group}.  Photometric redshifts are displayed on the top right of the second image.  \trr{The posited lens galaxy has a photo-$z$ of $0.269\pm 0.007$, and the posited lensed image has a photo-$z$ of $0.776\pm 0.118$.}}\label{fig:3042rgb}
\end{center}
\end{figure}

DESI-035.1374+00.4676 was discovered in \citet{storfer2022}, as a C-grade strong lensing candidate.  However, after viewing the Hyper Suprime-Cam (HSC; \citeauthor{HSC_DR2} \citeyear{HSC_DR2}) image (see Figure~\ref{fig:3042rgb}), we feel confident in moving this into the A-grade lens candidate category.
 
\begin{figure}[H]
\begin{center}
\includegraphics[width=135mm]{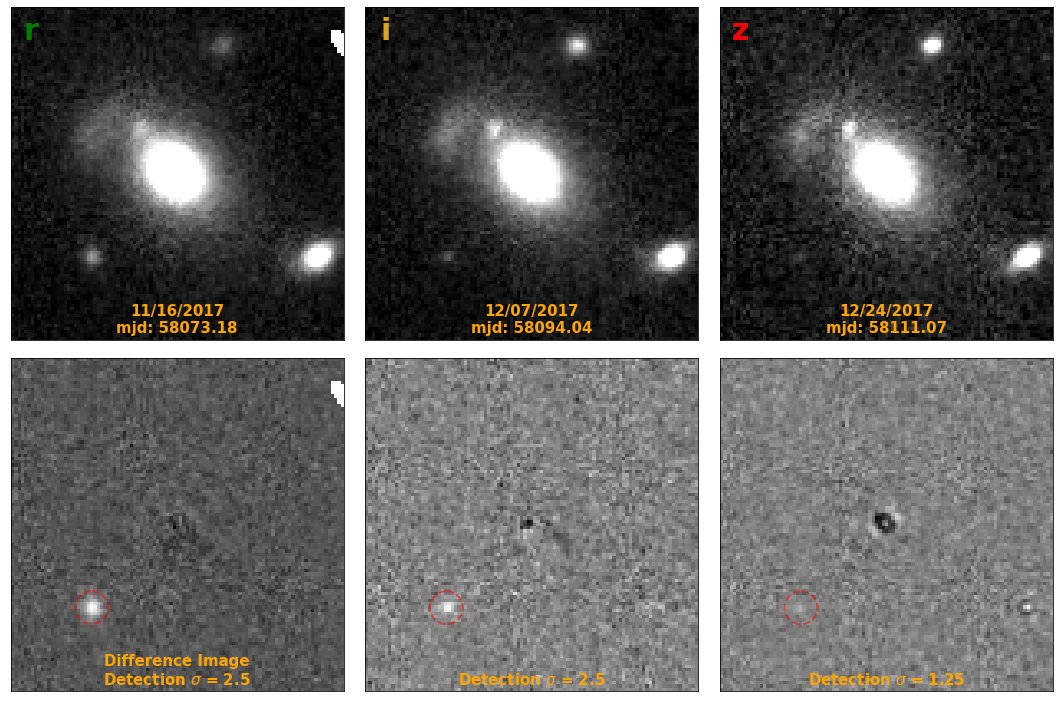}
\caption{Detection exposures for the transient in DESI-035.1374+00.4676 in chronological order. See caption of Figure~\ref{fig:3760dets} for the full description.}\label{fig:3042dets}
\end{center}
\end{figure}

\begin{figure}[H]
\begin{center}
\includegraphics[width=75mm]{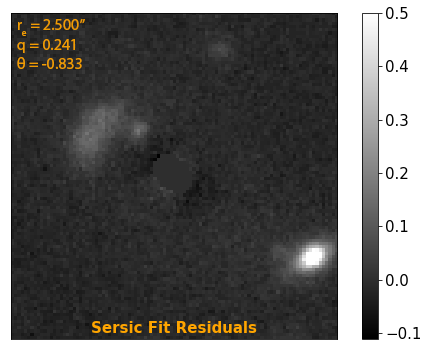}
\caption{Residual image after lens light subtraction for DESI-035.1374+00.4676, with the Sérsic light parameters (half-light radius, axis ratio, semi-major axis orientation from the y axis) in the top left (modelled in DECam \ipg~filter)\txr{, with the core of the lens galaxy masked out.}}\label{fig:3042subt}
\end{center}
\end{figure}

DESI-035.1374+00.4676 \txr{appears to be} a galaxy \txr{group-scale} strongly \txr{lensing} system, with the arc lying Northeast of the main lensing galaxy.  \trr{This is supported by the photo-$z$'s of the posited lensed and source galaxies ($0.296 \pm 0.007$ and $0.776 \pm 0.118$ respectively).}  The transient's location is somewhat far from both the lens and lensed image.  From the best-fit foreground galaxy light parameters shown in Figure~\ref{fig:3042subt}, the detection is approximately \txb{4} half-light radii away from the lensing galaxy, which does not exclude it from being the host galaxy of the transient.  On the other hand, if it is hosted by the lensed source galaxy, the distance between the transient and its center would be stretched along the tangential direction.  Without lens modeling, which would provide the delensed \txr{source, it} is difficult to estimate how far the transient is from the source galaxy center \txb{in meaningful terms (e.g., half-light radius or directional light radius)}.  Therefore neither is an impossible scenario \txb{based on the location}.  The possibility of a faint galaxy hosting the transient seems remote, as such a galaxy does not \txr{appear even} in the HSC image with superior seeing ($0.58''$ in the \ipi~band) and greater depth ($26.2$ \ipi~band limiting magnitude; see Figure~\ref{fig:3042rgb}).

\begin{figure}[H]
\begin{center}
\includegraphics[width=180mm]{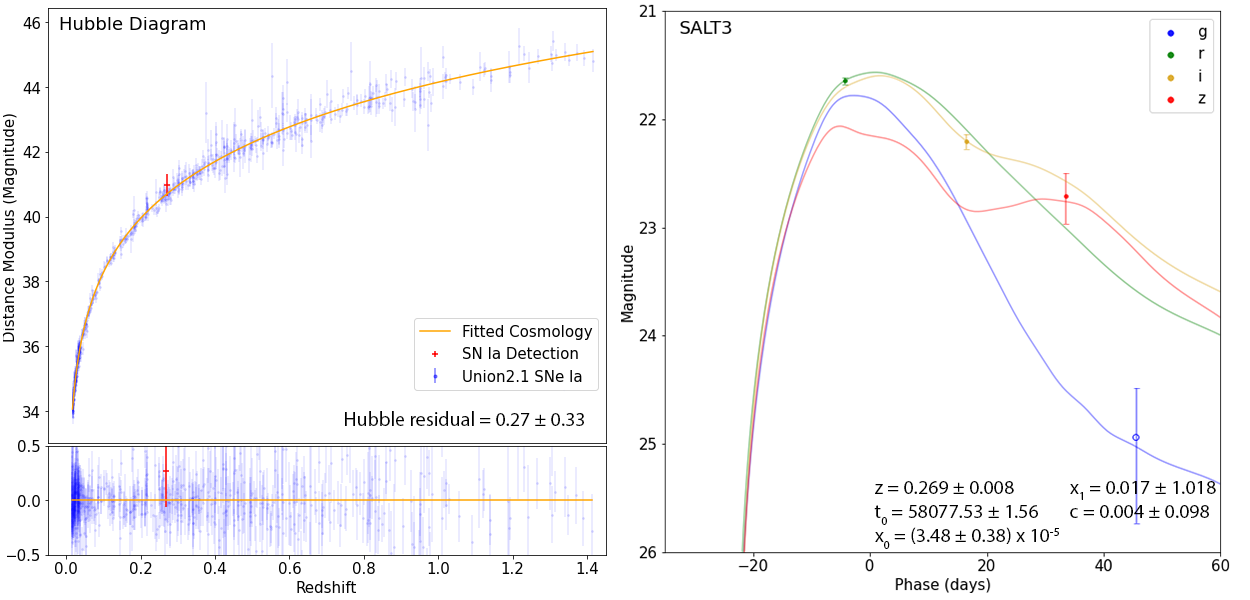}
\caption{Best-fit SALT3 model for DESI-035.1374+00.4676 with a lens photo-$z$ redshift prior of $0.269 \pm 0.007$.  For this and Figure~\ref{fig:3042p2}, solid photometry points correspond to the detection passes shown in Figure~\ref{fig:3042dets}.}\label{fig:3042p1}
\end{center}
\end{figure}

\newpage \paragraph{Postulation 1: uL-SN~Ia} Figure~\ref{fig:3042p1} shows the best-fit light curve model for the uL-SN~Ia scenario.  The SALT3 model fits the four photometric data points well, and its Hubble residual is consistent with the Union 2.1 best-fit cosmology.  We consider this to be a possible identity of this transient.

\begin{figure}[H]
\begin{center}
\includegraphics[width=110mm]{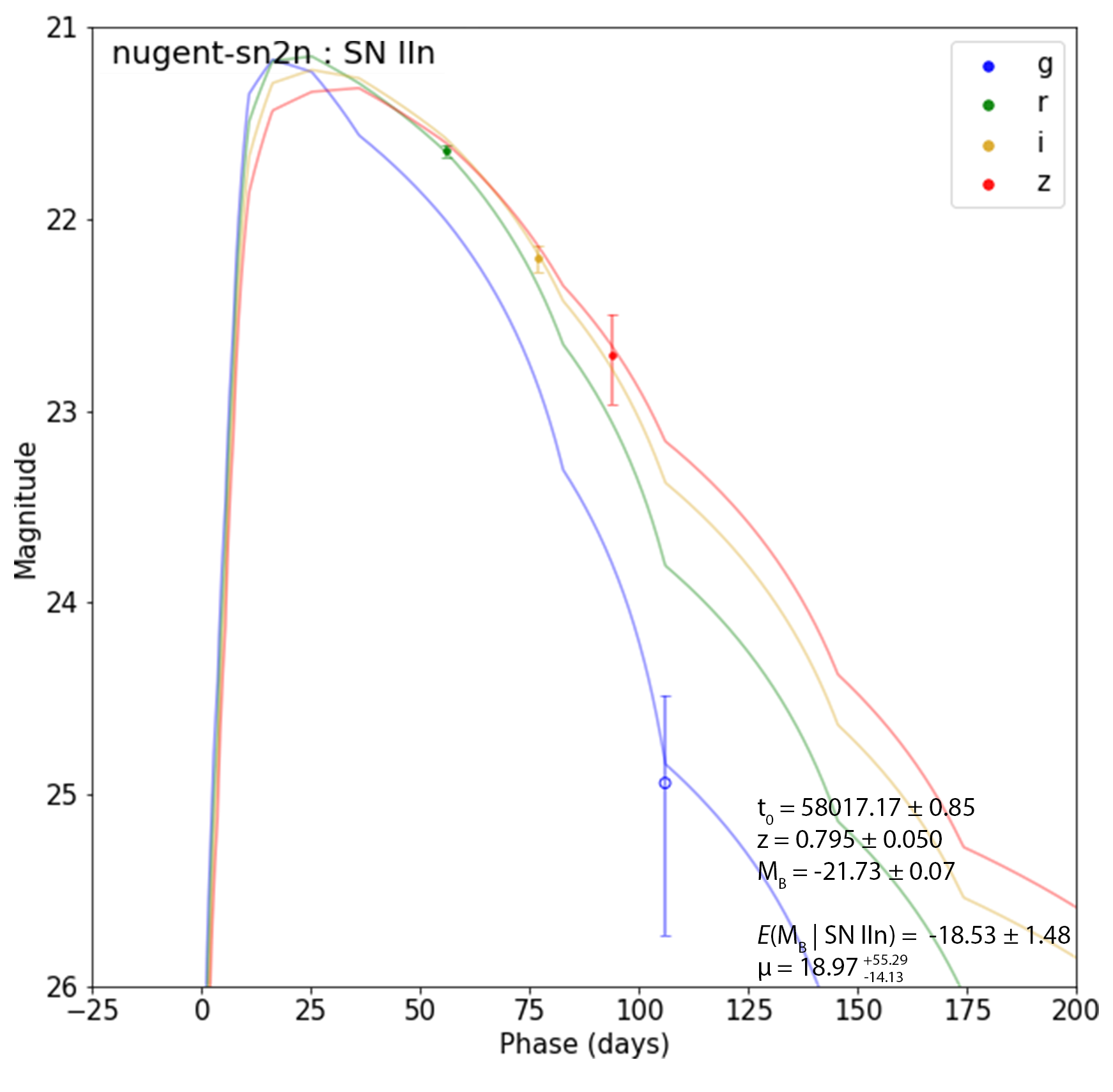}
\caption{Best-fit core collapse template model for DESI-035.1374+00.4676 with a source photo-$z$ redshift prior of $0.776 \pm 0.118$.}\label{fig:3042p2}
\end{center}
\end{figure}

\newpage \paragraph{Postulation 2: L-CC~SN} Figure~\ref{fig:3042p2} shows the best-fit light curve model for the L-CC~SN scenario.  This model also fits the available data well for a Type IIn SN template (``nugent-sn2n"), with an estimated amplification of $18.97^{+55.29}_{-14.13}$.  As this model is consistent with the data, we believe L-CC~SN to be a possible identity of the transient.

\paragraph{Conclusion} This transient in DESI-035.1374+00.4676 appears to be consistent with an uL-SN~Ia or a L-CC~SN.  With the data, it is difficult to discern which scenario is more likely.  In the case of a uL-SN~Ia, the supernova would have occurred at approximately \txb{four} effective radii away from the lensing galaxy.  On the other hand, while the detection is far from the center of the lensed galaxy, this separation may not rule out the background as the host due to the tangential ``stretching" from strong lensing.  Lens modeling (using HSC DR2 data or follow-up \txr{higher resolution} observations) may shed more light on this possibility.  If detected live, this detection would warrant follow-up observations.

\newpage \subsection{DESI-052.0083-37.2049}
 \label{subsubsec:DESI-052.0083-037.2049}
DESI-052.0083-37.2049 was discovered in \citet{storfer2022} as a C-grade strong lensing candidate.

\begin{figure}[H]
\begin{center}
\includegraphics[width=173mm]{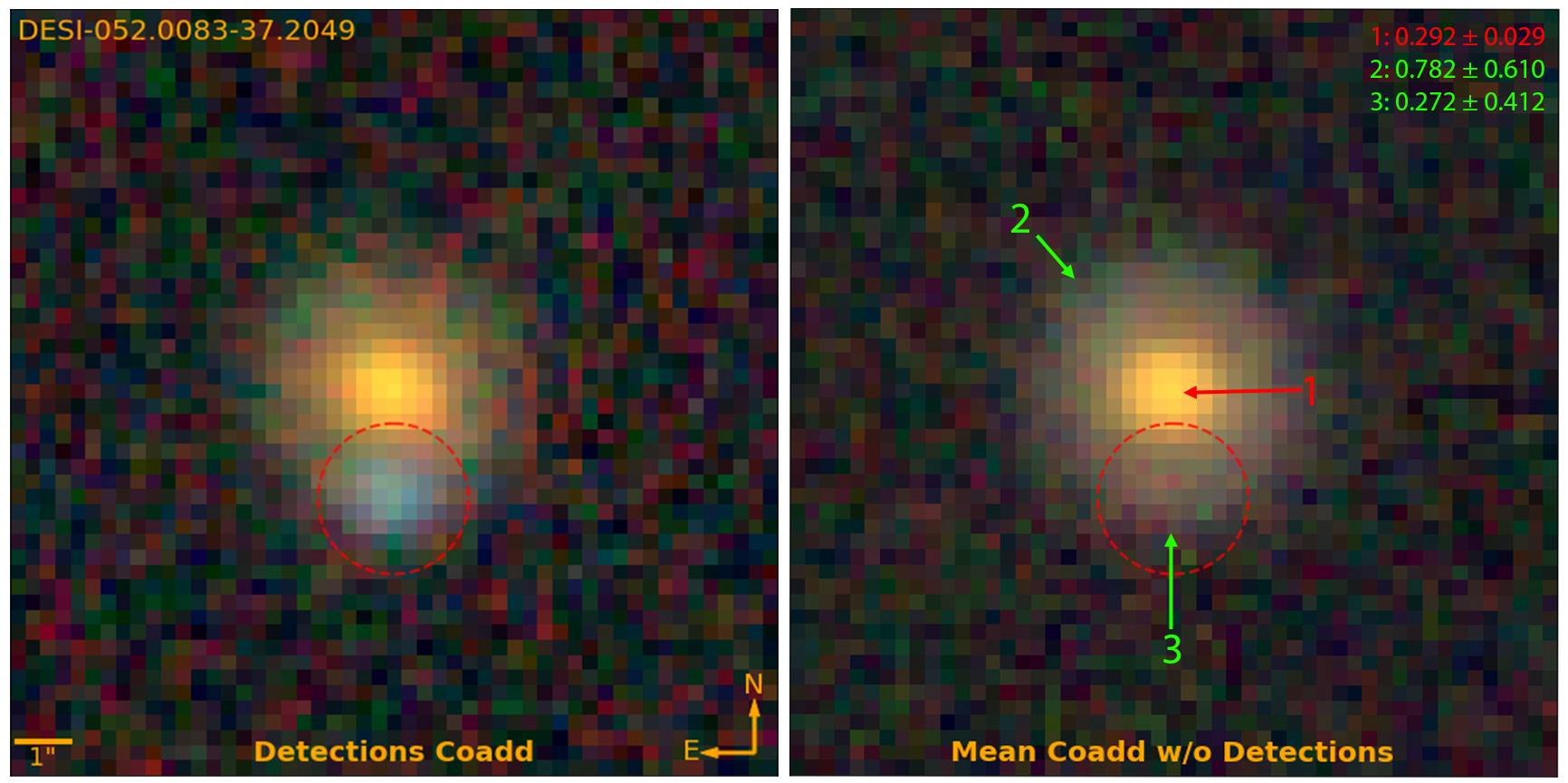}
\caption{Coadded RGB images (using \ipg, \ipr, \ipi, and \ipz~bands) of DESI-052.0083-37.2049 with and without the transient detection \txb{(red dotted circle)} exposures.  Labelled objects are color-coded as the following postulated scenario: red as the lens galaxy and green as the source galaxy.  Photometric redshifts are displayed on the top right.  \trr{The posited lens galaxy has a photo-$z$ of $0.292\pm 0.029$, the posited lensed images have photo-$z$'s of $0.782\pm 0.610$ and $0.272 \pm 0.412$.}}
\end{center}
\end{figure}

\begin{figure}[H]
\begin{center}
\includegraphics[width=173mm]{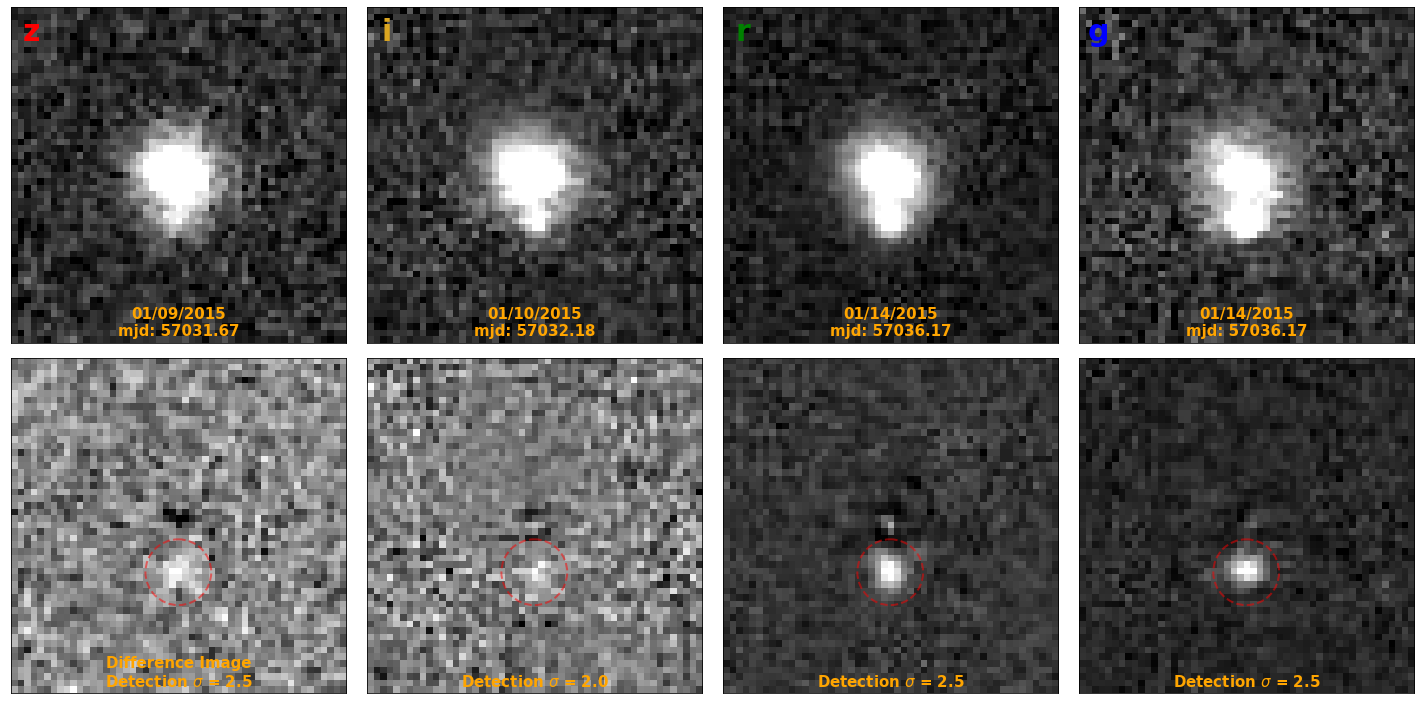}
\caption{Detection exposures for the transient in DESI-052.0083-37.2049 in chronological order. See caption of Figure~\ref{fig:3760dets} for the full description.}\label{fig:3172dets}
\end{center}
\end{figure}

DESI-052.0083-37.2049 is a possible strong lensing candidate, although a ring galaxy or a face-on spiral scenario is not ruled out.  \trr{The posited lens galaxy has a photo-$z$ of $0.292\pm 0.029$.  The photo-$z$'s of the posited source images have large uncertainties ($0.782\pm 0.610$ and $0.272 \pm 0.412$).}  The transient, however, is unmistakably present, with multiple detections lying directly on the arc-like structure.  

\begin{figure}[H]
\begin{center}
\includegraphics[width=180mm]{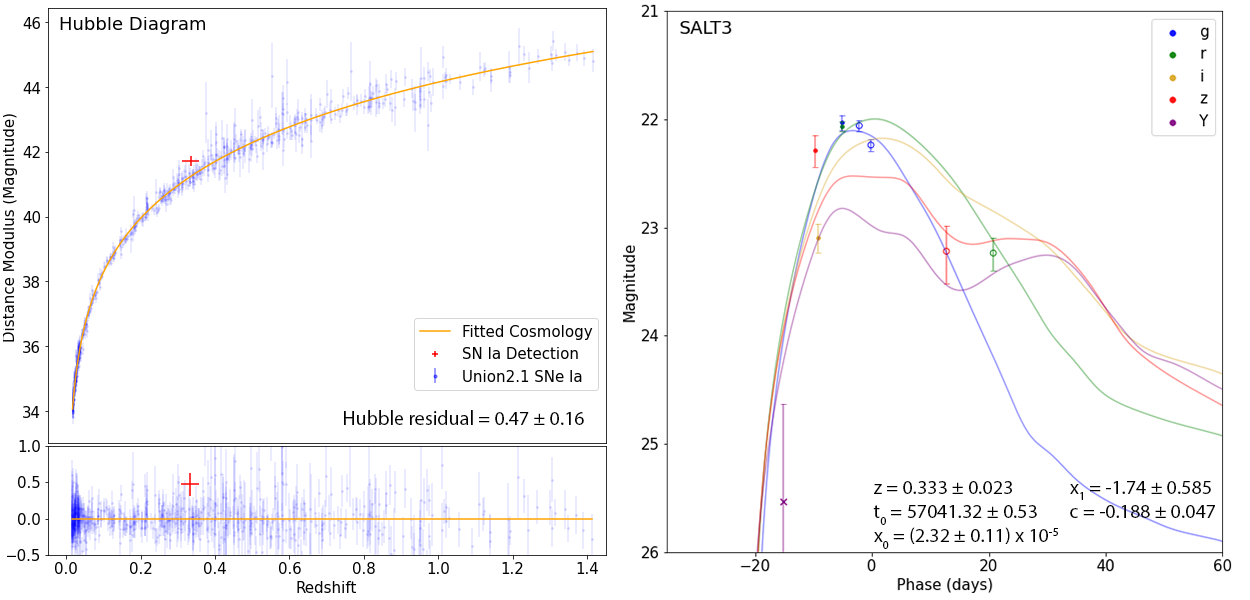}
\caption{Best-fit SALT3 model for DESI-052.0083-37.2049 with a lens photo-$z$ redshift prior of $0.292 \pm 0.029$.  \txr{For this and Figure~\ref{fig:3172p2}, solid photometry points correspond to the detection passes shown in Figure~\ref{fig:3172dets}, hollow points correspond to other exposures with PSF photometry, and crosses correspond to measurements using aperture photometry.}}\label{fig:3172p1}
\end{center}
\end{figure}

\paragraph{Postulation 1: uL-SN~Ia} Figure~\ref{fig:3172p1} shows the best-fit light curve model for the uL-SN~Ia scenario.  \txr{Note that the first \ipr~band point is near the peak, almost coincidental with a \ipg~band point.}  The SALT3 light curve model agrees reasonably well with the data.  The Hubble residual is somewhat large, but does not rule out this scenario.  Of note is that the first \ipz~band point appears to be too bright for this model.

\begin{figure}[H]
\begin{center}
\includegraphics[width=110mm]{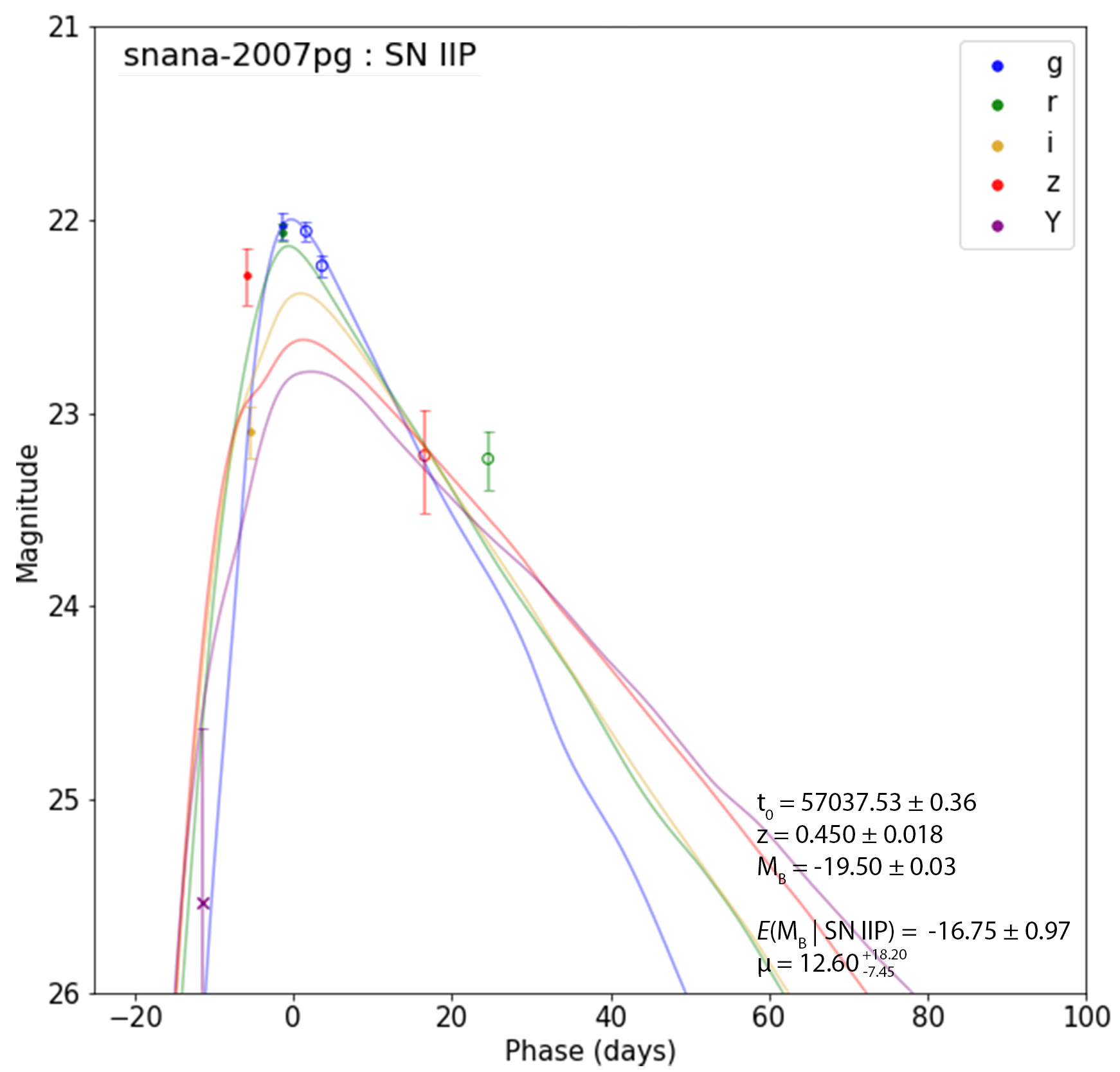}
\caption{Best-fit core collapse template model for DESI-052.0083-37.2049 with no redshift prior.}\label{fig:3172p2}
\end{center}
\end{figure}

\newpage \paragraph{Postulation 2: L-CC~SN} Figure~\ref{fig:3172p2} shows the best-fit light curve model for the L-CC~SN scenario.  The SN~IIP template provides a reasonable fit for the photometry, with an amplification of $12.60^{+18.20}_{-7.45}$.  Similarly to Postulation 1, the first \ipz~band point is not well fit by this model; nor is the second point of the \ipr~band.

\paragraph{Conclusion} The transient in DESI-052.0083-37.2049 appears to be either an uL-SN~Ia or a L-CC~SN.  If high resolution and/or spectroscopic observations reveal that the arc-like structure is part of a spiral or ring galaxy, that would obviously rule out the L-CC~SN scenario.  Conversely, if it is strongly lensed arc/Einstein ring formation, then the L-CC~SN possibility becomes quite possible, considering the location of the detection.  If \txb{found} live, this detection would warrant follow-up observations.

\newpage \subsection{DESI-084.8493-59.3586}
 \label{subsubsec:DESI-084.8493-059.3586}

DESI-084.8493-59.3586 was discovered in \citet{storfer2022}, labelled as a C-grade strong lensing candidate.
 
  \begin{figure}[H]
\begin{center}
\includegraphics[width=167mm]{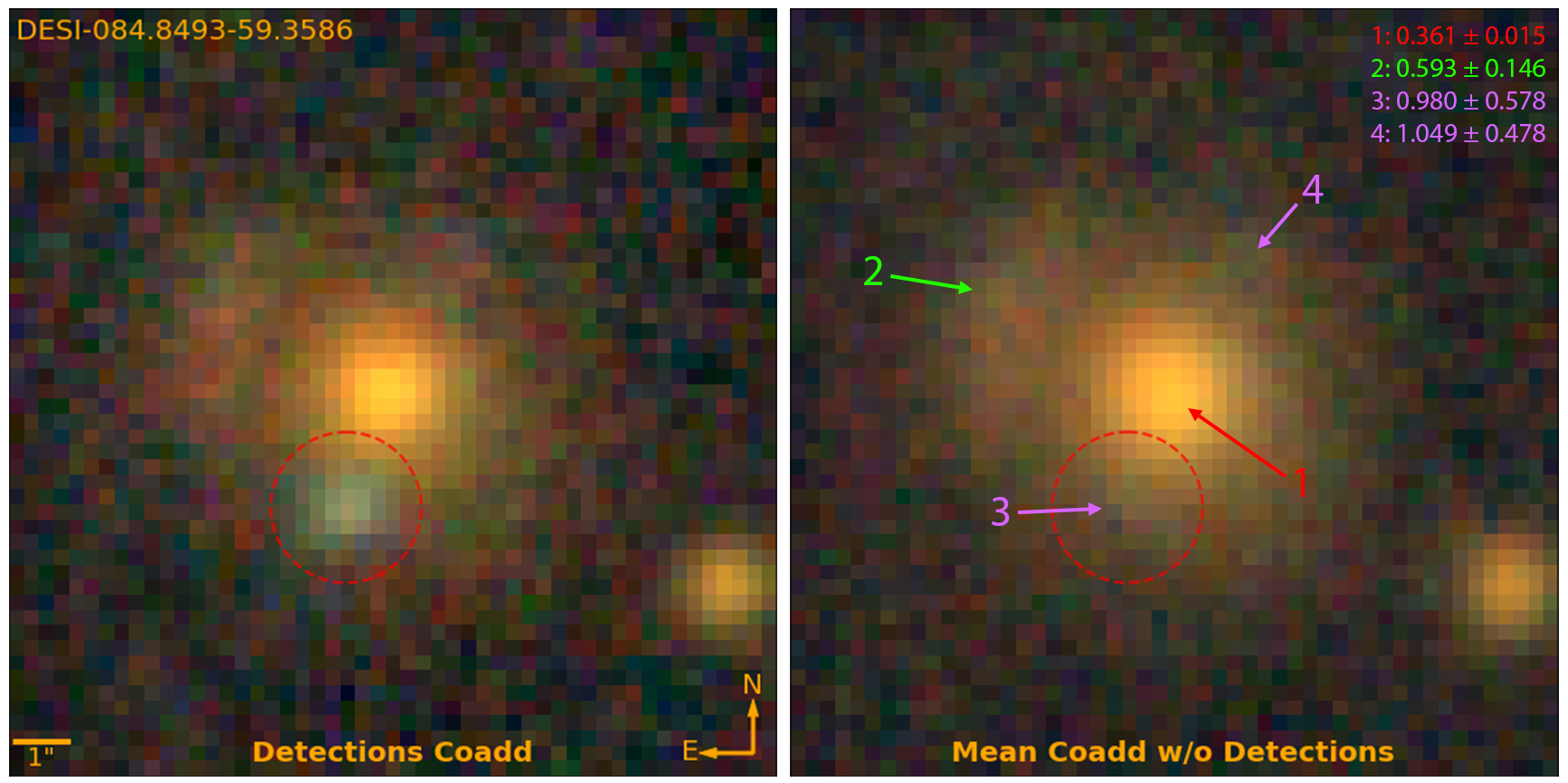}
\caption{Coadded RGB images (using \ipg, \ipr, \ipi, and \ipz~bands) of DESI-084.8493-59.3586 with and without the transient detection \txb{(red dotted circle)} exposures.  Labelled objects are color-coded as the following postulated scenario: red as the lens galaxy, green as a source galaxy, and purple as a second source.  Photometric redshifts are displayed on the top right.  \trr{The posited lens galaxy has a photo-$z$ of $0.361\pm 0.015$, the posited lensed image has a photo-$z$ of $0.593\pm 0.146$, and a second posited lensed source have photo-$z$'s of $0.980 \pm 0.578$ and $1.049 \pm 0.478$.}}\label{fig:3396rgb}
\end{center}
\end{figure}

\begin{figure}[H]
\begin{center}
\includegraphics[width=167mm]{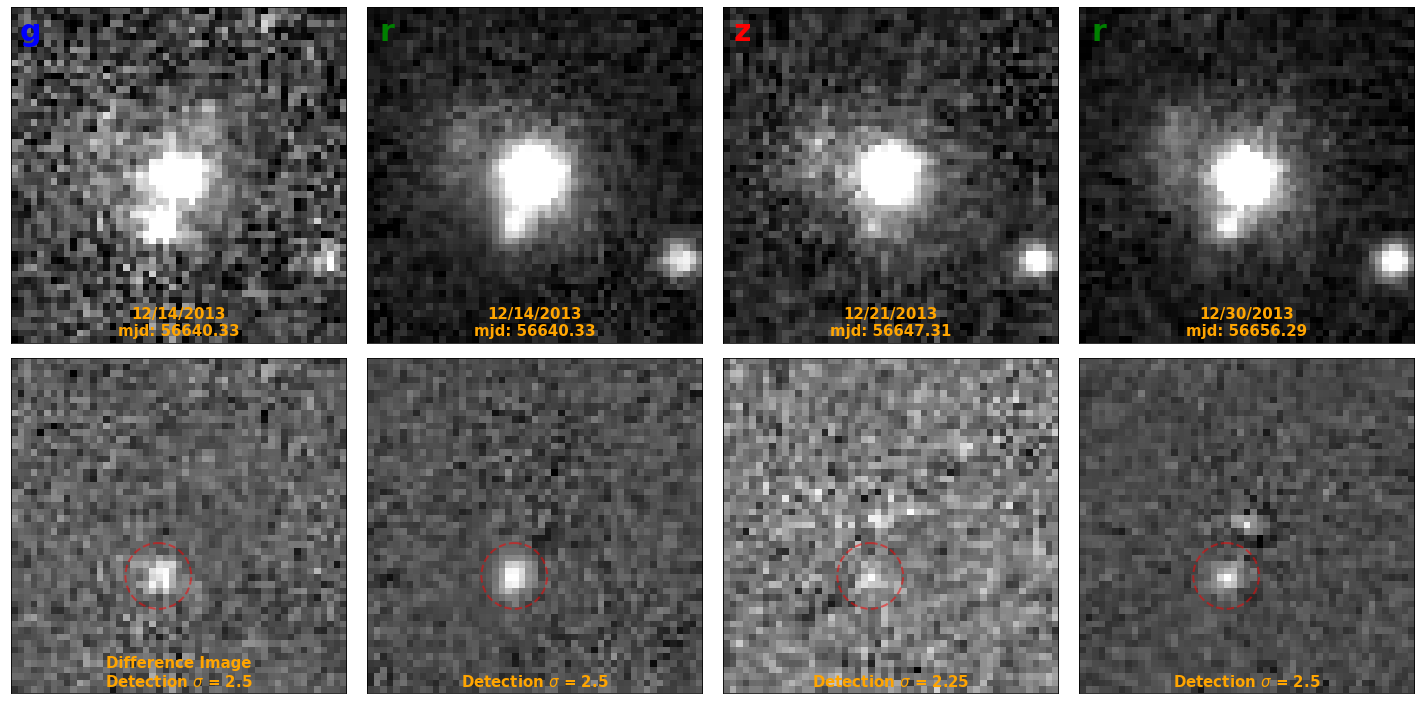}
\caption{Detection exposures for the transient in DESI-084.8493-59.3586 in chronological order. See caption of Figure~\ref{fig:3760dets} for the full description.}\label{fig:3396dets}
\end{center}
\end{figure}

DESI-084.8493-59.3586 is a single galaxy strongly lensed system.  There is a red lensed arc to the East of the lens \trr{(with a photo-$z$ of $0.593\pm 0.146$)}, and the transient detection lies South of the lens.   Additionally, \txr{it is possible} that objects \txr{3 and 4} (Figure~\ref{fig:3396rgb}) correspond to the same source galaxy, due to similarities in color and photo-$z$, \txb{with the possibility that object 3 is the host}.

\begin{figure}[H]
\begin{center}
\includegraphics[width=180mm]{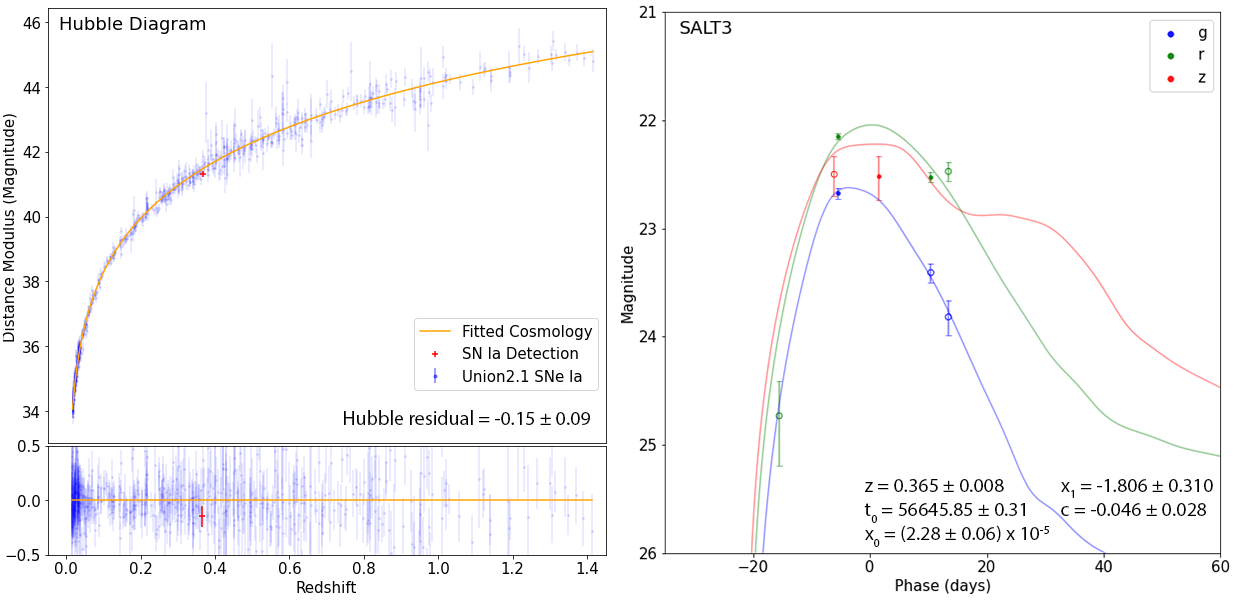}
\caption{Best-fit SALT3 model for DESI-084.8493-59.3586 with a lens photo-$z$ redshift prior of $0.361 \pm 0.015$.  For this and Figure~\ref{fig:3396p3}, solid photometry points correspond to the detection passes shown in Figure~\ref{fig:3396dets}.}\label{fig:3396p1}
\end{center}
\end{figure}

\paragraph{Postulation 1: uL-SN~Ia} Figure~\ref{fig:3396p1} shows the best-fit light curve model for the uL-SN~Ia scenario, with the foreground galaxy photo-$z$ used as the redshift prior.  The SALT3 model agrees well with the data, with reasonable light curve parameters and small Hubble residual.  

\begin{figure}[H]
\begin{center}
\includegraphics[width=110mm]{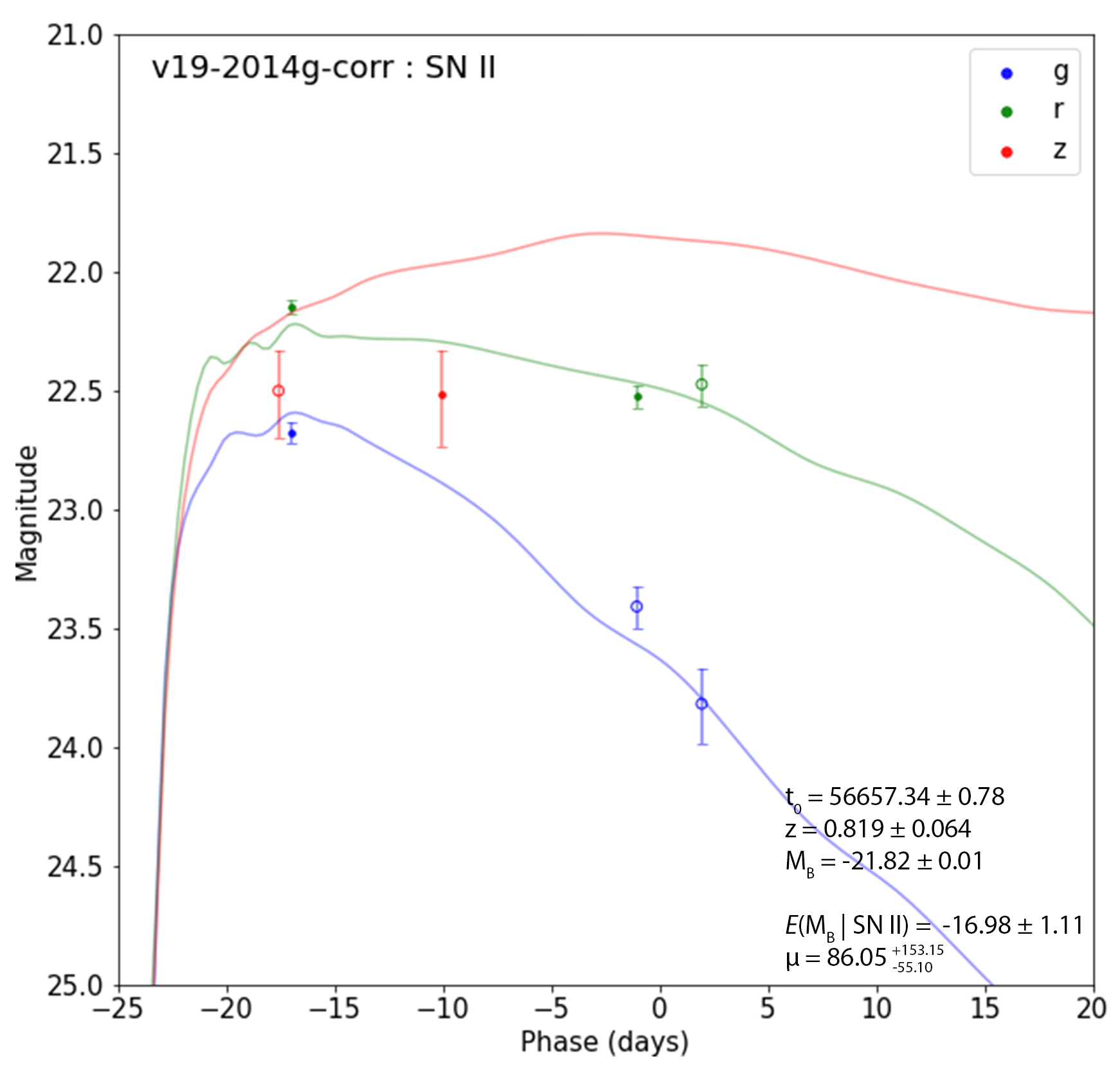}
\caption{Best-fit core collapse template model for DESI-084.8493-59.3586 with a source photo-$z$ redshift prior of $0.593 \pm 0.146$.}\label{fig:3396p3}
\end{center}
\end{figure}

\newpage \paragraph{Postulation 2: L-CC~SN} Figure~\ref{fig:3396p3} shows the best-fit light curve model for the L-CC~SN scenario.  The fit is significantly inferior to the previous postulation in the \ipz~band, requiring a high magnification of $86.05^{+153.15}_{-55.10}$, albeit with large uncertainties.

\paragraph{Conclusion} This transient in DESI-084.8493-59.3586 is likely an unlensed SN~Ia, though there is a \txr{small} possibility of it being a lensed CC~SN.  Thus, we give this system an appropriately low lensed SN grade of D.  If found live, follow-up spectroscopic observations at the transient location could easily distinguish these two scenarios.  

\newpage
\subsection{DESI-015.8465-50.5450}
 \label{subsubsec:DESI-015.8465-050.5450}

DESI-015.8465-50.5450 is a grade D$+$ strong lens candidate, discovered in \citet{storfer2022}.

  \begin{figure}[H]
\begin{center}
\includegraphics[width=173mm]{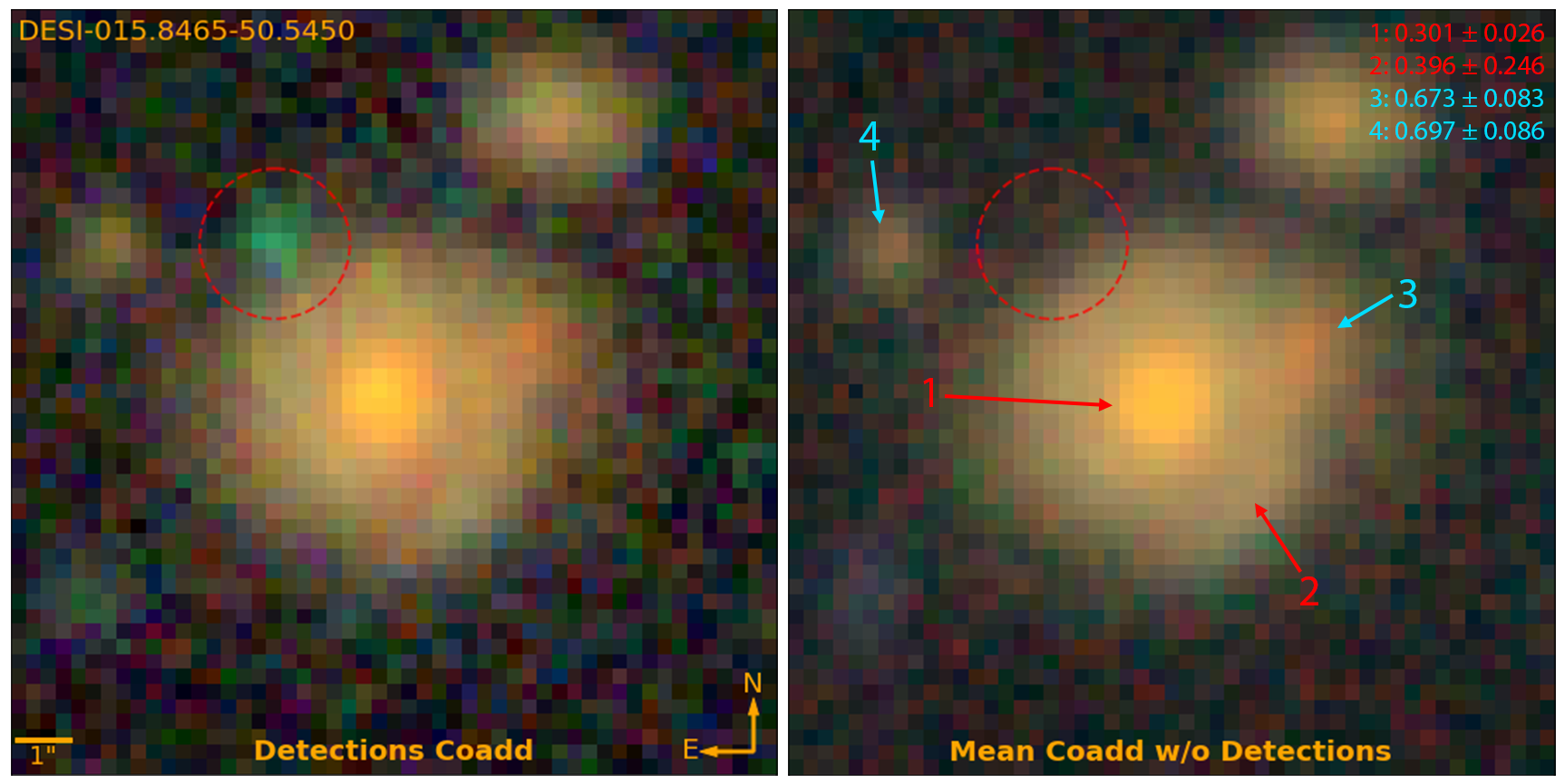}
\caption{Coadded RGB images (using \ipg, \ipr, \ipi, and \ipz~bands) of DESI-015.8465-50.5450 with and without the transient detection \txb{(red dotted circle)} exposures.  Labelled objects are color-coded as the following postulated scenario: red as the main galaxy and cyan as surrounding galaxies.  Photometric redshifts are displayed on the top right.  Photometric redshifts are displayed on the top right.  \trr{The posited host galaxy has photo-$z$'s of $0.301\pm 0.026$ and $0.396 \pm 0.246$.}}\label{fig:4004rgb}
\end{center}
\end{figure}

\begin{figure}[H]
\begin{center}
\includegraphics[width=84mm]{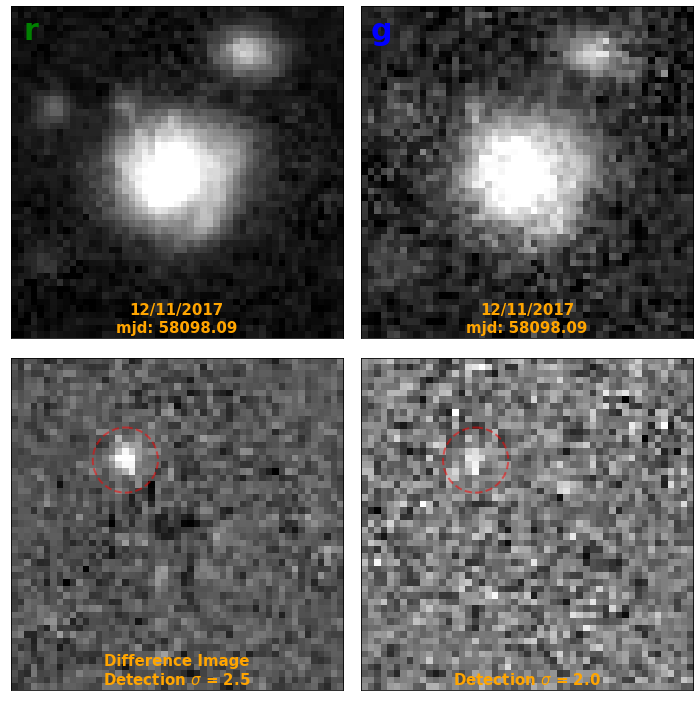}
\caption{Detection exposures for the transient in DESI-015.8465-50.5450 in chronological order. See caption of Figure~\ref{fig:3760dets} for the full description.}\label{fig:4004dets}
\end{center}
\end{figure}

Upon closer inspection of DESI-015.8465-50.5450, we believe it is more likely a face-on spiral galaxy.  In the detection coadd image in Figure~\ref{fig:4004rgb}, the evidence for the spiral pattern (rather than lensed arcs) is especially strong in the \ipg~band.  There are only two detections of this transient, observed two minutes apart.  However, we do not observe a shift in detection location above the level of \txr{noise, and so we do not consider asteroid as a likely scenario.}

\begin{figure}[H]
\begin{center}
\includegraphics[width=180mm]{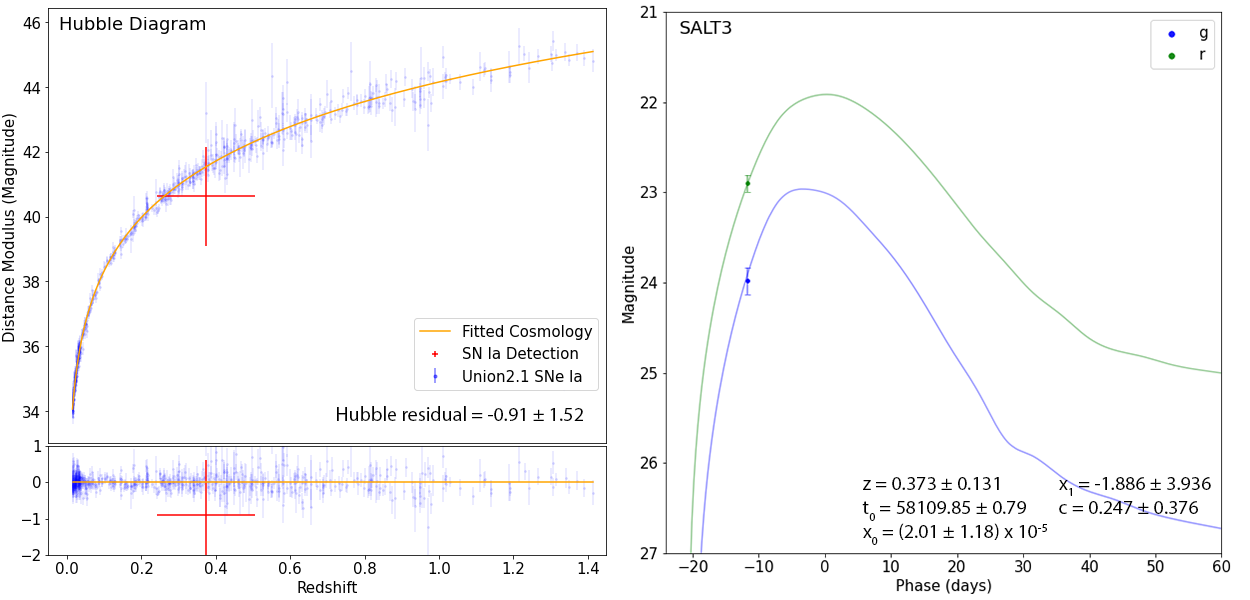}
\caption{Best-fit SALT3 model for DESI-015.8465-50.5450 with a lens photo-$z$ redshift prior of $0.301 \pm 0.026$.  For this and Figure~\ref{fig:4004p2}, photometry points correlate to the detections shown in Figure~\ref{fig:4004dets}.}\label{fig:4004p1}
\end{center}
\end{figure}

\paragraph{Postulation 1: uL-SN~Ia} Figure~\ref{fig:4004p1} shows the best-fit light curve model for the uL-SN~Ia scenario. Due to the \txr{sparsity} of the photometric data, the uncertainties of the SALT3 model parameters and Hubble residuals are large.  While there are only two detection exposures, the pipeline identified this transient \txr{with four} sub-detections.  As with all other light curves presented, the light curves below are constrained by both detection and non-detection exposures.

\begin{figure}[H]
\begin{center}
\includegraphics[width=110mm]{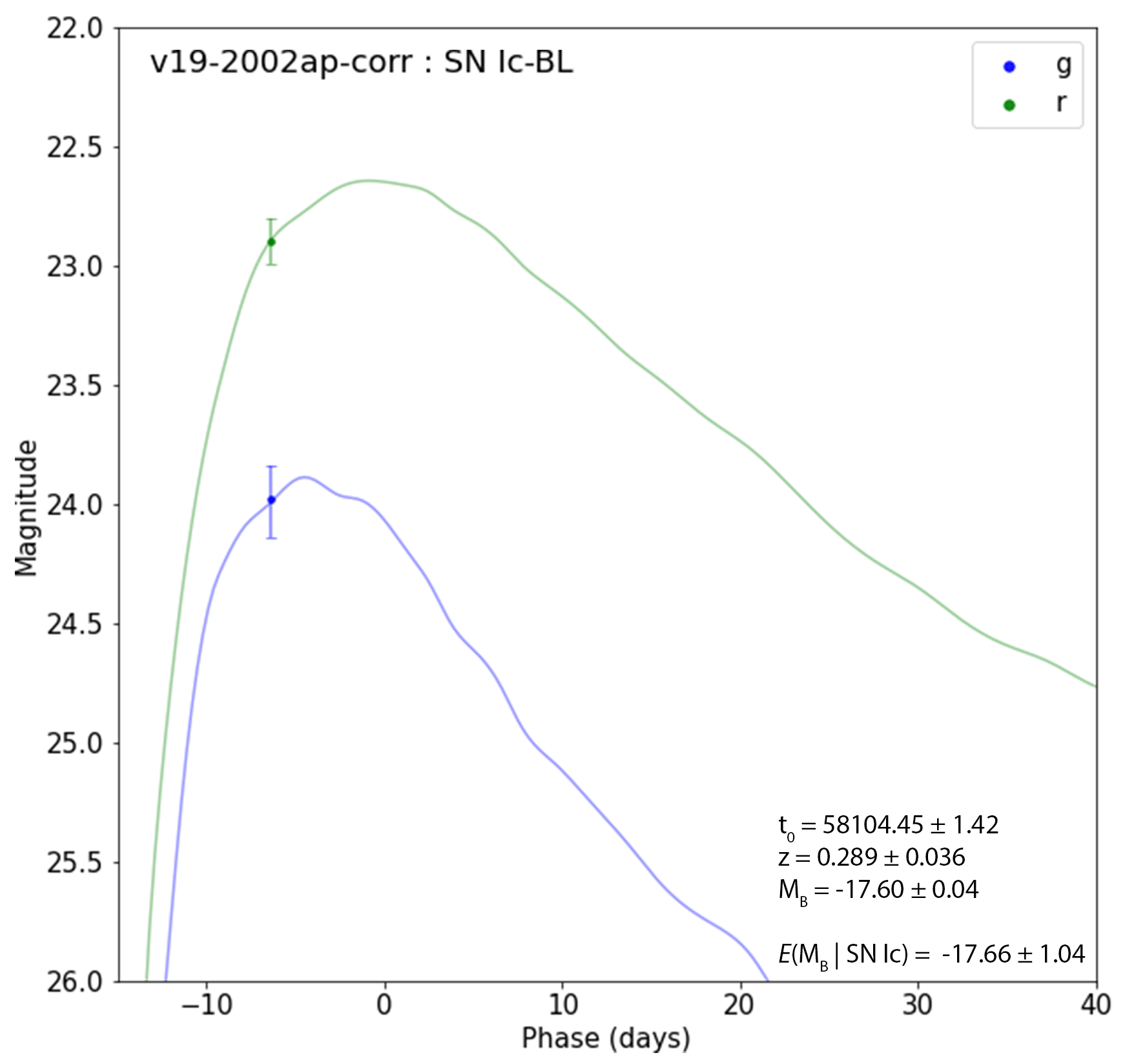}
\caption{Best-fit core collapse template model for DESI-015.8465-50.5450 with a lens photo-$z$ redshift prior of $0.301 \pm 0.026$.}\label{fig:4004p2}
\end{center}
\end{figure}

\newpage \paragraph{Postulation 2: uL-CC~SN} Figure~\ref{fig:4004p2} shows the best-fit light curve model for the uL-CC~SN scenario.  This SN~Ic-LB template (``v19-2002ap-coor") is one of many templates that are consistent with the data.

\paragraph{Conclusion} The most likely scenario for the transient in DESI-015.8465-50.5450 is an unlensed supernova, as the system is probably a spiral galaxy and not a strong lensing system.  With the sparsity of photometric data, it is not possible to determine the type.  It could also be the case that the host galaxy is object 4 \txr{(which could be lensed, possibly with object 3 as its counterimage)}, as opposed to object 1 (see Figure~\ref{fig:4004rgb}), but it is infeasible to determine with \txb{current data}.

\section{Asteroids}\label{appendix_asteroid}

We have found two asteroid candidates.  The detections are observed on the same night (for each respective system), separated by approximately one to two minutes.  The locations of the two detections in each case are spatially close enough for the pipeline to identify them as candidates (as a group of three to four sub-detections).  PSF fitting for the transient detections shows that the movements between detections for both systems are significant.  The approximate speeds of transients are consistent with that of a main-belt asteroid (roughly $0.5''$ per minute near opposition; \citealp{cicco}).   

\subsection{DESI-008.6173+02.4228}
 \label{subsubsec:DESI-008.6173+02.4228}
 \begin{figure}[H]
\begin{center}
\includegraphics[width=160mm]{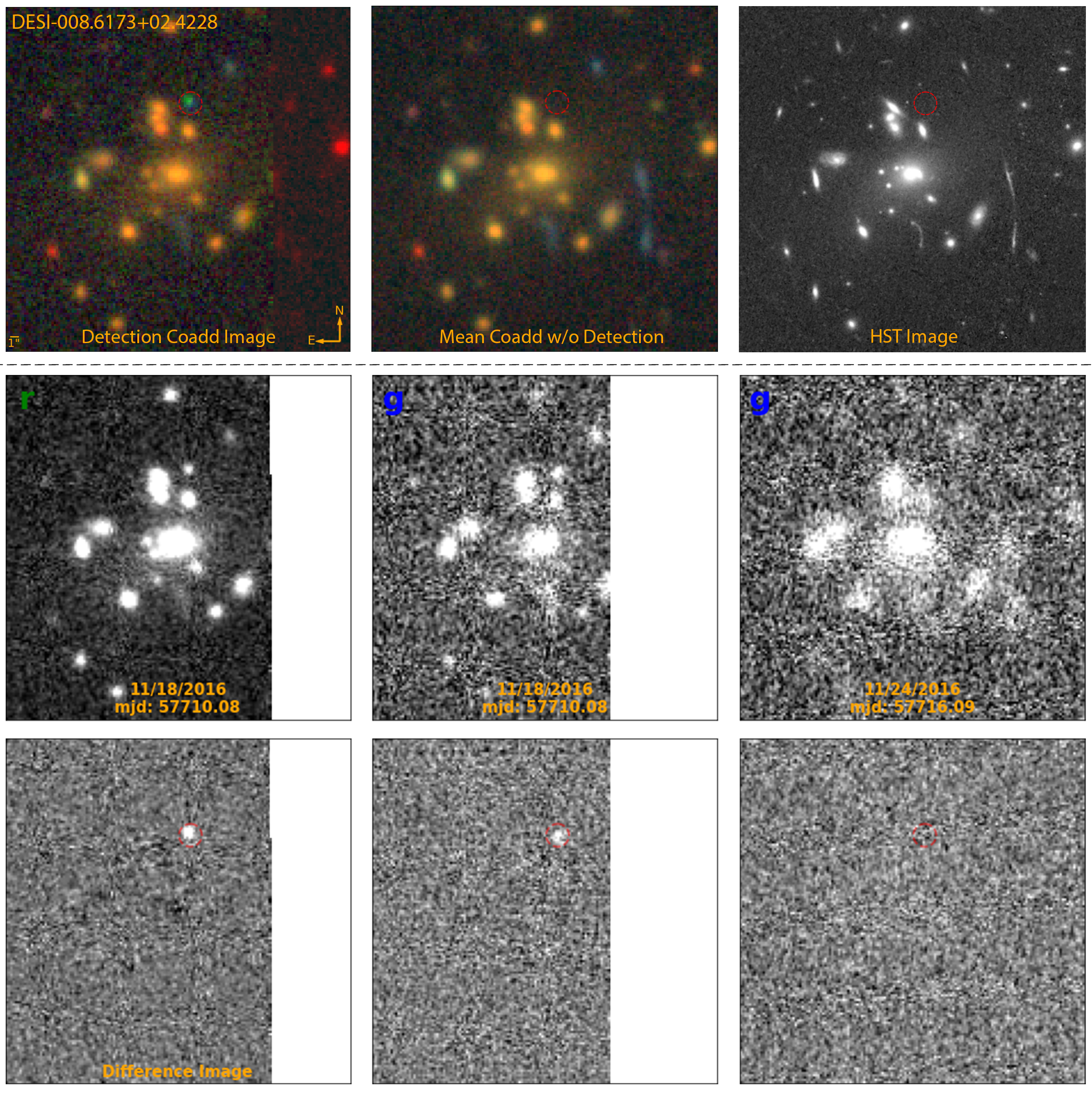}
\caption{Above the dotted line are coadded images of DESI-008.6173+02.4228 with and without the transient detection \txb{(red dotted circle)} exposures.  From left to right, the images show 1.) The coadded image generated from mean coadding from only the exposures with the transient detection, within each band.  2.) The coadded image, generated from mean coadding all exposures excluding the detection exposures within each band.  3.) The HST image of the system (HST Proposal ID: 12884; H. Ebeling).  The RGB image incorporates \ipg, \ipr, \ipi, and \ipz~bands.  Below the dotted line are detection exposures for the transient.  Each column is a single detection at the labelled date and band.  The top row is the single exposure image, whereas the bottom is the SFFT difference image.  The detection location is marked with a red dotted circle.  The first and second columns show \ipr~and \ipg~band detections on 01/18/2016 minutes apart, whereas the third column shows a set on nondetections \txb{six} days afterwards.  Note the slight but significant shift in the transient location between the two detections.  }\label{fig:astroid1}
\end{center}
\end{figure}

For DESI-008.6173+02.4228, we find that the transient has moved $0.475'' \pm 0.101$ between the \ipr~and \ipg~band detections; a movement of $>4\sigma$ significance.  As the exposures were taken $1.98$ minutes apart, the estimated speed of this asteroid is $0.240'' \pm 0.055$ per minute.  This is slower than the typical main-belt asteroid (near opposition) speed of $0.5''$ per minute, but not unreasonably so.  The coordinates and time observed does not correlate with any known asteroid in the \href{https://www.minorplanetcenter.net/iau/mpc.html}{IAU's Minor Planet Center} database. 

\subsection{DESI-150.4863+15.4209} 
 \label{subsubsec:DESI-150.4863+15.4209}

\begin{figure}[H]
\begin{center}
\includegraphics[width=160mm]{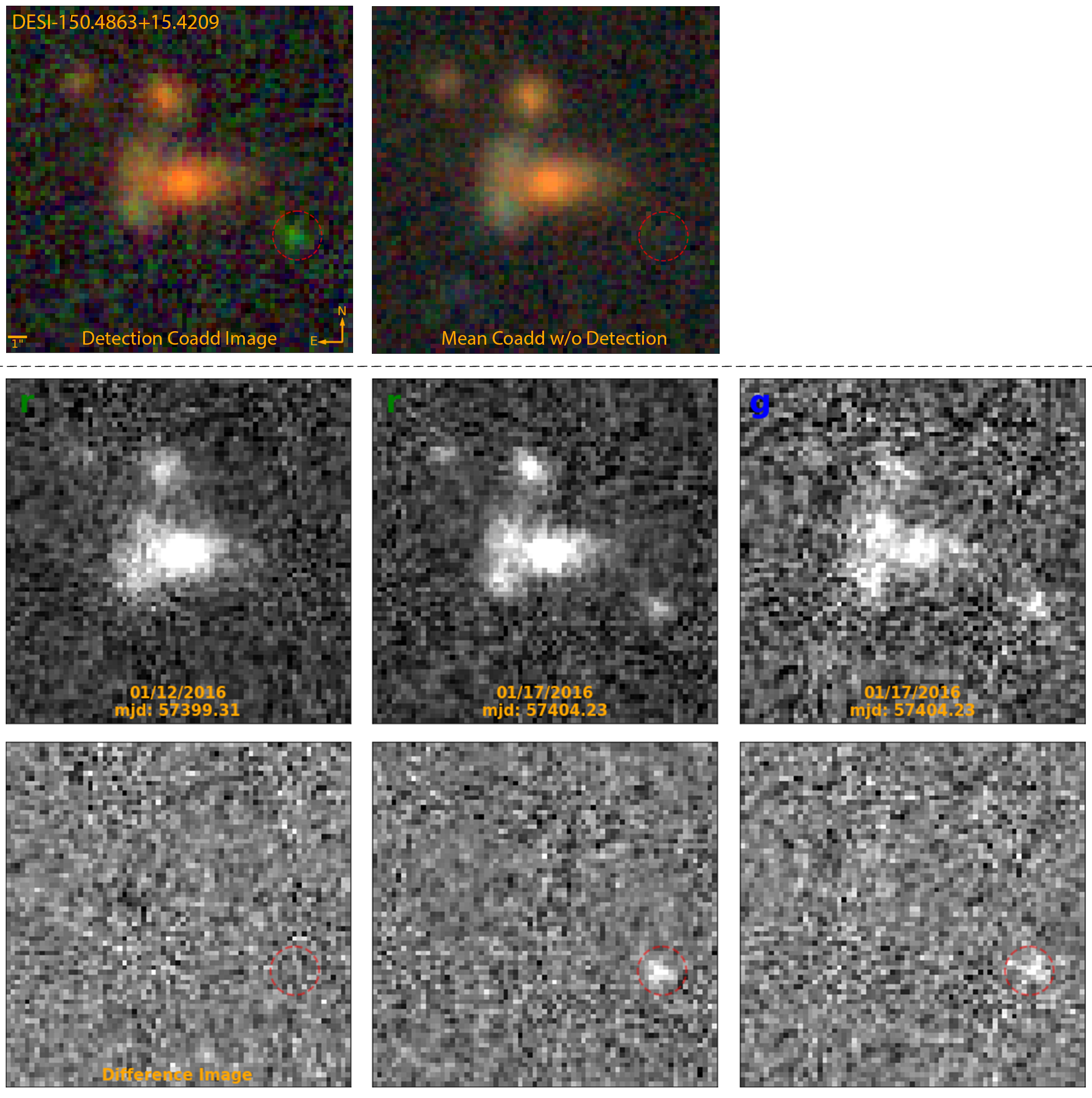}
\caption{Above the dotted line are coadded images of DESI-150.4863+15.4209 with and without the transient detection \txb{(red dotted circle)} exposures (see Figure~\ref{fig:astroid1} caption).  Below the dotted line are detection exposures for the transient.  The second and third columns show detections on 01/17/2016, while the first column shows a nondetection on 01/12/2016 (5 days prior to detection).  The detection location is marked with a red dotted circle.  As with the previous system, there is a significant shift in detection location between the exposures.  }
\end{center}
\end{figure}

In DESI-150.4863+15.4209, we find that the transient has moved $2.155'' \pm 0.372$ between the \ipr~and \ipg~band detections, with a significance of $>5\sigma$.  The exposures were taken $1.23$ minutes apart.  Thus, the estimated speed of this asteroid is $0.475'' \pm 0.082$ per minute, which is consistent with a main-belt asteroid near opposition.  The coordinates and time observed does not correlate with any known asteroid in the \href{https://www.minorplanetcenter.net/iau/mpc.html}{IAU's Minor Planet Center} database.


\bibliography{sample631}{}

\begin{thebibliography}{}
\expandafter\ifx\csname natexlab\endcsname\relax\def\natexlab#1{#1}\fi
\providecommand{\url}[1]{\href{#1}{#1}}
\providecommand{\dodoi}[1]{doi:~\href{http://doi.org/#1}{\nolinkurl{#1}}}
\providecommand{\doeprint}[1]{\href{http://ascl.net/#1}{\nolinkurl{http://ascl.net/#1}}}
\providecommand{\doarXiv}[1]{\href{https://arxiv.org/abs/#1}{\nolinkurl{https://arxiv.org/abs/#1}}}

\bibitem[{Abbott {et~al.}(2021)Abbott, Adam{\'{o} }w, Aguena, Allam, Amon,
  Annis, Avila, Bacon, Banerji, Bechtol, Becker, Bernstein, Bertin, Bhargava,
  Bridle, Brooks, Burke, Rosell, Kind, Carretero, Castander, Cawthon, Chang,
  Choi, Conselice, Costanzi, Crocce, da~Costa, Davis, Vicente, DeRose, Desai,
  Diehl, Dietrich, Drlica-Wagner, Eckert, Elvin-Poole, Everett, Evrard,
  Ferrero, Fert{\'{e}}, Flaugher, Fosalba, Friedel, Frieman,
  Garc{\'{\i}}a-Bellido, Gaztanaga, Gelman, Gerdes, Giannantonio, Gill, Gruen,
  Gruendl, Gschwend, Gutierrez, Hartley, Hinton, Hollowood, Honscheid, Huterer,
  James, Jeltema, Johnson, Kent, Kron, Kuehn, Kuropatkin, Lahav, Li, Lidman,
  Lin, MacCrann, Maia, Manning, Maloney, March, Marshall, Martini, Melchior,
  Menanteau, Miquel, Morgan, Myles, Neilsen, Ogando, Palmese,
  Paz-Chinch{\'{o}}n, Petravick, Pieres, Plazas, Pond, Rodriguez-Monroy, Romer,
  Roodman, Rykoff, Sako, Sanchez, Santiago, Scarpine, Serrano, Sevilla-Noarbe,
  Smith, Smith, Soares-Santos, Suchyta, Swanson, Tarle, Thomas, To, Tremblay,
  Troxel, Tucker, Turner, Varga, Walker, Wechsler, Weller, Wester, Wilkinson,
  Yanny, Zhang, Nikutta, Fitzpatrick, Jacques, Scott, Olsen, Huang, Herrera,
  Juneau, Nidever, Weaver, Adean, Correia, de~Freitas, Freitas, Singulani, \&
  Vila-Verde}]{DESDR2}
Abbott, T. M.~C., Adam{\'{o} }w, M., Aguena, M., {et~al.} 2021, The
  Astrophysical Journal Supplement, 255, 20, \dodoi{10.3847/1538-4365/ac00b3}

\bibitem[{Aihara {et~al.}(2019)Aihara, AlSayyad, Ando, Armstrong, Bosch, Egami,
  Furusawa, Furusawa, Goulding, Harikane, Hikage, Ho, Hsieh, Huang, Ikeda,
  Imanishi, Ito, Iwata, Jaelani, Kakuma, Kawana, Kikuta, Kobayashi, Koike,
  Komiyama, Li, Liang, Lin, Luo, Lupton, Lust, MacArthur, Matsuoka, Mineo,
  Miyatake, Miyazaki, More, Murata, Namiki, Nishizawa, Oguri, Okabe, Okamoto,
  Okura, Ono, Onodera, Onoue, Osato, Ouchi, Shibuya, Strauss, Sugiyama, Suto,
  Takada, Takagi, Takata, Takita, Tanaka, Terai, Toba, Uchiyama, Utsumi, Wang,
  Wang, \& Yamada}]{HSC_DR2}
Aihara, H., AlSayyad, Y., Ando, M., {et~al.} 2019, Publications of the
  Astronomical Society of Japan, 71, \dodoi{10.1093/pasj/psz103}

\bibitem[{{Alard}(2000)}]{alard2000}
{Alard}, C. 2000, \aaps, 144, 363, \dodoi{10.1051/aas:2000214}

\bibitem[{{Astropy Collaboration} {et~al.}(2013){Astropy Collaboration},
  {Robitaille}, {Tollerud}, {Greenfield}, {Droettboom}, {Bray}, {Aldcroft},
  {Davis}, {Ginsburg}, {Price-Whelan}, {Kerzendorf}, {Conley}, {Crighton},
  {Barbary}, {Muna}, {Ferguson}, {Grollier}, {Parikh}, {Nair}, {Unther},
  {Deil}, {Woillez}, {Conseil}, {Kramer}, {Turner}, {Singer}, {Fox}, {Weaver},
  {Zabalza}, {Edwards}, {Azalee Bostroem}, {Burke}, {Casey}, {Crawford},
  {Dencheva}, {Ely}, {Jenness}, {Labrie}, {Lim}, {Pierfederici}, {Pontzen},
  {Ptak}, {Refsdal}, {Servillat}, \& {Streicher}}]{astropy2013}
{Astropy Collaboration}, {Robitaille}, T.~P., {Tollerud}, E.~J., {et~al.} 2013,
  \aap, 558, A33, \dodoi{10.1051/0004-6361/201322068}

\bibitem[{{Astropy Collaboration} {et~al.}(2018){Astropy Collaboration},
  {Price-Whelan}, {Sip{\H{o}}cz}, {G{\"u}nther}, {Lim}, {Crawford}, {Conseil},
  {Shupe}, {Craig}, {Dencheva}, {Ginsburg}, {VanderPlas}, {Bradley},
  {P{\'e}rez-Su{\'a}rez}, {de Val-Borro}, {Aldcroft}, {Cruz}, {Robitaille},
  {Tollerud}, {Ardelean}, {Babej}, {Bach}, {Bachetti}, {Bakanov}, {Bamford},
  {Barentsen}, {Barmby}, {Baumbach}, {Berry}, {Biscani}, {Boquien}, {Bostroem},
  {Bouma}, {Brammer}, {Bray}, {Breytenbach}, {Buddelmeijer}, {Burke},
  {Calderone}, {Cano Rodr{\'\i}guez}, {Cara}, {Cardoso}, {Cheedella}, {Copin},
  {Corrales}, {Crichton}, {D'Avella}, {Deil}, {Depagne}, {Dietrich}, {Donath},
  {Droettboom}, {Earl}, {Erben}, {Fabbro}, {Ferreira}, {Finethy}, {Fox},
  {Garrison}, {Gibbons}, {Goldstein}, {Gommers}, {Greco}, {Greenfield},
  {Groener}, {Grollier}, {Hagen}, {Hirst}, {Homeier}, {Horton}, {Hosseinzadeh},
  {Hu}, {Hunkeler}, {Ivezi{\'c}}, {Jain}, {Jenness}, {Kanarek}, {Kendrew},
  {Kern}, {Kerzendorf}, {Khvalko}, {King}, {Kirkby}, {Kulkarni}, {Kumar},
  {Lee}, {Lenz}, {Littlefair}, {Ma}, {Macleod}, {Mastropietro}, {McCully},
  {Montagnac}, {Morris}, {Mueller}, {Mumford}, {Muna}, {Murphy}, {Nelson},
  {Nguyen}, {Ninan}, {N{\"o}the}, {Ogaz}, {Oh}, {Parejko}, {Parley}, {Pascual},
  {Patil}, {Patil}, {Plunkett}, {Prochaska}, {Rastogi}, {Reddy Janga},
  {Sabater}, {Sakurikar}, {Seifert}, {Sherbert}, {Sherwood-Taylor}, {Shih},
  {Sick}, {Silbiger}, {Singanamalla}, {Singer}, {Sladen}, {Sooley},
  {Sornarajah}, {Streicher}, {Teuben}, {Thomas}, {Tremblay}, {Turner},
  {Terr{\'o}n}, {van Kerkwijk}, {de la Vega}, {Watkins}, {Weaver}, {Whitmore},
  {Woillez}, {Zabalza}, \& {Astropy Contributors}}]{astropy2018}
{Astropy Collaboration}, {Price-Whelan}, A.~M., {Sip{\H{o}}cz}, B.~M., {et~al.}
  2018, \aj, 156, 123, \dodoi{10.3847/1538-3881/aabc4f}

\bibitem[{Barbary(2014)}]{sncosmo}
Barbary, K. 2014, sncosmo v0.4.2,  Zenodo, \dodoi{10.5281/zenodo.11938}

\bibitem[{{Barbary}(2018)}]{sep}
{Barbary}, K. 2018, {SEP: Source Extraction and Photometry}, Astrophysics
  Source Code Library, record ascl:1811.004.
\newblock \doeprint{1811.004}

\bibitem[{{Becker}(2015)}]{hotpants}
{Becker}, A. 2015, {HOTPANTS: High Order Transform of PSF ANd Template
  Subtraction}, Astrophysics Source Code Library, record ascl:1504.004.
\newblock \doeprint{1504.004}

\bibitem[{{Bell} {et~al.}(2007){Bell}, {Zheng}, {Papovich}, {Borch}, {Wolf}, \&
  {Meisenheimer}}]{bell2007}
{Bell}, E.~F., {Zheng}, X.~Z., {Papovich}, C., {et~al.} 2007, \apj, 663, 834,
  \dodoi{10.1086/518594}

\bibitem[{{Bertin} \& {Arnouts}(1996)}]{bertin1996}
{Bertin}, E., \& {Arnouts}, S. 1996, \aaps, 117, 393,
  \dodoi{10.1051/aas:1996164}

\bibitem[{Birrer {et~al.}(2022)Birrer, Dhawan, \& Shajib}]{birrer2021}
Birrer, S., Dhawan, S., \& Shajib, A.~J. 2022, The Astrophysical Journal, 924,
  2, \dodoi{10.3847/1538-4357/ac323a}

\bibitem[{Bramich(2008)}]{bramich2008}
Bramich, D.~M. 2008, Monthly Notices of the Royal Astronomical Society:
  Letters, 386, L77, \dodoi{10.1111/j.1745-3933.2008.00464.x}

\bibitem[{Ca{\~{n}}ameras {et~al.}(2020)Ca{\~{n}}ameras, Schuldt, Suyu,
  Taubenberger, Meinhardt, Leal-Taix{\'{e}}, Lemon, Rojas, \&
  Savary}]{canameras2020}
Ca{\~{n}}ameras, R., Schuldt, S., Suyu, S.~H., {et~al.} 2020, Astronomy \&
  Astrophysics, 644, A163, \dodoi{10.1051/0004-6361/202038219}

\bibitem[{Ca{\~{n}}ameras {et~al.}(2021)Ca{\~{n}}ameras, Schuldt, Shu, Suyu,
  Taubenberger, Meinhardt, Leal-Taix{\'{e}}, Chao, Inoue, Jaelani, \&
  More}]{canameras2021}
Ca{\~{n}}ameras, R., Schuldt, S., Shu, Y., {et~al.} 2021, Astronomy \&
  Astrophysics, 653, L6, \dodoi{10.1051/0004-6361/202141758}

\bibitem[{{Carrasco} {et~al.}(2017){Carrasco}, {Barrientos}, {Anguita},
  {Garcia-Vergara}, {Bayliss}, {Gladders}, {Gilbank}, {Yee}, \&
  {West}}]{carrasco2017}
{Carrasco}, M., {Barrientos}, L.~F., {Anguita}, T., {et~al.} 2017, VizieR
  Online Data Catalog, J/ApJ/834/210

\bibitem[{Chambers {et~al.}(2016)Chambers, Magnier, Metcalfe, Flewelling,
  Huber, Waters, Denneau, Draper, Farrow, Finkbeiner, Holmberg, Koppenhoefer,
  Price, Rest, Saglia, Schlafly, Smartt, Sweeney, Wainscoat, Burgett, Chastel,
  Grav, Heasley, Hodapp, Jedicke, Kaiser, Kudritzki, Luppino, Lupton, Monet,
  Morgan, Onaka, Shiao, Stubbs, Tonry, White, Bañados, Bell, Bender, Bernard,
  Boegner, Boffi, Botticella, Calamida, Casertano, Chen, Chen, Cole, Deacon,
  Frenk, Fitzsimmons, Gezari, Gibbs, Goessl, Goggia, Gourgue, Goldman, Grant,
  Grebel, Hambly, Hasinger, Heavens, Heckman, Henderson, Henning, Holman, Hopp,
  Ip, Isani, Jackson, Keyes, Koekemoer, Kotak, Le, Liska, Long, Lucey, Liu,
  Martin, Masci, McLean, Mindel, Misra, Morganson, Murphy, Obaika, Narayan,
  Nieto-Santisteban, Norberg, Peacock, Pier, Postman, Primak, Rae, Rai, Riess,
  Riffeser, Rix, Röser, Russel, Rutz, Schilbach, Schultz, Scolnic, Strolger,
  Szalay, Seitz, Small, Smith, Soderblom, Taylor, Thomson, Taylor, Thakar,
  Thiel, Thilker, Unger, Urata, Valenti, Wagner, Walder, Walter, Watters,
  Werner, Wood-Vasey, \& Wyse}]{panstarrs}
Chambers, K.~C., Magnier, E.~A., Metcalfe, N., {et~al.} 2016, The Pan-STARRS1
  Surveys,  arXiv, \dodoi{10.48550/ARXIV.1612.05560}

\bibitem[{Chen {et~al.}(2022)Chen, Kelly, Oguri, Broadhurst, Diego, Emami,
  Filippenko, Treu, \& Zitrin}]{kellysn2}
Chen, W., Kelly, P.~L., Oguri, M., {et~al.} 2022, Nature, 611, 256—259,
  \dodoi{10.1038/s41586-022-05252-5}

\bibitem[{{Cicco}(2006)}]{cicco}
{Cicco}, D. 2006, {Hunting Asteroids From Your Backyard}, Sky and Telescope

\bibitem[{Craig {et~al.}(2021)Craig, O'Connor, Chakrabarti, Rodney, Pierel,
  McCully, \& Perez-Fournon}]{craig2021}
Craig, P., O'Connor, K., Chakrabarti, S., {et~al.} 2021, arXiv e-prints,
  \dodoi{10.48550/ARXIV.2111.01680}

\bibitem[{{Dark Energy Survey Collaboration} {et~al.}(2016){Dark Energy Survey
  Collaboration}, {Abbott}, {Abdalla}, {Aleksi{\'c}}, {Allam}, {Amara},
  {Bacon}, {Balbinot}, {Banerji}, {Bechtol}, {Benoit-L{\'e}vy}, {Bernstein},
  {Bertin}, {Blazek}, {Bonnett}, {Bridle}, {Brooks}, {Brunner}, {Buckley-Geer},
  {Burke}, {Caminha}, {Capozzi}, {Carlsen}, {Carnero-Rosell}, {Carollo},
  {Carrasco-Kind}, {Carretero}, {Castander}, {Clerkin}, {Collett}, {Conselice},
  {Crocce}, {Cunha}, {D'Andrea}, {da Costa}, {Davis}, {Desai}, {Diehl},
  {Dietrich}, {Dodelson}, {Doel}, {Drlica-Wagner}, {Estrada}, {Etherington},
  {Evrard}, {Fabbri}, {Finley}, {Flaugher}, {Foley}, {Fosalba}, {Frieman},
  {Garc{\'\i}a-Bellido}, {Gaztanaga}, {Gerdes}, {Giannantonio}, {Goldstein},
  {Gruen}, {Gruendl}, {Guarnieri}, {Gutierrez}, {Hartley}, {Honscheid}, {Jain},
  {James}, {Jeltema}, {Jouvel}, {Kessler}, {King}, {Kirk}, {Kron}, {Kuehn},
  {Kuropatkin}, {Lahav}, {Li}, {Lima}, {Lin}, {Maia}, {Makler}, {Manera},
  {Maraston}, {Marshall}, {Martini}, {McMahon}, {Melchior}, {Merson}, {Miller},
  {Miquel}, {Mohr}, {Morice-Atkinson}, {Naidoo}, {Neilsen}, {Nichol}, {Nord},
  {Ogando}, {Ostrovski}, {Palmese}, {Papadopoulos}, {Peiris}, {Peoples},
  {Percival}, {Plazas}, {Reed}, {Refregier}, {Romer}, {Roodman}, {Ross},
  {Rozo}, {Rykoff}, {Sadeh}, {Sako}, {S{\'a}nchez}, {Sanchez}, {Santiago},
  {Scarpine}, {Schubnell}, {Sevilla-Noarbe}, {Sheldon}, {Smith}, {Smith},
  {Soares-Santos}, {Sobreira}, {Soumagnac}, {Suchyta}, {Sullivan}, {Swanson},
  {Tarle}, {Thaler}, {Thomas}, {Thomas}, {Tucker}, {Vieira}, {Vikram},
  {Walker}, {Wechsler}, {Weller}, {Wester}, {Whiteway}, {Wilcox}, {Yanny},
  {Zhang}, \& {Zuntz}}]{abbott2016}
{Dark Energy Survey Collaboration}, {Abbott}, T., {Abdalla}, F.~B., {et~al.}
  2016, \mnras, 460, 1270, \dodoi{10.1093/mnras/stw641}

\bibitem[{Dey {et~al.}(2016)Dey, Rabinowitz, Karcher, Bebek, Baltay,
  Sprayberry, Valdes, Stupak, Donaldson, Emmet, Hurteau, Abareshi, Marshall,
  Lang, Fitzpatrick, Daly, Joyce, Schlegel, Schweiker, Allen, Blum, \&
  Levi}]{dey2016}
Dey, A., Rabinowitz, D., Karcher, A., {et~al.} 2016, in Ground-based and
  Airborne Instrumentation for Astronomy VI, ed. C.~J. Evans, L.~Simard, \&
  H.~Takami, Vol. 9908, International Society for Optics and Photonics (SPIE),
  99082C, \dodoi{10.1117/12.2231488}

\bibitem[{Dey {et~al.}(2019)Dey, Schlegel, Lang, Blum, Burleigh, Fan, Findlay,
  Finkbeiner, Herrera, Juneau, Landriau, Levi, McGreer, Meisner, Myers,
  Moustakas, Nugent, Patej, Schlafly, Walker, Valdes, Weaver, Y{\`{e}}che, Zou,
  Zhou, Abareshi, Abbott, Abolfathi, Aguilera, Alam, Allen, Alvarez, Annis,
  Ansarinejad, Aubert, Beechert, Bell, BenZvi, Beutler, Bielby, Bolton,
  Brice{\~{n}}o, Buckley-Geer, Butler, Calamida, Carlberg, Carter, Casas,
  Castander, Choi, Comparat, Cukanovaite, Delubac, DeVries, Dey, Dhungana,
  Dickinson, Ding, Donaldson, Duan, Duckworth, Eftekharzadeh, Eisenstein,
  Etourneau, Fagrelius, Farihi, Fitzpatrick, Font-Ribera, Fulmer, Gänsicke,
  Gaztanaga, George, Gerdes, Gontcho, Gorgoni, Green, Guy, Harmer, Hernandez,
  Honscheid, Huang, James, Jannuzi, Jiang, Joyce, Karcher, Karkar, Kehoe,
  Jean-Paul, Kueter-Young, Lan, Lauer, Guillou, Suu, Lee, Lesser, Levasseur,
  Li, Mann, Marshall, Mart{\'{\i}}nez-V{\'{a}}zquez, Martini, du~Mas~des
  Bourboux, McManus, Meier, M{\'{e}}nard, Metcalfe,
  Mu{\~{n}}oz-Guti{\'{e}}rrez, Najita, Napier, Narayan, Newman, Nie, Nord,
  Norman, Olsen, Paat, Palanque-Delabrouille, Peng, Poppett, Poremba, Prakash,
  Rabinowitz, Raichoor, Rezaie, Robertson, Roe, Ross, Ross, Rudnick, Safonova,
  Saha, S{\'{a}}nchez, Savary, Schweiker, Scott, Seo, Shan, Silva, Slepian,
  Soto, Sprayberry, Staten, Stillman, Stupak, Summers, Tie, Tirado,
  Vargas-Maga{\~{n}}a, Vivas, Wechsler, Williams, Yang, Yang, Yapici, Zaritsky,
  Zenteno, Zhang, Zhang, Zhou, \& Zhou}]{DR9}
Dey, A., Schlegel, D.~J., Lang, D., {et~al.} 2019, The Astronomical Journal,
  157, 168, \dodoi{10.3847/1538-3881/ab089d}

\bibitem[{Dhawan {et~al.}(2019)Dhawan, Johansson, Goobar, Amanullah, Mörtsell,
  Cenko, Cooray, Fox, Goldstein, Kalender, Kasliwal, Kulkarni, Lee, Nayyeri,
  Nugent, Ofek, \& Quimby}]{dhawan19}
Dhawan, S., Johansson, J., Goobar, A., {et~al.} 2019, Monthly Notices of the
  Royal Astronomical Society, \dodoi{10.1093/mnras/stz2965}

\bibitem[{{Diehl} {et~al.}(2017){Diehl}, {Buckley-Geer}, {Lindgren}, {Nord},
  {Gaitsch}, {Gaitsch}, {Lin}, {Allam}, {Collett}, {Furlanetto}, {Gill},
  {More}, {Nightingale}, {Odden}, {Pellico}, {Tucker}, {da Costa}, {Fausti
  Neto}, {Kuropatkin}, {Soares-Santos}, {Welch}, {Zhang}, {Frieman}, {Abdalla},
  {Annis}, {Benoit-L{\'e}vy}, {Bertin}, {Brooks}, {Burke}, {Carnero Rosell},
  {Carrasco Kind}, {Carretero}, {Cunha}, {D'Andrea}, {Desai}, {Dietrich},
  {Drlica-Wagner}, {Evrard}, {Finley}, {Flaugher}, {Garc{\'\i}a-Bellido},
  {Gerdes}, {Goldstein}, {Gruen}, {Gruendl}, {Gschwend}, {Gutierrez}, {James},
  {Kuehn}, {Kuhlmann}, {Lahav}, {Li}, {Lima}, {Maia}, {Marshall}, {Menanteau},
  {Miquel}, {Nichol}, {Nugent}, {Ogando}, {Plazas}, {Reil}, {Romer}, {Sako},
  {Sanchez}, {Santiago}, {Scarpine}, {Schindler}, {Schubnell},
  {Sevilla-Noarbe}, {Sheldon}, {Smith}, {Sobreira}, {Suchyta}, {Swanson},
  {Tarle}, {Thomas}, {Walker}, \& {DES Collaboration}}]{diehl2017}
{Diehl}, H.~T., {Buckley-Geer}, E.~J., {Lindgren}, K.~A., {et~al.} 2017, The
  Astrophysical Journal Supplement, 232, 15, \dodoi{10.3847/1538-4365/aa8667}

\bibitem[{{Eldridge} {et~al.}(2013){Eldridge}, {Fraser}, {Smartt}, {Maund}, \&
  {Crockett}}]{eldridge2013}
{Eldridge}, J.~J., {Fraser}, M., {Smartt}, S.~J., {Maund}, J.~R., \&
  {Crockett}, R.~M. 2013, \mnras, 436, 774, \dodoi{10.1093/mnras/stt1612}

\bibitem[{{Fitzpatrick}(1999)}]{fitzpatrick1999}
{Fitzpatrick}, E.~L. 1999, \pasp, 111, 63, \dodoi{10.1086/316293}

\bibitem[{Flaugher {et~al.}(2015)Flaugher, Diehl, Honscheid, Abbott, Alvarez,
  Angstadt, Annis, Antonik, Ballester, Beaufore, Bernstein, Bernstein, Bigelow,
  Bonati, Boprie, Brooks, Buckley-Geer, Campa, Cardiel-Sas, Castander,
  Castilla, Cease, Cela-Ruiz, Chappa, Chi, Cooper, da~Costa, Dede, Derylo,
  DePoy, de~Vicente, Doel, Drlica-Wagner, Eiting, Elliott, Emes, Estrada, Neto,
  Finley, Flores, Frieman, Gerdes, Gladders, Gregory, Gutierrez, Hao, Holland,
  Holm, Huffman, Jackson, James, Jonas, Karcher, Karliner, Kent, Kessler,
  Kozlovsky, Kron, Kubik, Kuehn, Kuhlmann, Kuk, Lahav, Lathrop, Lee, Levi,
  Lewis, Li, Mandrichenko, Marshall, Martinez, Merritt, Miquel, Mu{\~{n} }oz,
  Neilsen, Nichol, Nord, Ogando, Olsen, Palaio, Patton, Peoples, Plazas, Rauch,
  Reil, Rheault, Roe, Rogers, Roodman, Sanchez, Scarpine, Schindler, Schmidt,
  Schmitt, Schubnell, Schultz, Schurter, Scott, Serrano, Shaw, Smith,
  Soares-Santos, Stefanik, Stuermer, Suchyta, Sypniewski, Tarle, Thaler, Tighe,
  Tran, Tucker, Walker, Wang, Watson, Weaverdyck, Wester, Woods, \&
  and}]{flaugher2015}
Flaugher, B., Diehl, H.~T., Honscheid, K., {et~al.} 2015, The Astronomical
  Journal, 150, 150, \dodoi{10.1088/0004-6256/150/5/150}

\bibitem[{Freedman(2021)}]{freedman2021}
Freedman, W.~L. 2021, The Astrophysical Journal, 919, 16,
  \dodoi{10.3847/1538-4357/ac0e95}

\bibitem[{{Gilliland} {et~al.}(1999){Gilliland}, {Nugent}, \&
  {Phillips}}]{nugent}
{Gilliland}, R.~L., {Nugent}, P.~E., \& {Phillips}, M.~M. 1999, The
  Astrophysical Journal, 521, 30, \dodoi{10.1086/307549}

\bibitem[{Goobar {et~al.}(2017)Goobar, Amanullah, Kulkarni, Nugent, Johansson,
  Steidel, Law, Mörtsell, Quimby, Blagorodnova, \& et~al.}]{goobar}
Goobar, A., Amanullah, R., Kulkarni, S.~R., {et~al.} 2017, Science, 356,
  291–295, \dodoi{10.1126/science.aal2729}

\bibitem[{{Goobar} {et~al.}(2022){Goobar}, {Johansson}, {Dhawan}, {Schulze},
  {Arendse}, {Carracedo}, {Joseph}, {Nordin}, \& {Townsend}}]{astronotegoobar}
{Goobar}, A., {Johansson}, J., {Dhawan}, S., {et~al.} 2022, Transient Name
  Server AstroNote, 180, 1

\bibitem[{Grillo {et~al.}(2020)Grillo, Rosati, Suyu, Caminha, Mercurio, \&
  Halkola}]{grillo2020}
Grillo, C., Rosati, P., Suyu, S.~H., {et~al.} 2020, The Astrophysical Journal,
  898, 87, \dodoi{10.3847/1538-4357/ab9a4c}

\bibitem[{Guy {et~al.}(2007)Guy, Astier, Baumont, Hardin, Pain, Regnault, Basa,
  Carlberg, Conley, Fabbro, Fouchez, Hook, Howell, Perrett, Pritchet, Rich,
  Sullivan, Antilogus, Aubourg, Bazin, Bronder, Filiol, Palanque-Delabrouille,
  Ripoche, \& Ruhlmann-Kleider}]{guyt2007}
Guy, J., Astier, P., Baumont, S., {et~al.} 2007, Astronomy \& Astrophysics,
  466, 11, \dodoi{10.1051/0004-6361:20066930}

\bibitem[{{Guy} {et~al.}(2010){Guy}, {Sullivan}, {Conley}, {Regnault},
  {Astier}, {Balland}, {Basa}, {Carlberg}, {Fouchez}, {Hardin}, {Hook},
  {Howell}, {Pain}, {Palanque-Delabrouille}, {Perrett}, {Pritchet}, {Rich},
  {Ruhlmann-Kleider}, {Balam}, {Baumont}, {Ellis}, {Fabbro}, {Fakhouri},
  {Fourmanoit}, {Gonz{\'a}lez-Gait{\'a}n}, {Graham}, {Hsiao}, {Kronborg},
  {Lidman}, {Mourao}, {Perlmutter}, {Ripoche}, {Suzuki}, \& {Walker}}]{guy2010}
{Guy}, J., {Sullivan}, M., {Conley}, A., {et~al.} 2010, \aap, 523, A7,
  \dodoi{10.1051/0004-6361/201014468}

\bibitem[{Harris {et~al.}(2020)Harris, Millman, van~der Walt, Gommers,
  Virtanen, Cournapeau, Wieser, Taylor, Berg, Smith, Kern, Picus, Hoyer, van
  Kerkwijk, Brett, Haldane, del R{\'{\i}}o, Wiebe, Peterson,
  G{\'{e}}rard-Marchant, Sheppard, Reddy, Weckesser, Abbasi, Gohlke, \&
  Oliphant}]{numpy}
Harris, C.~R., Millman, K.~J., van~der Walt, S.~J., {et~al.} 2020, Nature, 585,
  357, \dodoi{10.1038/s41586-020-2649-2}

\bibitem[{{Hounsell} {et~al.}(2018){Hounsell}, {Scolnic}, {Foley}, {Kessler},
  {Miranda}, {Avelino}, {Bohlin}, {Filippenko}, {Frieman}, {Jha}, {Kelly},
  {Kirshner}, {Mandel}, {Rest}, {Riess}, {Rodney}, \&
  {Strolger}}]{hounsell2017}
{Hounsell}, R., {Scolnic}, D., {Foley}, R.~J., {et~al.} 2018, The Astrophysical
  Journal, 867, 23, \dodoi{10.3847/1538-4357/aac08b}

\bibitem[{Hu {et~al.}(2022)Hu, Wang, Chen, \& Yang}]{hu2021}
Hu, L., Wang, L., Chen, X., \& Yang, J. 2022, The Astrophysical Journal, 936,
  157, \dodoi{10.3847/1538-4357/ac7394}

\bibitem[{{Huang} {et~al.}(2020){Huang}, {Storfer}, {Ravi}, {Pilon}, {Domingo},
  {Schlegel}, {Bailey}, {Dey}, {Gupta}, {Herrera}, {Juneau}, {Landriau},
  {Lang}, {Meisner}, {Moustakas}, {Myers}, {Schlafly}, {Valdes}, {Weaver},
  {Yang}, \& {Y{\`e}che}}]{huang2020}
{Huang}, X., {Storfer}, C., {Ravi}, V., {et~al.} 2020, The Astrophysical
  Journal, 894, 78, \dodoi{10.3847/1538-4357/ab7ffb}

\bibitem[{{Huang} {et~al.}(2021){Huang}, {Storfer}, {Gu}, {Ravi}, {Pilon},
  {Sheu}, {Venguswamy}, {Banka}, {Dey}, {Landriau}, {Lang}, {Meisner},
  {Moustakas}, {Myers}, {Sajith}, {Schlafly}, \& {Schlegel}}]{huang2021}
{Huang}, X., {Storfer}, C., {Gu}, A., {et~al.} 2021, The Astrophysical Journal,
  909, 27, \dodoi{10.3847/1538-4357/abd62b}

\bibitem[{Hunter(2007)}]{matplotlib}
Hunter, J.~D. 2007, Computing in Science \& Engineering, 9, 90,
  \dodoi{10.1109/MCSE.2007.55}

\bibitem[{{Ivezi{\'c}} {et~al.}(2019){Ivezi{\'c}}, {Kahn}, {Tyson}, {Abel},
  {Acosta}, {Allsman}, {Alonso}, {AlSayyad}, {Anderson}, {Andrew}, {Angel},
  {Angeli}, {Ansari}, {Antilogus}, {Araujo}, {Armstrong}, {Arndt}, {Astier},
  {Aubourg}, {Auza}, {Axelrod}, {Bard}, {Barr}, {Barrau}, {Bartlett}, {Bauer},
  {Bauman}, {Baumont}, {Bechtol}, {Bechtol}, {Becker}, {Becla}, {Beldica},
  {Bellavia}, {Bianco}, {Biswas}, {Blanc}, {Blazek}, {Blandford}, {Bloom},
  {Bogart}, {Bond}, {Booth}, {Borgland}, {Borne}, {Bosch}, {Boutigny},
  {Brackett}, {Bradshaw}, {Brandt}, {Brown}, {Bullock}, {Burchat}, {Burke},
  {Cagnoli}, {Calabrese}, {Callahan}, {Callen}, {Carlin}, {Carlson},
  {Chandrasekharan}, {Charles-Emerson}, {Chesley}, {Cheu}, {Chiang}, {Chiang},
  {Chirino}, {Chow}, {Ciardi}, {Claver}, {Cohen-Tanugi}, {Cockrum}, {Coles},
  {Connolly}, {Cook}, {Cooray}, {Covey}, {Cribbs}, {Cui}, {Cutri}, {Daly},
  {Daniel}, {Daruich}, {Daubard}, {Daues}, {Dawson}, {Delgado}, {Dellapenna},
  {de Peyster}, {de Val-Borro}, {Digel}, {Doherty}, {Dubois},
  {Dubois-Felsmann}, {Durech}, {Economou}, {Eifler}, {Eracleous}, {Emmons},
  {Fausti Neto}, {Ferguson}, {Figueroa}, {Fisher-Levine}, {Focke}, {Foss},
  {Frank}, {Freemon}, {Gangler}, {Gawiser}, {Geary}, {Gee}, {Geha}, {Gessner},
  {Gibson}, {Gilmore}, {Glanzman}, {Glick}, {Goldina}, {Goldstein}, {Goodenow},
  {Graham}, {Gressler}, {Gris}, {Guy}, {Guyonnet}, {Haller}, {Harris},
  {Hascall}, {Haupt}, {Hernandez}, {Herrmann}, {Hileman}, {Hoblitt}, {Hodgson},
  {Hogan}, {Howard}, {Huang}, {Huffer}, {Ingraham}, {Innes}, {Jacoby}, {Jain},
  {Jammes}, {Jee}, {Jenness}, {Jernigan}, {Jevremovi{\'c}}, {Johns}, {Johnson},
  {Johnson}, {Jones}, {Juramy-Gilles}, {Juri{\'c}}, {Kalirai}, {Kallivayalil},
  {Kalmbach}, {Kantor}, {Karst}, {Kasliwal}, {Kelly}, {Kessler}, {Kinnison},
  {Kirkby}, {Knox}, {Kotov}, {Krabbendam}, {Krughoff}, {Kub{\'a}nek},
  {Kuczewski}, {Kulkarni}, {Ku}, {Kurita}, {Lage}, {Lambert}, {Lange},
  {Langton}, {Le Guillou}, {Levine}, {Liang}, {Lim}, {Lintott}, {Long},
  {Lopez}, {Lotz}, {Lupton}, {Lust}, {MacArthur}, {Mahabal}, {Mandelbaum},
  {Markiewicz}, {Marsh}, {Marshall}, {Marshall}, {May}, {McKercher}, {McQueen},
  {Meyers}, {Migliore}, {Miller}, {Mills}, {Miraval}, {Moeyens}, {Moolekamp},
  {Monet}, {Moniez}, {Monkewitz}, {Montgomery}, {Morrison}, {Mueller},
  {Muller}, {Mu{\~n}oz Arancibia}, {Neill}, {Newbry}, {Nief}, {Nomerotski},
  {Nordby}, {O'Connor}, {Oliver}, {Olivier}, {Olsen}, {O'Mullane}, {Ortiz},
  {Osier}, {Owen}, {Pain}, {Palecek}, {Parejko}, {Parsons}, {Pease},
  {Peterson}, {Peterson}, {Petravick}, {Libby Petrick}, {Petry},
  {Pierfederici}, {Pietrowicz}, {Pike}, {Pinto}, {Plante}, {Plate}, {Plutchak},
  {Price}, {Prouza}, {Radeka}, {Rajagopal}, {Rasmussen}, {Regnault}, {Reil},
  {Reiss}, {Reuter}, {Ridgway}, {Riot}, {Ritz}, {Robinson}, {Roby}, {Roodman},
  {Rosing}, {Roucelle}, {Rumore}, {Russo}, {Saha}, {Sassolas}, {Schalk},
  {Schellart}, {Schindler}, {Schmidt}, {Schneider}, {Schneider}, {Schoening},
  {Schumacher}, {Schwamb}, {Sebag}, {Selvy}, {Sembroski}, {Seppala}, {Serio},
  {Serrano}, {Shaw}, {Shipsey}, {Sick}, {Silvestri}, {Slater}, {Smith},
  {Smith}, {Sobhani}, {Soldahl}, {Storrie-Lombardi}, {Stover}, {Strauss},
  {Street}, {Stubbs}, {Sullivan}, {Sweeney}, {Swinbank}, {Szalay}, {Takacs},
  {Tether}, {Thaler}, {Thayer}, {Thomas}, {Thornton}, {Thukral}, {Tice},
  {Trilling}, {Turri}, {Van Berg}, {Vanden Berk}, {Vetter}, {Virieux},
  {Vucina}, {Wahl}, {Walkowicz}, {Walsh}, {Walter}, {Wang}, {Wang}, {Warner},
  {Wiecha}, {Willman}, {Winters}, {Wittman}, {Wolff}, {Wood-Vasey}, {Wu},
  {Xin}, {Yoachim}, \& {Zhan}}]{LSST}
{Ivezi{\'c}}, {\v{Z}}., {Kahn}, S.~M., {Tyson}, J.~A., {et~al.} 2019, \apj,
  873, 111, \dodoi{10.3847/1538-4357/ab042c}

\bibitem[{Jacob {et~al.}(2010)Jacob, Katz, Berriman, Good, Laity, Deelman,
  Kesselman, Singh, Su, Prince, \& Williams}]{montage}
Jacob, J.~C., Katz, D.~S., Berriman, G.~B., {et~al.} 2010, arXiv e-prints,
  \dodoi{10.48550/ARXIV.1005.4454}

\bibitem[{{Jacobs} {et~al.}(2017){Jacobs}, {Glazebrook}, {Collett}, {More}, \&
  {McCarthy}}]{jacobs2017}
{Jacobs}, C., {Glazebrook}, K., {Collett}, T., {More}, A., \& {McCarthy}, C.
  2017, \mnras, 471, 167, \dodoi{10.1093/mnras/stx1492}

\bibitem[{{Jacobs} {et~al.}(2019){Jacobs}, {Collett}, {Glazebrook},
  {Buckley-Geer}, {Diehl}, {Lin}, {McCarthy}, {Qin}, {Odden}, {Caso Escudero},
  {Dial}, {Yung}, {Gaitsch}, {Pellico}, {Lindgren}, {Abbott}, {Annis}, {Avila},
  {Brooks}, {Burke}, {Carnero Rosell}, {Carrasco Kind}, {Carretero}, {da
  Costa}, {De Vicente}, {Fosalba}, {Frieman}, {Garc{\'\i}a-Bellido},
  {Gaztanaga}, {Goldstein}, {Gruen}, {Gruendl}, {Gschwend}, {Hollowood},
  {Honscheid}, {Hoyle}, {James}, {Krause}, {Kuropatkin}, {Lahav}, {Lima},
  {Maia}, {Marshall}, {Miquel}, {Plazas}, {Roodman}, {Sanchez}, {Scarpine},
  {Serrano}, {Sevilla-Noarbe}, {Smith}, {Sobreira}, {Suchyta}, {Swanson},
  {Tarle}, {Vikram}, {Walker}, {Zhang}, \& {DES Collaboration}}]{jacobs2019}
{Jacobs}, C., {Collett}, T., {Glazebrook}, K., {et~al.} 2019, The Astrophysical
  Journal Supplement, 243, 17, \dodoi{10.3847/1538-4365/ab26b6}

\bibitem[{{Kelly} {et~al.}(2022){Kelly}, {Zitrin}, {Oguri}, {Diego},
  {Williams}, {Broadhurst}, {Chen}, {Koekemoer}, {Pierel}, {Strolger}, \&
  {Treu}}]{astronotekelly}
{Kelly}, P., {Zitrin}, A., {Oguri}, M., {et~al.} 2022, Transient Name Server
  AstroNote, 169, 1

\bibitem[{Kelly {et~al.}(2015)Kelly, Rodney, Treu, Foley, Brammer, Schmidt,
  Zitrin, Sonnenfeld, Strolger, Graur, Filippenko, Jha, Riess, Bradac, Weiner,
  Scolnic, Malkan, von~der Linden, Trenti, Hjorth, Gavazzi, Fontana, Merten,
  McCully, Jones, Postman, Dressler, Patel, Cenko, Graham, \&
  Tucker}]{kelly2015}
Kelly, P.~L., Rodney, S.~A., Treu, T., {et~al.} 2015, Science, 347, 1123,
  \dodoi{10.1126/science.aaa3350}

\bibitem[{{Kelly} {et~al.}(2016){Kelly}, {Rodney}, {Treu}, {Strolger}, {Foley},
  {Jha}, {Selsing}, {Brammer}, {Brada{\v{c}}}, {Cenko}, {Graur}, {Filippenko},
  {Hjorth}, {McCully}, {Molino}, {Nonino}, {Riess}, {Schmidt}, {Tucker}, {von
  der Linden}, {Weiner}, \& {Zitrin}}]{kelly2016}
{Kelly}, P.~L., {Rodney}, S.~A., {Treu}, T., {et~al.} 2016, The Astrophysical
  Journal Letters, 819, L8, \dodoi{10.3847/2041-8205/819/1/L8}

\bibitem[{Kelly {et~al.}(2017)Kelly, Diego, Rodney, Kaiser, Broadhurst, Zitrin,
  Treu, Perez-Gonzalez, Morishita, Jauzac, Selsing, Oguri, Pueyo, Ross,
  Filippenko, Smith, Hjorth, Cenko, Wang, Howell, Richard, Frye, Jha, Foley,
  Norman, Bradac, Zheng, Brammer, Benito, Cava, Christensen, de~Mink, Graur,
  Grillo, Kawamata, Kneib, Matheson, McCully, Nonino, Perez-Fournon, Riess,
  Rosati, Schmidt, Sharon, \& Weiner}]{kelly2018}
Kelly, P.~L., Diego, J.~M., Rodney, S., {et~al.} 2017, Extreme magnification of
  a star at redshift 1.5 by a galaxy-cluster lens,  arXiv,
  \dodoi{10.48550/ARXIV.1706.10279}

\bibitem[{Kenworthy {et~al.}(2021)Kenworthy, Jones, Dai, Kessler, Scolnic,
  Brout, Siebert, Pierel, Dettman, Dimitriadis, Foley, Jha, Pan, Riess, Rodney,
  \& Rojas-Bravo}]{salt3}
Kenworthy, W.~D., Jones, D.~O., Dai, M., {et~al.} 2021, The Astrophysical
  Journal, 923, 265, \dodoi{10.3847/1538-4357/ac30d8}

\bibitem[{{Lang} {et~al.}(2016){Lang}, {Hogg}, \& {Mykytyn}}]{lang2016}
{Lang}, D., {Hogg}, D.~W., \& {Mykytyn}, D. 2016, {The Tractor: Probabilistic
  astronomical source detection and measurement}, Astrophysics Source Code
  Library, record ascl:1604.008.
\newblock \doeprint{1604.008}

\bibitem[{{Levan} {et~al.}(2005){Levan}, {Nugent}, {Fruchter}, {Burud},
  {Branch}, {Rhoads}, {Castro-Tirado}, {Gorosabel}, {Castro Cer{\'o}n},
  {Thorsett}, {Kouveliotou}, {Golenetskii}, {Fynbo}, {Garnavich}, {Holland},
  {Hjorth}, {M{\o}ller}, {Pian}, {Tanvir}, {Ulanov}, {Wijers}, \&
  {Woosley}}]{levan2005}
{Levan}, A., {Nugent}, P., {Fruchter}, A., {et~al.} 2005, The Astrophysical
  Journal, 624, 880, \dodoi{10.1086/428657}

\bibitem[{Madau \& Dickinson(2014)}]{madau2014}
Madau, P., \& Dickinson, M. 2014, Annual Review of Astronomy and Astrophysics,
  52, 415, \dodoi{10.1146/annurev-astro-081811-125615}

\bibitem[{{Moustakas}(2012)}]{moustakas2012}
{Moustakas}, L. 2012, in 10th Hellenic Astronomical Conference, ed.
  I.~{Papadakis} \& A.~{Anastasiadis}, 14--14

\bibitem[{Narayan \& Bartelmann(1996)}]{naryanandbart96}
Narayan, R., \& Bartelmann, M. 1996, Lectures on Gravitational Lensing,  arXiv,
  \dodoi{10.48550/ARXIV.ASTRO-PH/9606001}

\bibitem[{Oguri \& Marshall(2010)}]{oguri2010}
Oguri, M., \& Marshall, P.~J. 2010, Monthly Notices of the Royal Astronomical
  Society, 405, 2579, \dodoi{10.1111/j.1365-2966.2010.16639.x}

\bibitem[{{Perlmutter} {et~al.}(1999){Perlmutter}, {Aldering}, {Goldhaber},
  {Knop}, {Nugent}, {Castro}, {Deustua}, {Fabbro}, {Goobar}, {Groom}, {Hook},
  {Kim}, {Kim}, {Lee}, {Nunes}, {Pain}, {Pennypacker}, {Quimby}, {Lidman},
  {Ellis}, {Irwin}, {McMahon}, {Ruiz-Lapuente}, {Walton}, {Schaefer}, {Boyle},
  {Filippenko}, {Matheson}, {Fruchter}, {Panagia}, {Newberg}, {Couch}, \&
  {Project}}]{perlmutter99}
{Perlmutter}, S., {Aldering}, G., {Goldhaber}, G., {et~al.} 1999, The
  Astrophysical Journal, 517, 565, \dodoi{10.1086/307221}

\bibitem[{Pierel {et~al.}(2022)Pierel, Arendse, Ertl, Huang, Moustakas,
  Schuldt, Shajib, Shu, Birrer, Bronikowski, Hjorth, Suyu, Agarwal, Agnello,
  Bolton, Chakrabarti, Cold, Courbin, Della~Costa, Dhawan, Engesser, Fox, Gall,
  Gomez, Goobar, Jimenez, Johansson, Li, Marques-Chaves, Mao, Mazzali,
  Perez-Fournon, Petrushevska, Poidevin, Rest, Sheu, Shirley, Silver, Storfer,
  Treu, Wojtak, \& Zenati}]{zwicky_pierel}
Pierel, J. D.~R., Arendse, N., Ertl, S., {et~al.} 2022, arXiv e-prints,
  \dodoi{10.48550/ARXIV.2211.03772}

\bibitem[{{Planck Collaboration} {et~al.}(2020){Planck Collaboration},
  {Aghanim}, {Akrami}, {Ashdown}, {Aumont}, {Baccigalupi}, {Ballardini},
  {Banday}, {Barreiro}, {Bartolo}, {Basak}, {Battye}, {Benabed}, {Bernard},
  {Bersanelli}, {Bielewicz}, {Bock}, {Bond}, {Borrill}, {Bouchet}, {Boulanger},
  {Bucher}, {Burigana}, {Butler}, {Calabrese}, {Cardoso}, {Carron},
  {Challinor}, {Chiang}, {Chluba}, {Colombo}, {Combet}, {Contreras}, {Crill},
  {Cuttaia}, {de Bernardis}, {de Zotti}, {Delabrouille}, {Delouis}, {Di
  Valentino}, {Diego}, {Dor{\'e}}, {Douspis}, {Ducout}, {Dupac}, {Dusini},
  {Efstathiou}, {Elsner}, {En{\ss}lin}, {Eriksen}, {Fantaye}, {Farhang},
  {Fergusson}, {Fernandez-Cobos}, {Finelli}, {Forastieri}, {Frailis},
  {Fraisse}, {Franceschi}, {Frolov}, {Galeotta}, {Galli}, {Ganga},
  {G{\'e}nova-Santos}, {Gerbino}, {Ghosh}, {Gonz{\'a}lez-Nuevo}, {G{\'o}rski},
  {Gratton}, {Gruppuso}, {Gudmundsson}, {Hamann}, {Handley}, {Hansen},
  {Herranz}, {Hildebrandt}, {Hivon}, {Huang}, {Jaffe}, {Jones}, {Karakci},
  {Keih{\"a}nen}, {Keskitalo}, {Kiiveri}, {Kim}, {Kisner}, {Knox},
  {Krachmalnicoff}, {Kunz}, {Kurki-Suonio}, {Lagache}, {Lamarre}, {Lasenby},
  {Lattanzi}, {Lawrence}, {Le Jeune}, {Lemos}, {Lesgourgues}, {Levrier},
  {Lewis}, {Liguori}, {Lilje}, {Lilley}, {Lindholm}, {L{\'o}pez-Caniego},
  {Lubin}, {Ma}, {Mac{\'\i}as-P{\'e}rez}, {Maggio}, {Maino}, {Mandolesi},
  {Mangilli}, {Marcos-Caballero}, {Maris}, {Martin}, {Martinelli},
  {Mart{\'\i}nez-Gonz{\'a}lez}, {Matarrese}, {Mauri}, {McEwen}, {Meinhold},
  {Melchiorri}, {Mennella}, {Migliaccio}, {Millea}, {Mitra},
  {Miville-Desch{\^e}nes}, {Molinari}, {Montier}, {Morgante}, {Moss}, {Natoli},
  {N{\o}rgaard-Nielsen}, {Pagano}, {Paoletti}, {Partridge}, {Patanchon},
  {Peiris}, {Perrotta}, {Pettorino}, {Piacentini}, {Polastri}, {Polenta},
  {Puget}, {Rachen}, {Reinecke}, {Remazeilles}, {Renzi}, {Rocha}, {Rosset},
  {Roudier}, {Rubi{\~n}o-Mart{\'\i}n}, {Ruiz-Granados}, {Salvati}, {Sandri},
  {Savelainen}, {Scott}, {Shellard}, {Sirignano}, {Sirri}, {Spencer},
  {Sunyaev}, {Suur-Uski}, {Tauber}, {Tavagnacco}, {Tenti}, {Toffolatti},
  {Tomasi}, {Trombetti}, {Valenziano}, {Valiviita}, {Van Tent}, {Vibert},
  {Vielva}, {Villa}, {Vittorio}, {Wandelt}, {Wehus}, {White}, {White},
  {Zacchei}, \& {Zonca}}]{planck2018}
{Planck Collaboration}, {Aghanim}, N., {Akrami}, Y., {et~al.} 2020, \aap, 641,
  A6, \dodoi{10.1051/0004-6361/201833910}

\bibitem[{{Pourrahmani} {et~al.}(2018){Pourrahmani}, {Nayyeri}, \&
  {Cooray}}]{pourrahmani2018}
{Pourrahmani}, M., {Nayyeri}, H., \& {Cooray}, A. 2018, The Astrophysical
  Journal, 856, 68, \dodoi{10.3847/1538-4357/aaae6a}

\bibitem[{Quimby {et~al.}(2014)Quimby, Oguri, More, More, Moriya, Werner,
  Tanaka, Folatelli, Bersten, Maeda, \& Nomoto}]{quimby2014}
Quimby, R.~M., Oguri, M., More, A., {et~al.} 2014, Science, 344, 396,
  \dodoi{10.1126/science.1250903}

\bibitem[{Richardson {et~al.}(2014)Richardson, III, Wright, \&
  Maddox}]{richardson2014}
Richardson, D., III, R. L.~J., Wright, J., \& Maddox, L. 2014, The Astronomical
  Journal, 147, 118, \dodoi{10.1088/0004-6256/147/5/118}

\bibitem[{Riess {et~al.}(2021)Riess, Casertano, Yuan, Bowers, Macri, Zinn, \&
  Scolnic}]{reiss2021}
Riess, A.~G., Casertano, S., Yuan, W., {et~al.} 2021, The Astrophysical Journal
  Letters, 908, L6, \dodoi{10.3847/2041-8213/abdbaf}

\bibitem[{{Riess} {et~al.}(1998){Riess}, {Filippenko}, {Challis},
  {Clocchiatti}, {Diercks}, {Garnavich}, {Gilliland}, {Hogan}, {Jha},
  {Kirshner}, {Leibundgut}, {Phillips}, {Reiss}, {Schmidt}, {Schommer},
  {Smith}, {Spyromilio}, {Stubbs}, {Suntzeff}, \& {Tonry}}]{riess98}
{Riess}, A.~G., {Filippenko}, A.~V., {Challis}, P., {et~al.} 1998, \aj, 116,
  1009, \dodoi{10.1086/300499}

\bibitem[{Rodney {et~al.}(2021)Rodney, Brammer, Pierel, Richard, Toft,
  O'Connor, Akhshik, \& Whitaker}]{rodney2021}
Rodney, S.~A., Brammer, G.~B., Pierel, J. D.~R., {et~al.} 2021, Nature
  Astronomy

\bibitem[{Rodney {et~al.}(2016)Rodney, Strolger, Kelly, Brada{\v{c} }, Brammer,
  Filippenko, Foley, Graur, Hjorth, Jha, McCully, Molino, Riess, Schmidt,
  Selsing, Sharon, Treu, Weiner, \& Zitrin}]{rodney2016}
Rodney, S.~A., Strolger, L.-G., Kelly, P.~L., {et~al.} 2016, The Astrophysical
  Journal, 820, 50, \dodoi{10.3847/0004-637x/820/1/50}

\bibitem[{Sako {et~al.}(2018)Sako, Bassett, Becker, Brown, Campbell, Wolf,
  Cinabro, D'Andrea, Dawson, DeJongh, Depoy, Dilday, Doi, Filippenko, Fischer,
  Foley, Frieman, Galbany, Garnavich, Goobar, Gupta, Hill, Hayden, Hlozek,
  Holtzman, Hopp, Jha, Kessler, Kollatschny, Leloudas, Marriner, Marshall,
  Miquel, Morokuma, Mosher, Nichol, Nordin, Olmstead, Östman, Prieto,
  Richmond, Romani, Sollerman, Stritzinger, Schneider, Smith, Wheeler, Yasuda,
  \& Zheng}]{sako2014}
Sako, M., Bassett, B., Becker, A.~C., {et~al.} 2018, Publications of the
  Astronomical Society of the Pacific, 130, 064002,
  \dodoi{10.1088/1538-3873/aab4e0}

\bibitem[{{Schlafly} \& {Finkbeiner}(2011)}]{schlafly2011}
{Schlafly}, E.~F., \& {Finkbeiner}, D.~P. 2011, The Astrophysical Journal, 737,
  103, \dodoi{10.1088/0004-637X/737/2/103}

\bibitem[{Scolnic {et~al.}(2021)Scolnic, Brout, Carr, Riess, Davis, Dwomoh,
  Jones, Ali, Charvu, Chen, Peterson, Popovic, Rose, Wood, Brown, Chambers,
  Coulter, Dettman, Dimitriadis, Filippenko, Foley, Jha, Kilpatrick, Kirshner,
  Pan, Rest, Rojas-Bravo, Siebert, Stahl, \& Zheng}]{scolnic2022}
Scolnic, D., Brout, D., Carr, A., {et~al.} 2021, The Pantheon+ Analysis: The
  Full Dataset and Light-Curve Release,  arXiv,
  \dodoi{10.48550/ARXIV.2112.03863}

\bibitem[{{Shu} {et~al.}(2018){Shu}, {Bolton}, {Mao}, {Kang}, {Li}, \&
  {Soraisam}}]{shu2018}
{Shu}, Y., {Bolton}, A.~S., {Mao}, S., {et~al.} 2018, The Astrophysical
  Journal, 864, 91, \dodoi{10.3847/1538-4357/aad5ea}

\bibitem[{Shu {et~al.}(2021)Shu, Bolton, Mao, Kang, Li, \& Soraisam}]{shu2021}
Shu, Y., Bolton, A.~S., Mao, S., {et~al.} 2021, The Astrophysical Journal, 919,
  67, \dodoi{10.3847/1538-4357/ac24a4}

\bibitem[{Shu {et~al.}(2022)Shu, Ca{\~{n}}ameras, Schuldt, Suyu, Taubenberger,
  Inoue, \& Jaelani}]{shu2022}
Shu, Y., Ca{\~{n}}ameras, R., Schuldt, S., {et~al.} 2022, Astronomy \&
  Astrophysics, 662, A4, \dodoi{10.1051/0004-6361/202243203}

\bibitem[{Shu {et~al.}(2017)Shu, Brownstein, Bolton, Koopmans, Treu,
  Montero-Dorta, Auger, Czoske, Gavazzi, Marshall, \& Moustakas}]{shu2017}
Shu, Y., Brownstein, J.~R., Bolton, A.~S., {et~al.} 2017, The Astrophysical
  Journal, 851, 48, \dodoi{10.3847/1538-4357/aa9794}

\bibitem[{Smit {et~al.}(2012)Smit, Bouwens, Franx, Illingworth, Labb{\'{e} },
  Oesch, \& van Dokkum}]{smit2012}
Smit, R., Bouwens, R.~J., Franx, M., {et~al.} 2012, The Astrophysical Journal,
  756, 14, \dodoi{10.1088/0004-637x/756/1/14}

\bibitem[{Smith {et~al.}(2020)Smith, D'Andrea, Sullivan, Möller, Nichol,
  Thomas, Kim, Sako, Castander, Filippenko, Foley, Galbany, Gonz{\'{a}
  }lez-Gait{\'{a}}n, Kasai, Kirshner, Lidman, Scolnic, Brout, Davis, Gupta,
  Hinton, Kessler, Lasker, Macaulay, Wolf, Zhang, Asorey, Avelino, Bassett,
  Calcino, Carollo, Casas, Challis, Childress, Clocchiatti, Crawford,
  Frohmaier, Glazebrook, Goldstein, Graham, Hoormann, Kuehn, Lewis, Mandel,
  Morganson, Muthukrishna, Nugent, Pan, Pursiainen, Sharp, Sommer, Swann,
  Thomas, Tucker, Uddin, Wiseman, Zheng, Abbott, Annis, Avila, Bechtol,
  Bernstein, Bertin, Brooks, Burke, Rosell, Kind, Carretero, Cunha, da~Costa,
  Davis, Vicente, Diehl, Eifler, Estrada, Frieman, Garc{\'{\i}}a-Bellido,
  Gaztanaga, Gerdes, Gruen, Gruendl, Gschwend, Gutierrez, Hartley, Hollowood,
  Honscheid, Hoyle, James, Johnson, Johnson, Kuropatkin, Li, Lima, Maia, March,
  Marshall, Martini, Menanteau, Miller, Miquel, Neilsen, Ogando, Plazas, Romer,
  Sanchez, Scarpine, Schubnell, Serrano, Sevilla-Noarbe, Soares-Santos,
  Sobreira, Suchyta, Tarle, Tucker, \& Wester}]{Andrea2018}
Smith, M., D'Andrea, C.~B., Sullivan, M., {et~al.} 2020, The Astronomical
  Journal, 160, 267, \dodoi{10.3847/1538-3881/abc01b}

\bibitem[{Sobral {et~al.}(2012)Sobral, Smail, Best, Geach, Matsuda, Stott,
  Cirasuolo, \& Kurk}]{sobral2012}
Sobral, D., Smail, I., Best, P.~N., {et~al.} 2012, Monthly Notices of the Royal
  Astronomical Society, 428, 1128, \dodoi{10.1093/mnras/sts096}

\bibitem[{{Sonnenfeld} \& {Leauthaud}(2018)}]{sonnenfeld2018}
{Sonnenfeld}, A., \& {Leauthaud}, A. 2018, \mnras, 477, 5460,
  \dodoi{10.1093/mnras/sty935}

\bibitem[{{Spergel} {et~al.}(2015){Spergel}, {Gehrels}, {Baltay}, {Bennett},
  {Breckinridge}, {Donahue}, {Dressler}, {Gaudi}, {Greene}, {Guyon}, {Hirata},
  {Kalirai}, {Kasdin}, {Macintosh}, {Moos}, {Perlmutter}, {Postman},
  {Rauscher}, {Rhodes}, {Wang}, {Weinberg}, {Benford}, {Hudson}, {Jeong},
  {Mellier}, {Traub}, {Yamada}, {Capak}, {Colbert}, {Masters}, {Penny},
  {Savransky}, {Stern}, {Zimmerman}, {Barry}, {Bartusek}, {Carpenter}, {Cheng},
  {Content}, {Dekens}, {Demers}, {Grady}, {Jackson}, {Kuan}, {Kruk}, {Melton},
  {Nemati}, {Parvin}, {Poberezhskiy}, {Peddie}, {Ruffa}, {Wallace}, {Whipple},
  {Wollack}, \& {Zhao}}]{Roman}
{Spergel}, D., {Gehrels}, N., {Baltay}, C., {et~al.} 2015, arXiv e-prints,
  arXiv:1503.03757.
\newblock \doarXiv{1503.03757}

\bibitem[{Stein {et~al.}(2022)Stein, Blaum, Harrington, Medan, \&
  Luki{\'{c}}}]{stein2022}
Stein, G., Blaum, J., Harrington, P., Medan, T., \& Luki{\'{c}}, Z. 2022, The
  Astrophysical Journal, 932, 107, \dodoi{10.3847/1538-4357/ac6d63}

\bibitem[{Storfer {et~al.}(2022)Storfer, Huang, Gu, Sheu, Banka, Dey, Jain,
  Kwon, Lang, Lee, Meisner, Moustakas, Myers, Tabares-Tarquinio, Schlafly, \&
  Schlegel}]{storfer2022}
Storfer, C., Huang, X., Gu, A., {et~al.} 2022, arXiv e-prints,
  \dodoi{10.48550/ARXIV.2206.02764}

\bibitem[{Suyu {et~al.}(2020)Suyu, Huber, Ca{\~{n} }ameras, Kromer, Schuldt,
  Taubenberger, Y{\i}ld{\i}r{\i}m, Bonvin, Chan, Courbin, Nöbauer, Sim, \&
  Sluse}]{holismokes}
Suyu, S.~H., Huber, S., Ca{\~{n} }ameras, R., {et~al.} 2020, Astronomy \&
  Astrophysics, 644, A162, \dodoi{10.1051/0004-6361/202037757}

\bibitem[{Suzuki {et~al.}(2012)Suzuki, Rubin, Lidman, Aldering, Amanullah,
  Barbary, Barrientos, Botyanszki, Brodwin, Connolly, Dawson, Dey, Doi,
  Donahue, Deustua, Eisenhardt, Ellingson, Faccioli, Fadeyev, Fakhouri,
  Fruchter, Gilbank, Gladders, Goldhaber, Gonzalez, Goobar, Gude, Hattori,
  Hoekstra, Hsiao, Huang, Ihara, Jee, Johnston, Kashikawa, Koester, Konishi,
  Kowalski, Linder, Lubin, Melbourne, Meyers, Morokuma, Munshi, Mullis, Oda,
  Panagia, Perlmutter, Postman, Pritchard, Rhodes, Ripoche, Rosati, Schlegel,
  Spadafora, Stanford, Stanishev, Stern, Strovink, Takanashi, Tokita, Wagner,
  Wang, Yasuda, \& and}]{suzuki2012}
Suzuki, N., Rubin, D., Lidman, C., {et~al.} 2012, The Astrophysical Journal,
  746, 85, \dodoi{10.1088/0004-637x/746/1/85}

\bibitem[{Vincenzi {et~al.}(2019)Vincenzi, Sullivan, Firth, Guti{\'{e} }rrez,
  Frohmaier, Smith, Angus, \& Nichol}]{vincenzi2019}
Vincenzi, M., Sullivan, M., Firth, R.~E., {et~al.} 2019, Monthly Notices of the
  Royal Astronomical Society, 489, 5802, \dodoi{10.1093/mnras/stz2448}

\bibitem[{Welch {et~al.}(2022)Welch, Coe, Diego, Zitrin, Zackrisson, Dimauro,
  Jim{\'{e}}nez-Teja, Kelly, Mahler, Oguri, Timmes, Windhorst, Florian,
  de~Mink, Avila, Anderson, Bradley, Sharon, Vikaeus, McCandliss, Brada{\v{c}},
  Rigby, Frye, Toft, Strait, Trenti, Sharma, Andrade-Santos, \&
  Broadhurst}]{welch2022}
Welch, B., Coe, D., Diego, J.~M., {et~al.} 2022, Nature, 603, 815,
  \dodoi{10.1038/s41586-022-04449-y}

\bibitem[{Williams {et~al.}(2004)Williams, Olszewski, Lesser, \&
  Burge}]{williams2004}
Williams, G.~G., Olszewski, E., Lesser, M.~P., \& Burge, J.~H. 2004, in
  Ground-based Instrumentation for Astronomy, ed. A.~F.~M. Moorwood \& M.~Iye,
  Vol. 5492, International Society for Optics and Photonics (SPIE), 787 -- 798,
  \dodoi{10.1117/12.552189}

\bibitem[{Wong {et~al.}(2018)Wong, Sonnenfeld, Chan, Rusu, Tanaka, Jaelani,
  Lee, More, Oguri, Suyu, \& Komiyama}]{wong2018}
Wong, K.~C., Sonnenfeld, A., Chan, J. H.~H., {et~al.} 2018, The Astrophysical
  Journal, 867, 107, \dodoi{10.3847/1538-4357/aae381}

\bibitem[{Wong {et~al.}(2019)Wong, Suyu, Chen, Rusu, Millon, Sluse, Bonvin,
  Fassnacht, Taubenberger, Auger, Birrer, Chan, Courbin, Hilbert, Tihhonova,
  Treu, Agnello, Ding, Jee, Komatsu, Shajib, Sonnenfeld, Blandford, Koopmans,
  Marshall, \& Meylan}]{wong2019}
Wong, K.~C., Suyu, S.~H., Chen, G. C.-F., {et~al.} 2019, Monthly Notices of the
  Royal Astronomical Society, 498, 1420, \dodoi{10.1093/mnras/stz3094}

\bibitem[{Zackay {et~al.}(2016)Zackay, Ofek, \& Gal-Yam}]{zogy}
Zackay, B., Ofek, E.~O., \& Gal-Yam, A. 2016, The Astrophysical Journal, 830,
  27, \dodoi{10.3847/0004-637x/830/1/27}

\bibitem[{Zhou {et~al.}(2020)Zhou, Newman, Mao, Meisner, Moustakas, Myers,
  Prakash, Zentner, Brooks, Duan, Landriau, Levi, Prada, \& Tarle}]{zhou2020}
Zhou, R., Newman, J.~A., Mao, Y.-Y., {et~al.} 2020, Monthly Notices of the
  Royal Astronomical Society, 501, 3309, \dodoi{10.1093/mnras/staa3764}

\end{thebibliography}
\bibliographystyle{aasjournal}



\end{document}